\documentclass[12pt,preprint]{aastex}

\usepackage{epsfig}
\usepackage{amsmath}
\usepackage{natbib}
\usepackage{longtable}
\usepackage{graphicx}
\usepackage{epstopdf}

\usepackage{natbib}
\usepackage{hyperref}

\shorttitle{Ultraviolet, Optical and Near-infrared Light Curves of SN~2012fr}
\shortauthors{Contreras et al.}

\newcommand{\dm}{\Delta{\rm m}_{15}}
\newcommand{\dmB}{\Delta{\rm m}_{15}(B)}
\newcommand{\tmax}{$t_{B_{max}}$}
\newcommand{\kms}{${\rm km~s}^{-1}$}
\begin{document}

\title{SN 2012fr: Ultraviolet, Optical, and Near-Infrared Light Curves of a Type~Ia Supernova Observed 
Within a Day of Explosion\altaffilmark{1}}

\author{
Carlos~Contreras\altaffilmark{2,3},
M.~M.~Phillips\altaffilmark{3},
Christopher~R.~Burns\altaffilmark{4},
Anthony~L.~Piro\altaffilmark{4},
B.~J.~Shappee\altaffilmark{4},
Maximilian~D.~Stritzinger\altaffilmark{2},
C.~Baltay\altaffilmark{5},
Peter J. Brown,\altaffilmark{6},
Emmanuel Conseil\altaffilmark{7},
Alain Klotz\altaffilmark{8},
Peter~E.~Nugent\altaffilmark{9, 10},
Damien Turpin\altaffilmark{8},
Stu~Parker\altaffilmark{11},
D.~Rabinowitz\altaffilmark{5},
Eric~Y.~Hsiao\altaffilmark{2,3,12},
Nidia~Morrell\altaffilmark{3},
Abdo~Campillay\altaffilmark{3, 13},
Sergio~Castell\'on\altaffilmark{3},
Carlos~Corco\altaffilmark{3, 14},
Consuelo~Gonz\'{a}lez\altaffilmark{3},
Kevin Krisciunas\altaffilmark{6},
Jacqueline~Ser\'{o}n,\altaffilmark{3,15},
Brad~E.~Tucker\altaffilmark{16,17},
E.~S.~Walker\altaffilmark{5,18},
E.~Baron\altaffilmark{19},
C.~Cain\altaffilmark{19},
Michael~J.~Childress\altaffilmark{16,17,20},
Gast\'on~Folatelli\altaffilmark{21},
Wendy~L.~Freedman\altaffilmark{4,22},
Mario~Hamuy\altaffilmark{23},
P.~Hoeflich\altaffilmark{12},
S.~E.~Persson\altaffilmark{4},
Richard~Scalzo\altaffilmark{16,17,24},
Brian~Schmidt\altaffilmark{16},
Nicholas~B.~Suntzeff\altaffilmark{6}
}

\altaffiltext{1}{This paper includes data gathered with the 6.5 m Magellan  Baade Telescope, located at Las Campanas Observatory, Chile.}
\altaffiltext{2}{Department of Physics and Astronomy, Aarhus University, Ny Munkegade 120, DK-8000 Aarhus C, Denmark.}
\altaffiltext{3}{Las Campanas Observatory, Carnegie Observatories, Casilla 601, La Serena, Chile;
{carlos.astro@gmail.com}.}
\altaffiltext{4}{Observatories of the Carnegie Institution for
 Science, 813 Santa Barbara St., Pasadena, CA 91101, USA}
\altaffiltext{5}{Physics Department, Yale University, 217 Prospect Street, New Haven, CT 06511-8499, USA}
\altaffiltext{6}{George P. and Cynthia Woods Mitchell Institute for Fundamental Physics and Astronomy, 
Department of Physics and Astronomy, Texas A\&M University, College Station, TX 77843, USA}
\altaffiltext{7}{Association Fran\c{c}aise des Observateurs d'\'{E}toiles Variables (AFOEV), 11, rue de l'universit\'{e} 
67000 Strasbourg, France}
\altaffiltext{8}{Universit\'{e} de Toulouse, UPS-OMP, IRAP, 14, avenue Edouard Belin, F-31400 Toulouse, France}
\altaffiltext{9}{Lawrence Berkeley National Laboratory, Department of Physics, 1 Cyclotron Road, Berkeley, CA 94720, USA}
\altaffiltext{10}{Astronomy Department, University of California at Berkeley, Berkeley, CA 94720, USA}
\altaffiltext{11}{Parkdale Observatory, Backyard Observatory Supernova Search (BOSS), \url{http://bosssupernova.com/}}
\altaffiltext{12}{Department of Physics, Florida State University, Tallahassee, FL 32306, USA}
\altaffiltext{13}{Departamento de F\'{i}sica y Astronom\'{i}a, Universidad de La Serena, Av. Cisternas 1200, La Serena, Chile}
\altaffiltext{14}{SOAR Telescope, Casilla 603, La Serena, Chile}
\altaffiltext{15}{Cerro Tololo Interamerican Observatory, Casilla 603, La Serena, Chile}
\altaffiltext{16}{Research School of Astronomy and Astrophysics, Australian National University, Canberra, ACT 2611, Australia}
\altaffiltext{17}{ARC Centre of Excellence for All-sky Astrophysics (CAASTRO)}
\altaffiltext{18}{Qriously Corporation, 1 Hartwick St, London EC1R4RB, U.K.}
\altaffiltext{19}{Homer L. Dodge Department of Physics and Astronomy, 440 W. Brooks St., Rm 100, Norman, OK 73019, USA}
\altaffiltext{20}{School of Physics and Astronomy, University of Southampton, Southampton, SO17 1BJ, UK}
\altaffiltext{21}{Facultad de Ciencias Astron\'{o}micas y Geof\'{i}sicas, Universidad Nacional de La Plata, 
Instituto de Astrof\'{i}sica de La Plata (IALP), CONICET, Paseo del Bosque S/N, B1900FWA La Plata, 
Argentina}
\altaffiltext{22}{Department of Astronomy and Astrophysics, University of Chicago, 5640 South Ellis Avenue, Chicago, IL 60637, USA}
\altaffiltext{23}{Departamento de Astronom\'ia, Universidad de Chile, Casilla 36-D, Santiago, Chile}
\altaffiltext{24}{Centre for Translational Data Science, University of Sydney, Darlington, NSW 2008, Australia}

\begin{abstract}
We present detailed ultraviolet, optical  and near-infrared
light curves of the Type~Ia supernova (SN) 2012fr, which
exploded in the Fornax cluster member NGC~1365.
These precise high-cadence light curves provide a dense
coverage of the flux evolution from $-$12 to $+$140 days with respect
to the epoch of $B$-band maximum (\tmax).
Supplementary imaging at the earliest epochs reveals an
initial slow, nearly linear rise in luminosity with a duration of $\sim$2.5~days, followed by a faster rising phase that is well
reproduced by an explosion model with a moderate
amount of $^{56}$Ni mixing in the ejecta.
From an analysis of the light curves, we conclude: $(i)$ explosion
occurred $< 22$~hours before the first detection of the
supernova, $(ii)$ the rise time to peak
bolometric ($\lambda > 1800~$\AA) luminosity was $16.5 \pm 0.6$ days, $(iii)$ the
supernova suffered little or no host-galaxy dust reddening,
$(iv)$ the peak luminosity in both the optical and near-infrared was
consistent with the bright end of normal Type~Ia diversity, and
$(v)$ $0.60 \pm 0.15~M_{\odot}$ of $^{56}$Ni was
synthesized in the explosion.
Despite its normal luminosity, SN~2012fr displayed unusually prevalent
high-velocity \ion{Ca}{2} and \ion{Si}{2} absorption features,
and a nearly constant photospheric velocity of the
\ion{Si}{2} $\lambda$6355 line at $\sim$12,000~\kms\ beginning
$\sim$5~days before \tmax.
Other peculiarities in the early phase photometry and
the spectral evolution are highlighted.
SN~2012fr also adds to a growing number of Type~Ia supernovae
hosted by galaxies with direct  Cepheid distance measurements.

\end{abstract}

\section{INTRODUCTION}
\label{sec:intro}

Type~Ia supernovae (SNe~Ia) are the major producers of iron in the Galaxy
\citep{tinsley79}, and thus are intimately tied to its chemical evolution 
\citep[e.g., see][]{mcwilliam97}.  The fact that they are among the brightest and most homogenous 
of the supernovae (SNe) has  allowed them to be used with great success as
distance indicators for measuring the expansion history of the Universe
\citep[e.g., see][]{betoule14}.  Nevertheless, progress has remained disappointingly 
slow in identifying the progenitor systems of these objects and understanding the details 
of their explosion mechanisms. More than 50~years ago \citet{hoyle60} recognized
that SNe~Ia were thermonuclear disruptions of a 
white dwarfs in binary systems, but there is still considerable debate as to whether 
the companion is a main sequence or giant star (``single-degenerate'' model) or another 
white dwarf (``double-degenerate'' model).  The collision of two white dwarfs (as opposed
to the merger) has recently garnered interest as yet a third possible way of producing
SNe~Ia \citep{raskin09,rosswog09,thompson11,kushnir13}. As to the explosion mechanism, the
deflagration (subsonic burning) of a Chandrasekhar mass white dwarf
which at some point transitions to a supersonic detonation \citep{khokhlov91} provides the best
match to the observational properties of SNe~Ia \citep{hoeflich96,hoeflich96b,hoeflich17}, 
but the details of how or why this occurs are still a mystery.
On the other hand, detonations of sub-Chandrasekhar-mass white dwarfs also have many attractive 
properties for explaining SNe Ia \citep{sim10,shen18}.  These can be triggered in a number of ways, 
including the double detonation mechanism \citep[e.g.,][]{woosley94} from the accretion of a helium 
shell \citep[e.g.,][]{fink10}, a detonation in an accretion stream 
\citep{guillochon10,dan12} in a violent merger involving a massive WD \citep{pakmor12}, 
or even a more long-term evolution of a merger remnant \citep{shen12}.

As emphasized by \cite{howell11}, large samples of SNe~Ia are helping to improve our
understanding of the progenitors and explosion mechanisms.  For example, measurements
of the relative rates in different types of galaxies can be used to infer the
delay time distribution, which in turn can serve as a discriminant between different
progenitor scenarios \citep{totani08,moaz10}.  High precision photometry and spectroscopy
of individual SNe Ia also provides insight regarding the progenitors and the
physics of the explosion mechanism \citep[e.g., see][]{hoeflich10,hsiao13}.
Photometric observations of SNe~Ia at the very earliest epochs following explosion
can be used to provide important constraints on the initial radius of the primary star
\citep{piro10,bloom12}, the size of its companion star \citep{kasen10,hosseinzadeh17},
and the distribution of $^{56}$Ni in the outermost ejecta and/or the possible presence of 
circumstellar material \citep{piro16}.

\begin{figure}[h]
\centering
\includegraphics[width=7.in]{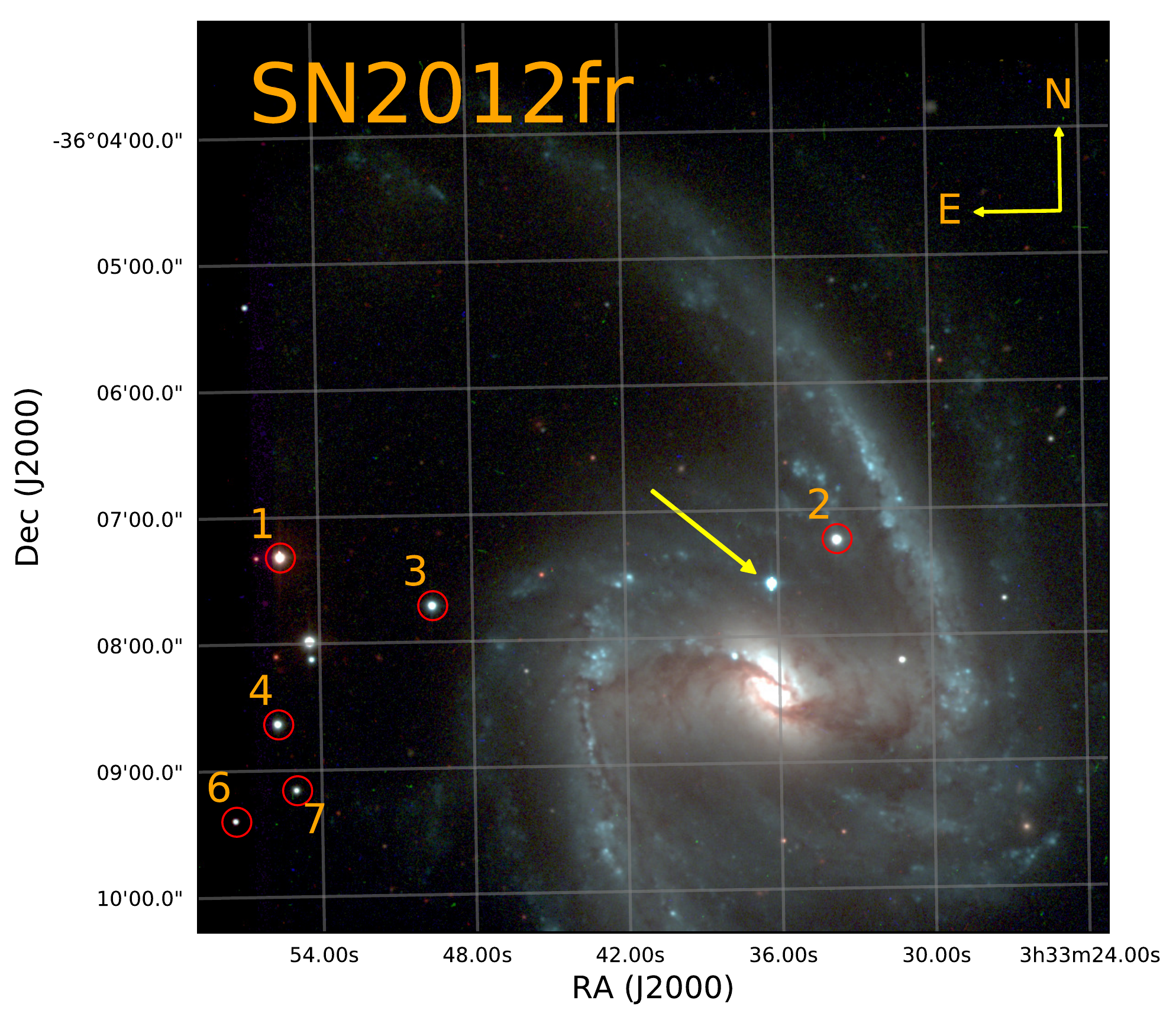}
\caption[CSP]{Swope telescope color image of NGC~1365 with local sequence stars and SN~2012fr indicated.}
     \label{fig:fchart} 
\end{figure}

SN~2012fr was discovered by the TAROT (T\'elescopes \`a Action Rapide pour les Objets 
Transitoires) collaboration in images taken with their robotic telescope at the La Silla 
Observatory, Chile \citep{klotz12}. With J2000.0 coordinates of  $\alpha$ $=$ 03$^{\rm h}$33$^{\rm m}$35$\fs$99 and 
$\delta = -$36$^{\circ}$07\arcmin37\farcs7, the transient was located 3\arcsec\ west and 52\arcsec\ north 
from the center of the SBb host-galaxy NGC~1365 (see Figure~\ref{fig:fchart}).  Within a day 
and a half of discovery, SN~2012fr was spectroscopically  classified  as a young, normal
SN~Ia, caught well before maximum light \citep{childress12}. 
\citet{childress13} analyzed an extensive set of optical spectral data  of SN~2012fr 
and documented 
the presence of high-velocity Si~II and Ca~II absorption components in the first spectrum obtained more than
14~days before \tmax which persisted for the two weeks leading up to maximum. 
While the presence of such high-velocity features (HVFs) at early epochs is common, it is 
relatively unusual
for them to remain visible at maximum light \citep{mazzali05}. The evolution of the photospheric velocity
as inferred from the Si~II $\lambda$6355 absorption was also unusually flat.
Nevertheless, the overall spectral evolution of SN~2012fr was that of a ``normal''
SN~Ia.  \cite{maund13} published spectropolarimetry at four epochs between 11~days before and
24~days after \tmax.  They found that although the continuum polarization of the SN was
low ($< 0.01$\%) throughout this period, the Si~II and Ca~II HVFs
were highly polarized at the earliest epoch.
\citet{zhang14} also published extensive ultraviolet and optical photometry of SN~2012fr
along with optical spectroscopy, and conjectured that SN~2012fr might be related to the 1991T-like
events, but viewed from a direction where the ejecta presented a clumpy structure.
More recently, \citet{graham17} presented Las Cumbres Observatory early-time $gri$ and $BVRI$
photometry of SN~2012fr, and \citet{childress15} and \citet{graham17} published nebular-phase
spectra.

Between 2004 and 2009, the Carnegie Supernova Project (``CSP-I'', \citet{hamuy06}) carried out precise 
optical and near-infrared (NIR) photometry 
of more than 100 nearby SNe Ia using the facilities at the Las Campanas Observatory (LCO). 
A second phase of the CSP (the ``CSP-II'') was initiated in 2011 in order to obtain optical and NIR light curves 
of  $\sim$100~SNe~Ia located further in the Hubble flow at $z \simeq$ 0.03--0.08
(Phillips et al., in preparation).
In addition, the CSP-II obtained light curves of a sample of $\sim$100 more nearby SNe~Ia at $z \leq 0.04$ for which 
detailed NIR spectroscopic time-series observations could be made.
In this paper we present comprehensive optical and NIR light curves of one of the latter objects,
SN~2012fr.  We also analyze the earliest images  of the
SN, including pre-discovery observations obtained by the La Silla Quest survey less than one day
after explosion.

The host-galaxy of SN~2012fr, NGC~1365, is a member of the Fornax cluster and  has numerous published 
Tully-Fisher and Cepheid distance measurements. 
In what follows we adopt the \citet{freedman01} Cepheid distance modulus of $\mu = 31.27\pm0.05$~mag
because of its internal consistency with the SNooPy analysis tool that we employ in this 
paper.\footnote[25]{The \citet{freedman01} distance modulus for NGC~1365 is consistent within
the errors with those obtained more recently by \citet{riess16} from Cepheid variables and by \citet{jang18}
from the tip of the red giant branch method.}
The Galactic reddening in the direction of NGC~1365
as computed from the \citet{schlafly11} recalibration of the \citet{schlegel98}
infrared dust maps 
is $E(B-V)_{\rm MW} = 0.018 \pm 0.003$ mag.   When adopting 
a \citet{fitzpatrick99} reddening law characterized by $R_{V} = 3.1$, this 
corresponds to  $A_{V} = 0.056 \pm 0.009$ mag. 
As discussed in \S~\ref{sec:hostred}, an analysis of the optical and NIR light curves
indicates that SN~2012fr suffered little or no host-galaxy reddening. 

The organization of this paper is as follows: 
\S~\ref{sec:data} presents the observational data, consisting of the broad-band ultraviolet (UV), 
optical and NIR photometry.  In \S~\ref{sec:analysis}, these observations are analyzed to
derive light curve parameters, the host-galaxy reddening, an independent estimate of the distance
based on the SN properties, a bolometric light curve, the rise time
to maximum, and the mass of the $^{56}$Ni produced in the explosion. 
This is followed in \S~\ref{sec:discussion} with a discussion of the early light curve and 
an assessment of the ``normality'' of SN~2012fr.  Finally, the main 
conclusions of this study are summarized in \S~\ref{sec:summary}.

\section{OBSERVATIONS}
\label{sec:data}

\subsection{Ultraviolet Photometry}
\label{sec:uv}

Imaging of SN~2012fr was performed from space with the {\em Swift} UltraViolet Optical Telescope (UVOT).
In this paper, we present the $uvw2$, $uvm2$, and $uvw1$ passbands only.
According to \citet{brown16} the effective wavelengths of these filters (convolved with a 
SN Ia spectrum of SN~1992A, at $+5$ epoch) are 2010~\AA, 2360~\AA, and 2890~\AA, respectively.

Photometry of SN~2012fr was computed from these images following the techniques 
described in  detail by \citet[][see their \S~2]{brown09}.  In short, 
for each image  a 3\arcsec\ to 5\arcsec\ source aperture was used,
depending on the signal-to-noise ratio, to measure counts at the position of the SN.  
The count rate from the underlying galaxy was subtracted from these counts using images without the 
supernova present.  The corrected counts were then converted to the UVOT 
photometric system \citep{poole08}, adopting the zero points  provided  
 by \citet{breeveld11}.  
The measured instrumental magnitudes are corrected for a coincidence loss correction factor based on 
measurements made with a 5\arcsec\ aperture, and an aperture correction is also applied based on an average  
point-spread-function (PSF) in the {\em Swift} calibration database. 
 
\begin{figure}[h]
\centering
\includegraphics[width=15cm]{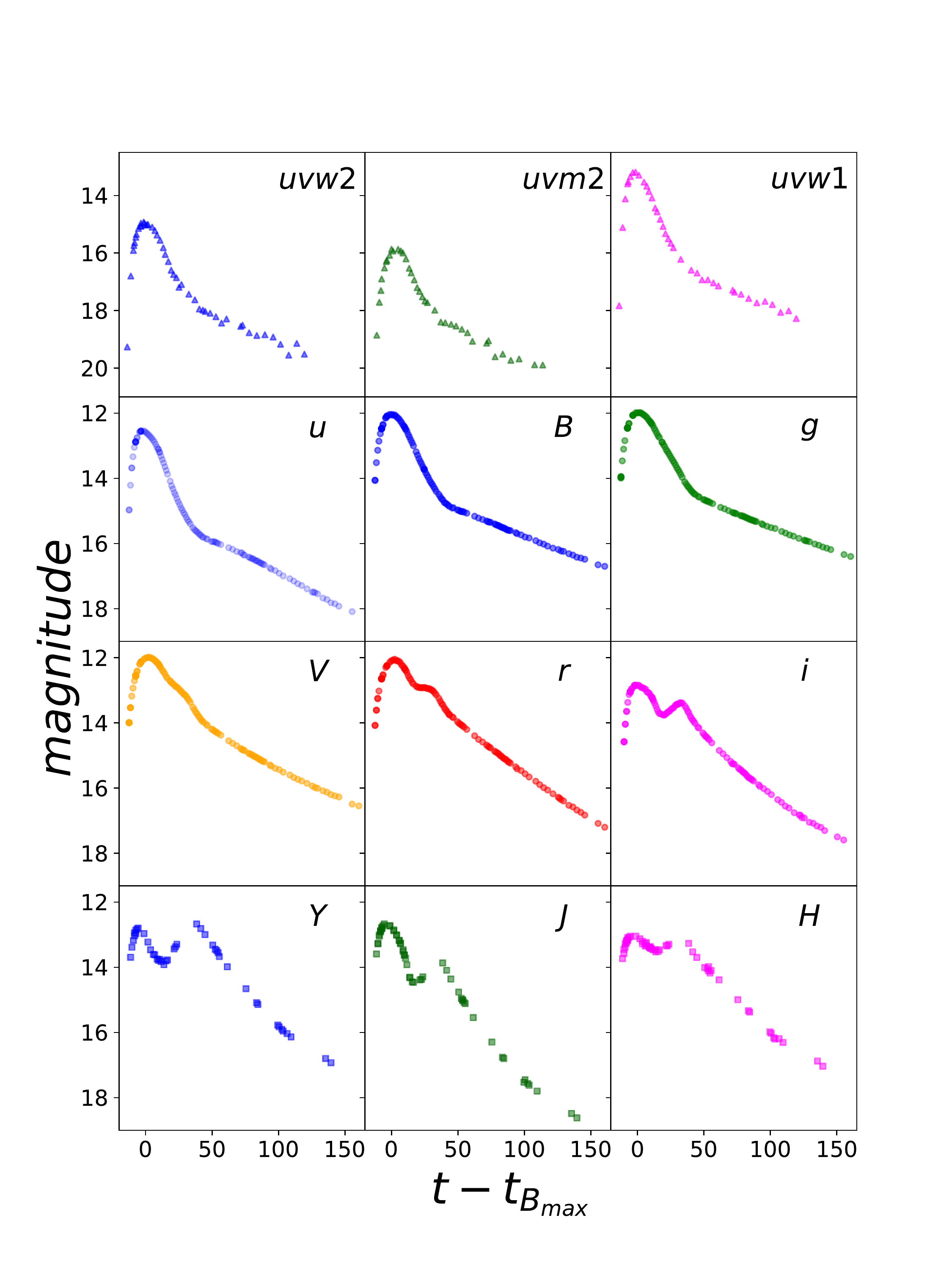}
\caption[]{UV, optical and NIR light curves of SN~2012fr. 
For the {\em Swift} UV bands, the error bars are omitted for clarity; for 
the optical and NIR bands, the size of the error bars are smaller than the symbols.}
     \label{fig:phot} 
\end{figure}

Covering $\sim$40 epochs ranging from $-$13~days to $+$120~days relative to the epoch of 
\tmax (see $\S$~\ref{sec:analysis}), 
the high-cadence UV light curves of SN~2012fr are plotted in Figure~\ref{fig:phot}. 
The corresponding photometry is listed in Table~\ref{tab:uv}.

\subsection{Optical Photometry}
\label{sec:optical}

\subsubsection{CSP}
\label{sec:csp}

Optical imaging of SN~2012fr was obtained with the  1~m Henrietta Swope telescope 
located at LCO,  equipped with the same Johnson ($BV$) and Sloan ($ugri$) filter set 
and CCD detector used for the CSP-I \citep{contreras10,stritzinger11,krisciunas17}. 
All images were reduced in the manner described by \citet{contreras10} and
\citet{krisciunas17}, including the subtraction of host-galaxy reference images obtained
after the SN had fully disappeared.
PSF photometry of the SN was computed with respect to a local sequence of 
stars calibrated to \citet{landolt92}  and \citet{smith02} standard fields 
observed over the course of more than 60 photometric nights. 
Photometry of the local sequence in the standard system is given in Table~\ref{tab:locseq}.

\begin{figure}[h]
\centering
\includegraphics[width=16cm]{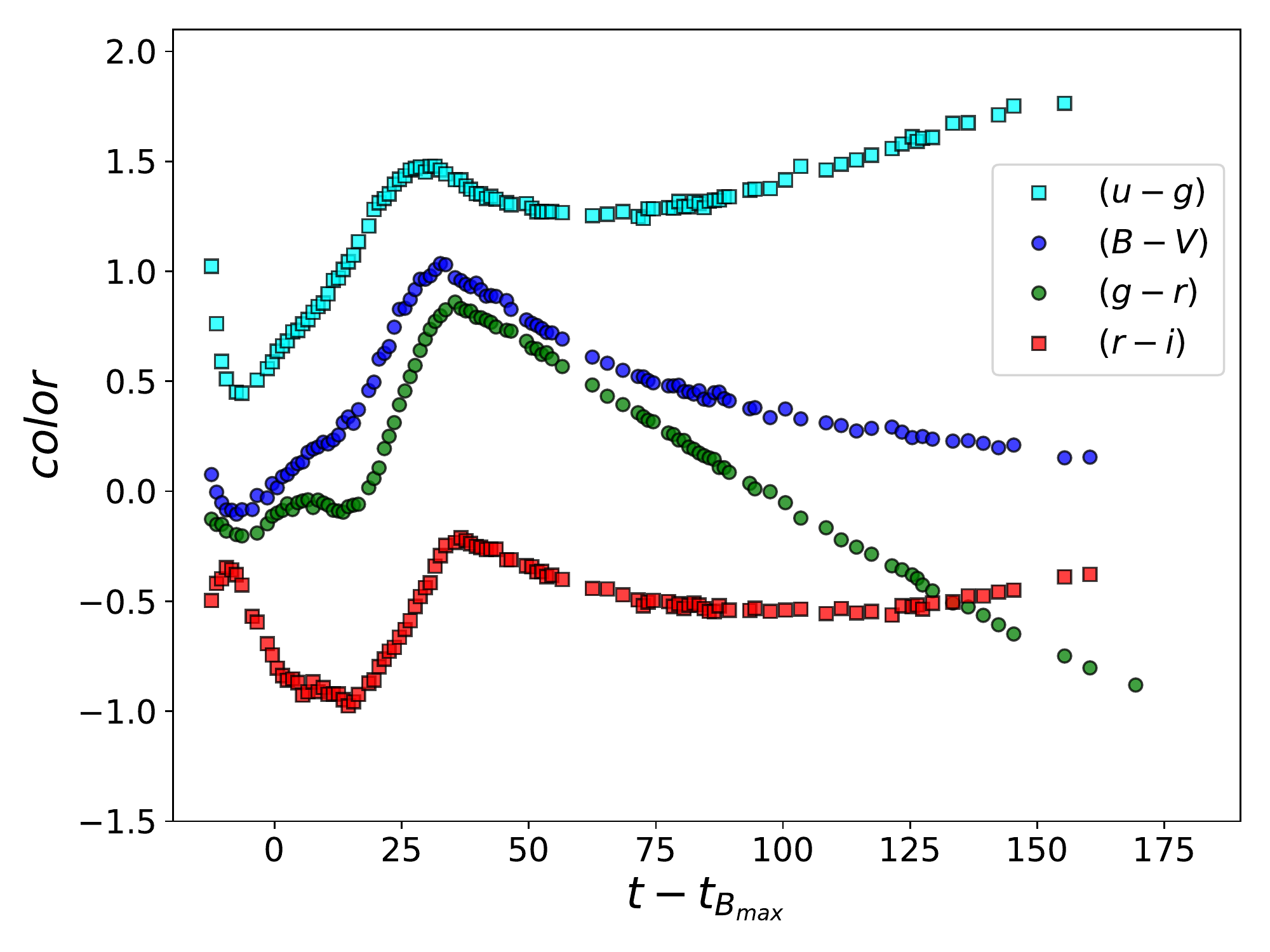}
\caption[]{Optical colors of SN~2012fr. Error bars are omitted for clarity}     \label{fig:color} 
\end{figure}

Definitive $uBgVri$-band photometry of SN~2012fr in the Swope {\em natural} system \citep{krisciunas17} is given 
in Table~\ref{tab:opt}, and the corresponding light curves are plotted in individual panels contained 
within  Figure~\ref{fig:phot}. Color curves are also plotted in Figure~\ref{fig:color} showing the high 
quality of the data set. Comprising  $\sim$120 epochs, the light curves track the flux evolution from 
$-$12~days to  $+$160 days  relative to  \tmax, representing one of the most comprehensive data 
sets yet obtained of a Type~Ia~SN.

As discussed in \citet{contreras10}, photometry in the natural system is the ``purest'' form of the data, 
and provides the most transparent way to combine CSP photometry with data sets from other groups. 
Nevertheless, as requested by the referee, we provide S-corrections in Table~\ref{tab:sc} which, when added to the 
corresponding optical natural photometry magnitudes in Table~\ref{tab:opt}, convert the photometry to the standard 
systems --- i.e. \citet{landolt92} for $B$ and $V$ , and \citet{smith02} for $u¿$, $g¿$, $r¿$, and $i¿$.  
The S-corrections were calculated using the \citet{hsiao07} spectral template, and SN~2012fr 
spectra \citep{childress13} when available (values in parenthesis). The S-corrections are measured
after the spectra are mangled to match the photometric colors of the corresponding phases.

\subsubsection{TAROT}
\label{sec:tarot}

SN~2012fr was discovered in images obtained with the robotic telescope operated by the TAROT 
collaboration at the La Silla Observatory.  The discovery image was taken on
27.05 October 2012 UT  (JD 2456227.55) with an ``open'' filter
(i.e., without a filter), and follow-up $BVR$ images were obtained beginning two nights later.
The telescope optics, filters, and detector are described in detail by \citet{klotz08}.  In
Appendix~\ref{sec:filter_functions}, throughput functions are presented for the ``open'' and $V$
filters, and magnitudes for SN~2012fr and the local standards in the field are derived 
in the natural systems of these two filters.

\subsubsection{La Silla-QUEST}
\label{sec:quest}

\begin{figure}
\centering
\leavevmode
\begin{tabular}{cc} 
\includegraphics[width=7.5cm]{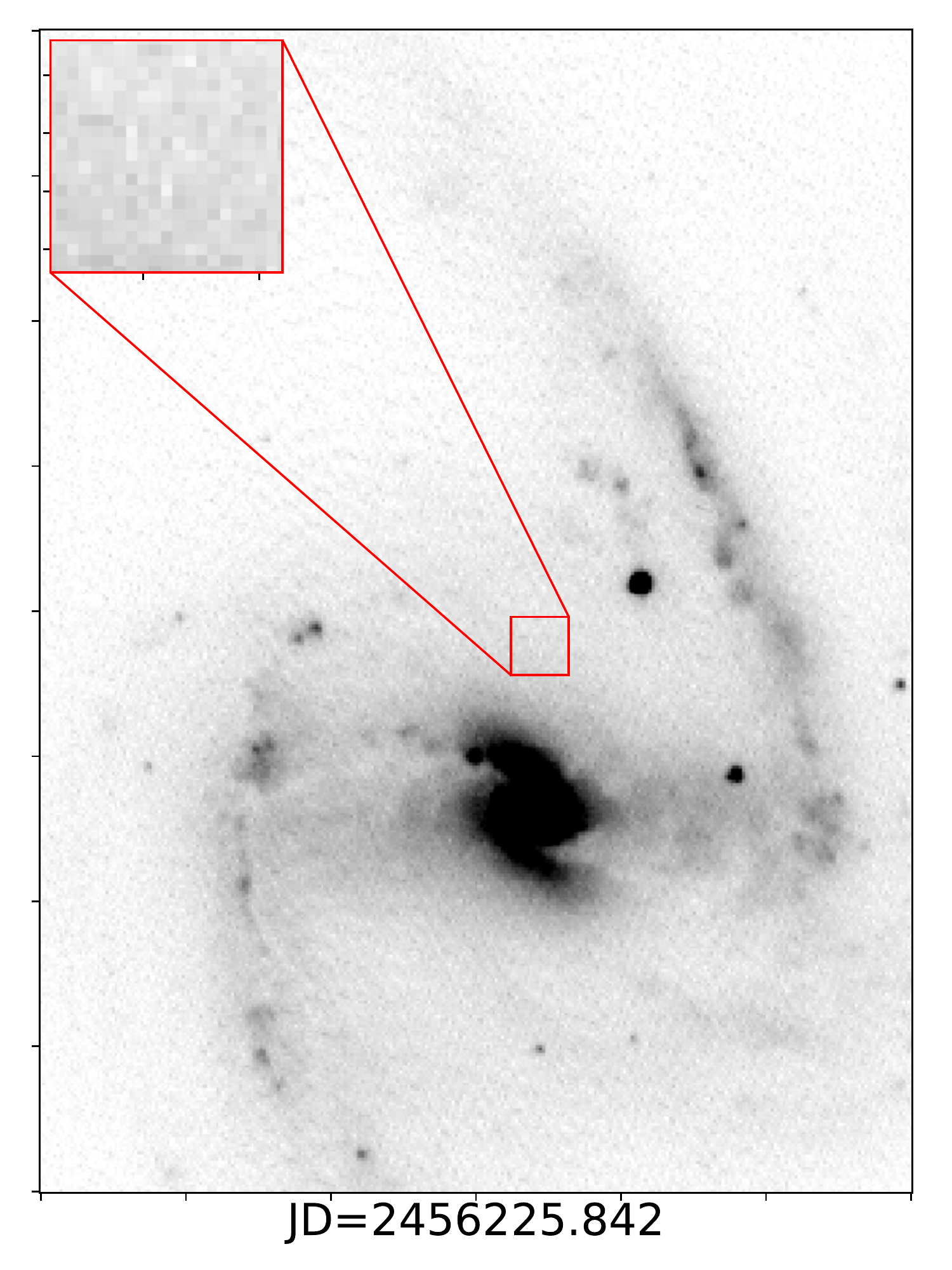} & \includegraphics[width=7.5cm]{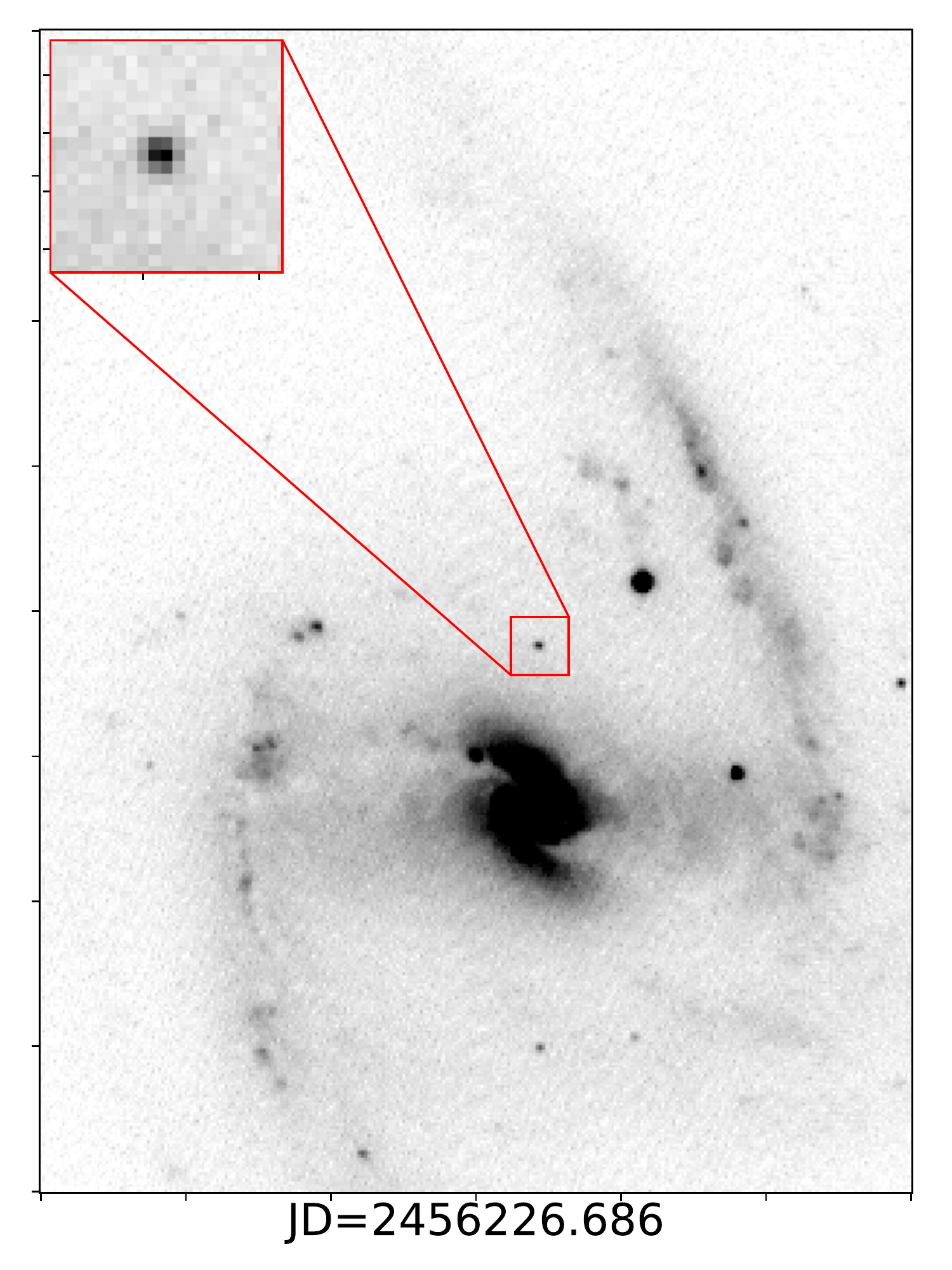}  \\
\end{tabular}
\caption{LSQ Survey images of NGC~1365 taken on 25.34~Oct~2012 UT and 26.19~Oct~2012 UT.
The position of SN~2012fr is magnified in the upper left-hand corner of both images.}
\label{fig:ds9}
\end{figure}

In an attempt to better constrain the rise time of the SN, we examined the observing log of the
La Silla-QUEST (LSQ) Low Redshift Supernova Survey \citep{baltay13}, which went into routine 
operations approximately one year before the discovery of SN~2012fr.  The log indicated that the 
field of NGC~1365 was observed through a wide-band ``$gr$'' filter 
every night from 23 to 27~October~2012 UT.  Subsequent 
examination of the images revealed that the SN was clearly visible 
on 26.19~October~2012 UT (see Figure~\ref{fig:ds9}), but was absent in the images obtained
before this date.  Unfortunately, the image of the SN was saturated in all images obtained
after 26~October~2012 UT.   Details
of the telescope and detector system employed for the LSQ survey are described by  
\citet{baltay07, baltay13}.  In Appendix~\ref{sec:filter_functions} the throughput function of the
$gr$ filter is derived and natural system magnitudes for the SN and the local standards are presented.

\subsubsection{Slooh}
\label{sec:slooh}

As reported by \citet{klotz12b}, SN~2012fr was confirmed by E.~Conseil from images obtained 
with the 0.5~m Slooh Space Camera robotic telescope at Mt. Teide on the island of Tenerife.
These observations were made with a set of Astrodon Tru-Balance $LRGB$ E-Series filters
using an FLI PL09000 CCD camera on 27.13 October 2012 UT --- i.e., 
less than two hours after the TAROT discovery image.  The Slooh observations are extremely 
important since they provide information on both the brightness of the SN and the color  
shortly after discovery.  In Appendix~\ref{sec:filter_functions}, the throughput functions of the 
$LRGB$ filters are estimated and natural system magnitudes are derived for the SN and the
local standards in the field of SN~2012fr.
 
\subsubsection{BOSS}
\label{sec:boss}

Confirming images of SN~2012fr were taken by Stu Parker at Parkdale Observatory in New Zealand
starting 1.3~days after discovery by the TAROT collaboration \citep{parker12}.  Parker, who is a member
of the Backyard Observatory Supernova Search (BOSS), used a 0.36~m Celestron telescope with an
SBIG ST-10XME CCD camera and no filter for these observations.  In Appendix~\ref{sec:filter_functions},
the throughput function of this combination is calculated, which we henceforth refer to as the BOSS ``open'' 
system, and natural system magnitudes for the SN and the local standards are derived.

\subsection{NIR Photometry}
\label{sec:nir_phot}

\begin{figure}[h]
\centering
\includegraphics[width=6.7in]{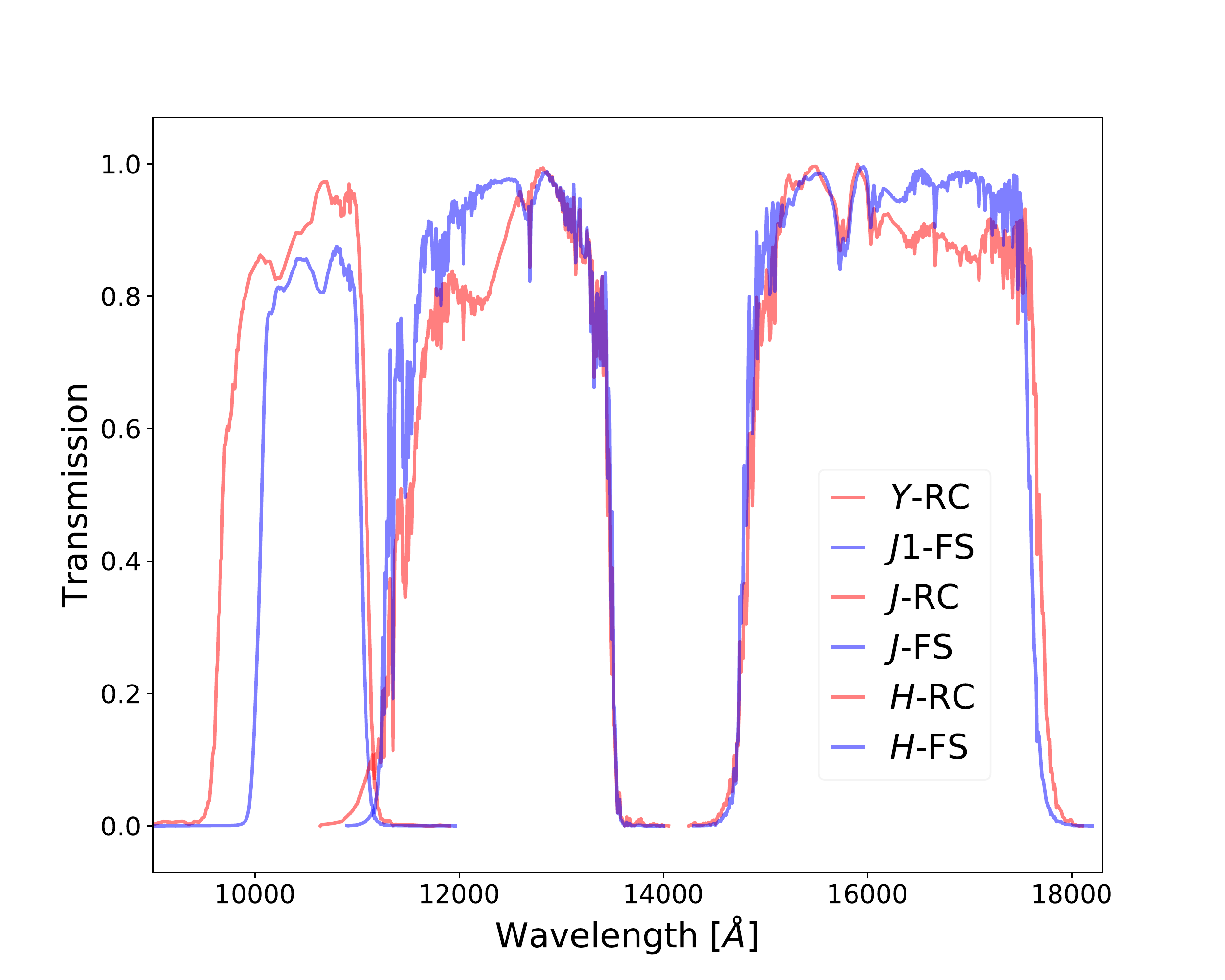}
\caption{Comparison between the transmission functions of the Swope + RetroCam $Y_{RC}$, $J_{RC}$, 
and $H_{RC}$ filters and the Magellan Baade + FourStar $J1_{FS}$, $J_{FS}$, and $H_{FS}$ filters.}
     \label{fig:J1}
\end{figure}

The majority of our NIR imaging was obtained on the LCO du Pont 2.5~m telescope 
with RetroCam, a $YJH$-band imager employing a
Hawaii-1 HgCdTe $1024 \times1024$ pixel array.  The field of view of 
RetroCam at the du Pont telescope is 3\farcm5$\times 3$\farcm5, with a pixel scale of 
0\farcs20.  Additional NIR imaging of SN~2012fr was obtained in $J1$, $J$ and $H$ filters using
the FourStar camera attached to the 6.5~m Magellan Baade telescope \citep{persson13}.
FourStar consists of a mosaic of four Hawaii-2RG HgCdTe detectors, with each chip yielding a 
field of view of 5\arcmin$\times$5\arcmin. 
The RetroCam and FourStar bandpasses are illustrated in Figure~\ref{fig:J1}.  As is
seen, 
the FourStar $J1$ filter covers $\sim$ 75\% of the wavelength range of the RetroCam $Y$ filter.
 
The NIR images were reduced in the standard manner following the steps described 
by \citet{contreras10}. In short, images were dark subtracted, flat fielded (and 
sky~$+$~fringing subtracted in the case of RetroCam images), and each dithered frame was 
aligned and combined.  Host galaxy reference images were subtracted from each combined image on
a 15\arcsec\ radius circle around the SN.
 
The NIR photometry of SN~2012fr was computed differentially with respect to a local sequence 
of stars defined using RetroCam observations. The local sequence was calibrated in the 
\citet{persson98} $JH$ photometric 
system using standard star fields observed during 10 photometric nights. 
For the $Y$ band, the  local sequence stars were calibrated using the magnitudes 
for a subset of \citeauthor{persson98} standards published by \citet{krisciunas17}. 
The final  NIR $Y$, $J$ and $H$ magnitudes for the local sequence stars  
in these standard systems are listed in Table~\ref{tab:locseq}.

To compute magnitudes for the local sequence stars in the natural system of the three FourStar filters
(hereafter referred to as $J1_{FS}$, $J_{FS}$, and $H_{FS}$), the following
set of transformation relations was obtained  via synthetic photometry of \citet{castelli03} model atmospheres: 
 
\begin{equation}
Y_{std} = J1_{FS} + 0.1064 (J-H)_{std} + 0.0052 \;,
\end{equation}
\begin{equation}
J_{std} = J_{FS} + 0.0008 (J-H)_{std} + 0.0047 \;,
\end{equation}
\begin{equation}
H_{std} = H_{FS} - 0.0395 (J-H)_{std} +0.0062 \;.
\end{equation}

\noindent In these equations, $J_{std}$, and $H_{std}$ are the magnitudes in the standard system
of \citeauthor{persson98}; $Y_{std}$ is the magnitude in the RetroCam $Y$-band
standard system defined by \citet{krisciunas17}; $J_{FS}$, $J1_{FS}$, and $H_{FS}$ are the
magnitudes in natural system of FourStar; and $(J-H)_{std}$ is the color index of the star 
in \citeauthor{persson98} system.  The \citeauthor{persson98} and FourStar filter functions used to derive
the synthetic magntiudes are available on the CSP website.\footnote[26]{\url{http://csp.obs.carnegiescience.edu/}}

Making use of the local sequence, we proceeded to compute the definitive NIR photometry 
of SN~2012fr in the {\it natural system} of the RetroCam $YJH$ filters 
\citep[hereafter referred to as $Y_{RC}$, $J_{RC}$, and $H_{RC}$, as in ][]{krisciunas17}
that is listed in Table~\ref{tab:nir}.

This photometry is plotted in individual panels contained within 
Figure~\ref{fig:phot}. Consisting of more than 40 epochs, the light curves follow the flux 
evolution from $-$11 to $+$140~days with respect to \tmax. 
The FourStar $J1_{FS}$, $J_{FS}$, and $H_{FS}$ photometry was 
S-corrected \citep{suntzeff00,stritzinger02} to match the RetroCam $Y_{RC}$, $J_{RC}$, and 
$H_{RC}$ system using the following spectrophotometric prescription: 

\begin{itemize}

\item{Spectral templates \citep{hsiao07} matching the phases of each FourStar photometry epoch
were employed.  The template spectrum for each FourStar epoch was color matched separately to
 the photometry in the $J1_{FS}$, $J_{FS}$, and $H_{FS}$ bands, respectively.}

\item{For the $J1_{FS}$ band, the color-matching function was
a 2nd order polynomial  calculated using CSP $i$ and FourStar $J1_{FS}$ and $J_{FS}$ 
photometry.  Likewise, for the $J_{FS}$ band, a
2nd order polynomial was derived from the $J1_{FS}$, $J_{FS}$, and $H_{FS}$ photometry. 
For the $H_{FS}$ band, a linear polynomial was employed to color match the template to the 
$J_{FS}$ and $H_{FS}$ magnitudes.}

\item{The S-correction was then derived as the difference between the synthetic RetroCam
and FourStar magnitudes derived from the mangled spectrum.
Specifically: $S_{J1_{FS}} = Y_{RC} - J1_{FS}$; $S_{J_{FS}} = J_{RC} - J_{FS}$; and
$S_{H_{FS}} = H_{RC} - H_{FS}$}.

\end{itemize}

\begin{figure}[h]
\centering
\begin{tabular}{ccc} 
\includegraphics[width=7.0in]{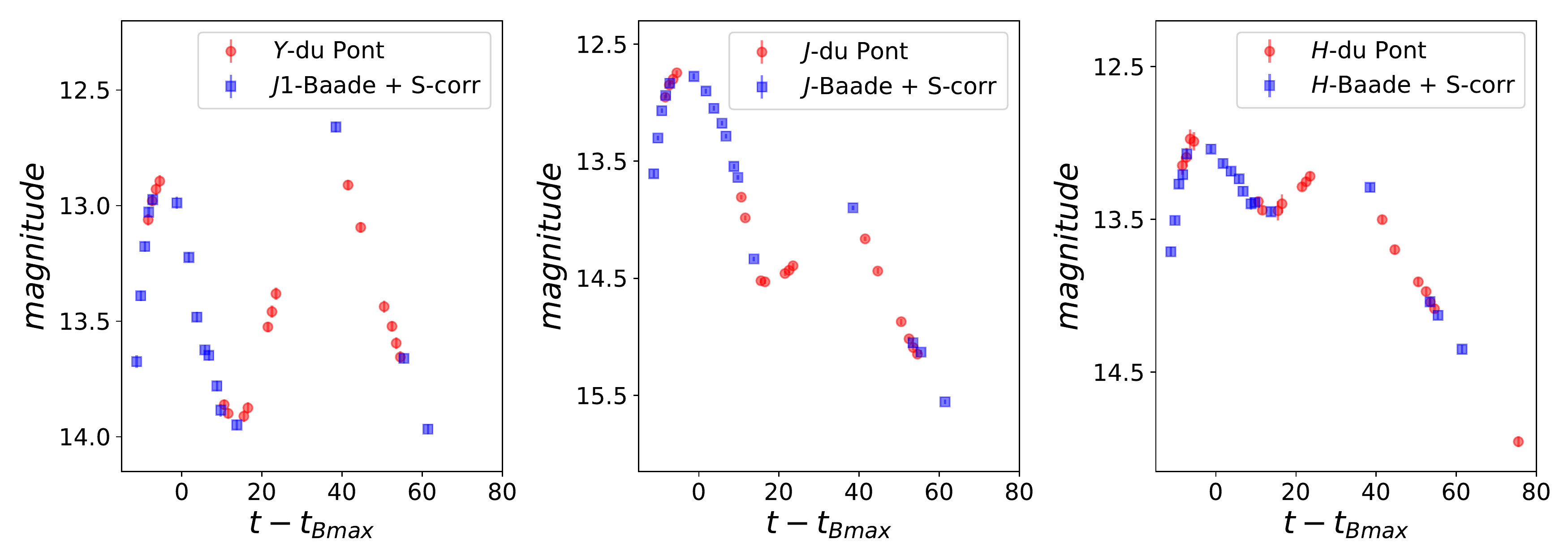}\\
\end{tabular}
\caption{$YJH$  light curves of SN~2012fr from $-$12 to $+$60~days relative to \tmax, acquired with the du Pont ($+$ RetroCam) and
the Magellan Baade ($+$ FourStar) telescopes. The photometry obtained between the two facilities matches exceptionally
well. Note in particular the excellent agreement between the FourStar S-corrected $J1$$\rightarrow$$Y$-band
and RetroCam $Y$-band photometry.}
\label{fig:scorr}
\end{figure}

The resulting S-corrections are listed in Table~\ref{tab:scorr}. 
The excellent consistency between the $S$-corrected FourStar photometry and the RetroCam observations
is illustrated in Figure~\ref{fig:scorr}.  This is confirmed by looking at nights where nearly simultaneous 
measurements were made with both instruments.  For the $Y$ band at phases of $-8.3$ and $-7.3$~days,
we find differences between the S-corrected FourStar and RetroCam
photometry of $-0.022 \pm 0.030$ and $-0.006 \pm 0.031$~mag, respectively.  
For the $J$ band at phases of $-8.3$, $-7.3$, and $+53.5$~days, 
the differences are $-0.016 \pm 0.028$, $-0.017 \pm 0.028$, and $+0.035 \pm 0.041$~mag, 
respectively, while for the $H$ band at the same phases, the differences are
$-0.061 \pm 0.068$, $-0.020 \pm 0.060$, and $0.000 \pm 0.001$~mag.

\section{ANALYSIS}
\label{sec:analysis}

\subsection{Light Curve Parameters}
\label{sec:lcpars}

The densely-sampled light curves of SN~2012fr allow us to accurately measure the apparent
magnitude at maximum and the decline-rate parameter, $\Delta$m$_{15}(X)$,\footnote[27]{The decline-rate parameters,
$\Delta m_{15}(X)$, are here defined as the magnitude difference of the light curve in filter $X$ from peak to 15 
days later. Historically, the value of $\dmB$ was shown by \citet{phillips93} to correlate 
with the absolute peak magnitude in such a way that more 
luminous objects show slower light curve decline rates.} in each filter, $X$, using 
a smooth Gaussian process's fitting curve.
The  values are summarized  
in Table~\ref{tab:pars}. K-corrections were ignored as the small redshift ($z_{helio} = 0.005457$
according to NED\footnote[28]{NED is the NASA/IPAC Extragalactic Database.}) of the host-galaxy, 
NGC~1365, has very little effect on these quantities.
In addition, absolute magnitudes are also provided as computed 
using the adopted average Cepheid distance discussed in \S~\ref{sec:distance}
and assuming a host-galaxy dust reddening of $E(B-V)_{host} = 0.03 \pm 0.03$ mag (see \S~\ref{sec:hostred}).
The fit to the $B$-band light curve indicates that \tmax\ occurred on 
JD~$2456243.1\pm0.3$ and $\dmB =  0.82 \pm 0.03$ mag. 
This rather slow decline-rate implies that SN~2012fr should be   
moderately over-luminous compared to a ``standard" SN~Ia with
 $\dmB = 1.1$ mag and an absolute $B$-band magnitude  $M_{B} = -19.1$ mag
 \citep{folatelli10}.
Indeed, as shown in Table~\ref{tab:pars}, SN~2012fr was $\sim$0.3 mag more luminous than
this value.

As expected for a normal SN~Ia, the NIR light 
curves reached primary maxima  $\sim$4--5 days prior to \tmax.   
The absolute magnitudes in $Y$, $J$, and $H$, calculated using the Cepheid distance to 
NGC~1365, are given in Table~\ref{tab:pars}.  These are also fully consistent with the
average values for slow-to-mid-decliners given in Table~7 of \citet{kattner12}. 

\subsection{Host Galaxy Reddening}
\label{sec:hostred}

To determine if SN~2012fr suffered any  reddening due to dust external to the Milky Way 
we make use of three methods. In the first of these, the ($B-V$) color curve of SN~2012fr  is 
compared to the Lira relation \citep{lira95,phillips99}.
The Lira relation relies on the fact that the ($B-V$) color of normal 
un-reddened SNe~Ia follows a linear evolution with minimal scatter between 30$-$90 days past maximum
\citep[see][for a description of the physics underlying the Lira relation]{hoeflich17}. 
Therefore, comparing the ($B-V$) color curve of any given SN~Ia, once corrected for Galactic reddening, 
to the Lira relation provides an indication of the amount of dust reddening external to the Milky Way. 
\citet{burns14} provide the following fit to the Lira relation for a SN with $\dmB$ = 0.82 mag 
based on the CSP-I data releases 1 and 2 \citep{contreras10,stritzinger11}:

\begin{eqnarray}
\label{eqn2}
	(B-V)_{Lira} = 0.78(\pm0.04) -0.0094(\pm0.0005) [t-t_{B_{max}} - 45].
\end{eqnarray}

\noindent
This calibration differs somewhat from an analysis done previously by \citet{folatelli10}, who used 
a smaller sample of SNe~Ia presumed to be un-reddened.  The approach taken by \citet{burns14} was to use 
all SNe in the CSP-I sample with good photometric coverage extending beyond 40~days past \tmax. 
The late-time slopes of the $(B-V)$ light curves were measured as well 
as the value of $(B-V)$ at 45~days after \tmax\ for each object separately. The median slope is used 
for the Lira law, and the median absolute deviation is taken as its uncertainty. The observed distribution 
of $(B-V)$ colors at day $+45$ is modeled in a similar fashion to \citet{jha07} as the convolution of an intrinsic Gaussian 
distribution and exponential tail. The maximum and standard deviation of the resulting Gaussian 
distribution are taken as the Lira intercept and uncertainty, respectively.

\begin{figure}[h]
\centering
\includegraphics[width=6.5in]{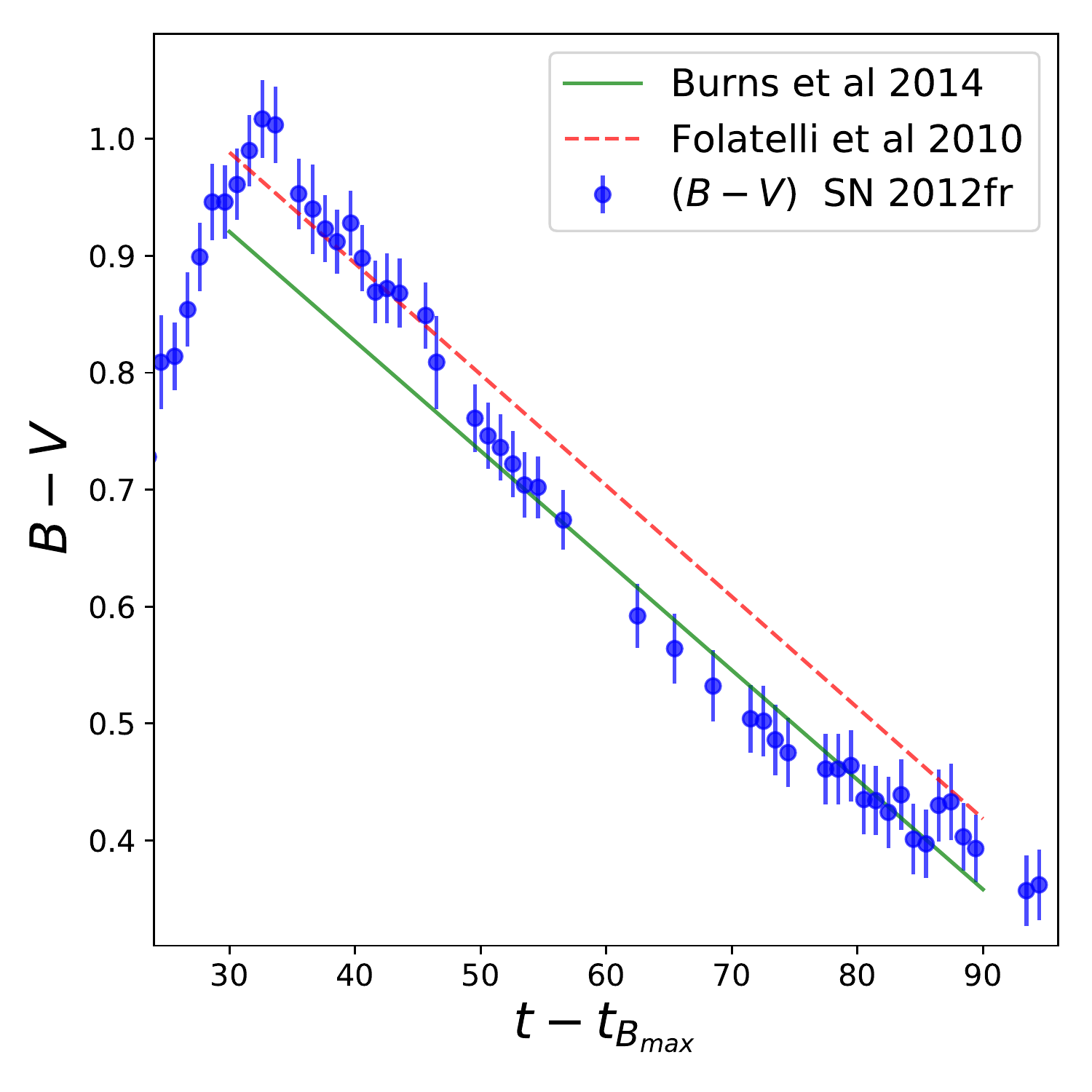}
\caption[CSP]{Galactic extinction-corrected  $(B-V)$ color evolution of SN~2012fr. Over-plotted as a dashed line is the Lira relation as 
determined by \citet{folatelli10}, while the solid line is the re-calibration presented by \citet{burns14}. In both cases,
if the overall range of the Lira relation is considered, the implication is that SN~2012fr suffered little 
or no host-galaxy reddening.} 
\label{fig:lira}
\end{figure}

Plotted in Figure~\ref{fig:lira} is the Galactic reddening-corrected ($B-V$) color evolution of SN~2012fr
from 25 to 95 days past \tmax.  Over-plotted as a solid line is 
the Lira relation from \citet{burns14} as defined in Eq.~\ref{eqn2}.  
Also shown for reference  is the relation given by \citet{folatelli10}.
The comparison between the color evolution and the Lira relation from \citeauthor{burns14} implies 
SN~2012fr suffered minimal host-galaxy reddening, $E(B-V)_{host} = 0.03 \pm 0.04$ mag.
However, the excellent precision and sampling of
the observations clearly reveals that the $(B-V)$ temporal evolution of SN~2012fr 
was not linear at these epochs.  From 35--60 days past $t_{B_{max}}$, the slope is 
$\sim-0.014~{\rm mag~day}^{-1}$,
whereas from 60--95 days it is $\sim-0.008~{\rm mag~day}^{-1}$.  As shown in Figure~12 of 
\cite{burns14}, these values cover the range of slopes displayed by normal SNe~Ia.  Nevertheless,
it is unusual to observe such a large change of slope in any single event.

The second  method adopted to estimate the host-galaxy $E(B-V)_{host}$ color excess
relies on well-defined relations between maximum light intrinsic
pseudo-colors\footnote[29]{A {\em pseudo-color} is
defined to be the difference between  peak magnitudes of two passbands.
In the case of SNe~Ia, the time of peak brightness can vary by up to a few days from passband to
passband.}
 and the decline rate derived from a large  sample of SNe~Ia.
This method
is more often applicable to SNe~Ia observations as it makes use of maximum light observations,
which are normally more readily  available than the post-maximum
regime required for an accurate Lira relation analysis. The method is fully detailed in
\citet{burns14} and we briefly describe it here.

The observed colors $(B-X)$, where $X$ represents each filter except for $B$, are modeled as the
sum of an intrinsic color that depends on the decline rate parameter of the SN, the color excess $E(B-X)_{MW}$
from the Milky-Way dust, and the color excess $E(B-X)_{host}$ from the host-galaxy ISM. The intrinsic
colors are derived from an MCMC analysis of a training sample of SNe~Ia from the CSP-I \citep{burns14}.
The Milky-Way component of the reddening is determined by assuming a value $E(B-V)_{MW}$ from
the \citet{schlafly11} dust maps, a fixed value for the ratio of total-to-selective absorption
$R_{V}^{MW} = 3.1$, and the reddening law of \citet{fitzpatrick99}. Finally, the host-galaxy
component of the reddening is modeled with the \citet{fitzpatrick99} reddening law and two
free parameters:  $E(B-V)_{host}$ and $R_V^{host}$, which are determined using MCMC methods.
The resulting reddening $E(B-V)_{host} = 0.06 \pm 0.02$ mag is quite low and as a result, the posterior
of $R_V^{host} = 4.2 \pm 1.3$ reflects more the population distribution of $R_V$ from the training sample
rather than the observed colors of SN~2012fr.

The \ion{Na}{1}~D interstellar absorption lines can be used to provide a third estimate of the host-galaxy
reddening of SN~2012fr.  For this purpose, we use the Keck HIRES echelle spectrum 
published by \citet{childress12}.  These authors measured weak absorption in the D1 and D2 lines
due to gas in the Milky Way at a combined equivalent width (EW) of $118 \pm 19$~m\AA.  Employing 
the fit to EW(\ion{Na}{1}~D) vs. $A_V$ for Galactic stars using the \citet{munari97} relation shown in 
Figure~9 of \citet{phillips13}, this value implies $A_V = 0.05 \pm 0.03$ mag.  This is in excellent agreement
with the \citet{schlafly11} value given in \S \ref{sec:intro}.  No \ion{Na}{1}D absorption at the redshift of
NGC~1365 is visible in the echelle spectrum of \citeauthor{childress12} at a 3$\sigma$ upper limit of
EW(\ion{Na}{1}~D)$ = 82.8$~m\AA, implying $A_V \leq 0.04 \pm 0.03$~mag, or $E(B-V)_{host} = 0.01 \pm 0.01$~mag
if the interstellar gas has similar properties to that in the solar neighborhood.

Taken together, these three estimates are consistent with zero or negligible host-galaxy reddening
of SN~2012fr.  In the remainder of this paper, we adopt a value of $E(B-V)_{host} = 0.03 \pm 0.03$ mag,
consistent with all three estimates.  We also assume $R_{V} = 3.1$, but the exact value is not 
critical since the reddening is low.

\subsection{Distance to NGC~1365 as Derived from SN~2012fr}
\label{sec:distance}

From the fits to the observed light curves of SN~2012fr using the SNooPy  ``EBV" method \citep{burns11}, 
the SN-based distance to NGC~1365 is found to be 
$\mu_{\circ} =  31.25 \pm 0.01_{stat} \pm 0.08_{syst}$ mag.  SNooPy also offers an 
alternative way to estimate the distance to the host-galaxy using 
the broad-band light curves through the \citet{tripp98} method. 
The functional form of the Tripp method relates the distance modulus of a SN~Ia 
to  its decline rate and color via:

\begin{equation}
\mu_\circ = m_{X}^{\rm max}  -  M_{X}(0)  - b_X \cdot[\dm - 1.1] -          
\beta^{YZ}_{X} \cdot (Y-Z).
\label{beqn7}
\end{equation}

\noindent Here $m_{X}^{\rm max}$ is the observed K-corrected and Galactic-reddening corrected magnitude at maximum,
 $M_{X}^{0}$ is the peak absolute magnitude of SNe~Ia with $\dm = 1.1$ and
zero dust extinction, $b{_X}$ is the slope of the luminosity vs. decline rate relation,
$\beta^{YZ}_{X}$ is the slope of the luminosity-color relationship, and
($Y - Z$) is a pseudo-color at maximum.
Note that the SNooPy parameter, $\dm$, is the template-derived value of the decline rate parameter, 
which correlates strongly with the directly measured value, $\dmB$, but with some random and systematic 
deviations \citep[see Figure 6 of][]{burns11}.

Use of the Tripp method requires an accurate 
calibration between the relations of  peak  absolute magnitude vs. $\dm$
and pseudo-color. 
Here the calibrations presented by \citet[][Table~8, lines 2 and 6]{folatelli10} are adopted,
which are based on 26 well-observed SN~Ia light curves published by \citet{contreras10} for $B$ band,
and 21 well-observed SN~Ia light curves for $J$ band of the same paper.
The resulting estimates of distance modulus for NGC~1365 based on the Tripp method 
are $\mu_B = 31.14\pm0.15$~mag and $\mu_J = 31.34\pm0.14$~mag.

These results are consistent with the Cepheid distance modulus of $\mu=31.27\pm0.05$~mag from \citet{freedman01}
adopted in this work, indicating that SN~2012fr had a luminosity consistent with a normal SNe~Ia with 
$\dmB =  0.82$~mag.

\subsection{Bolometric Light Curve of SN~2012fr}

The extended wavelength coverage of SN~2012fr afforded by observations spanning from 
UV through NIR wavelengths allows for the  construction of an essentially  complete 
ultraviolet-optical-infrared bolometric light curve, usually termed as UVOIR bolometric curve in the literature.
However, as pointed out by \citet{brown16}, this does not clearly specifiy the boundaries
of the wavelength domain over which the flux estimate is done.
For this reason in this paper we define the bolometric luminosity, $L_{Bol}$, as the luminosity 
between wavelengths of $1800$~\AA\ and infinity,
corresponding to the sum of $L_{UVOIR}\,(3000-16600\,$\AA) flux plus the 
contribution, $L_{uvm2}\,(1800-3000\,$\AA), deduced from the SWIFT photometry, and
the unobserved far-infrared $L_{\lambda>\lambda_H}\,(16600-\infty\,$\AA).

\begin{figure}[p]
\centering
\includegraphics[width=16cm]{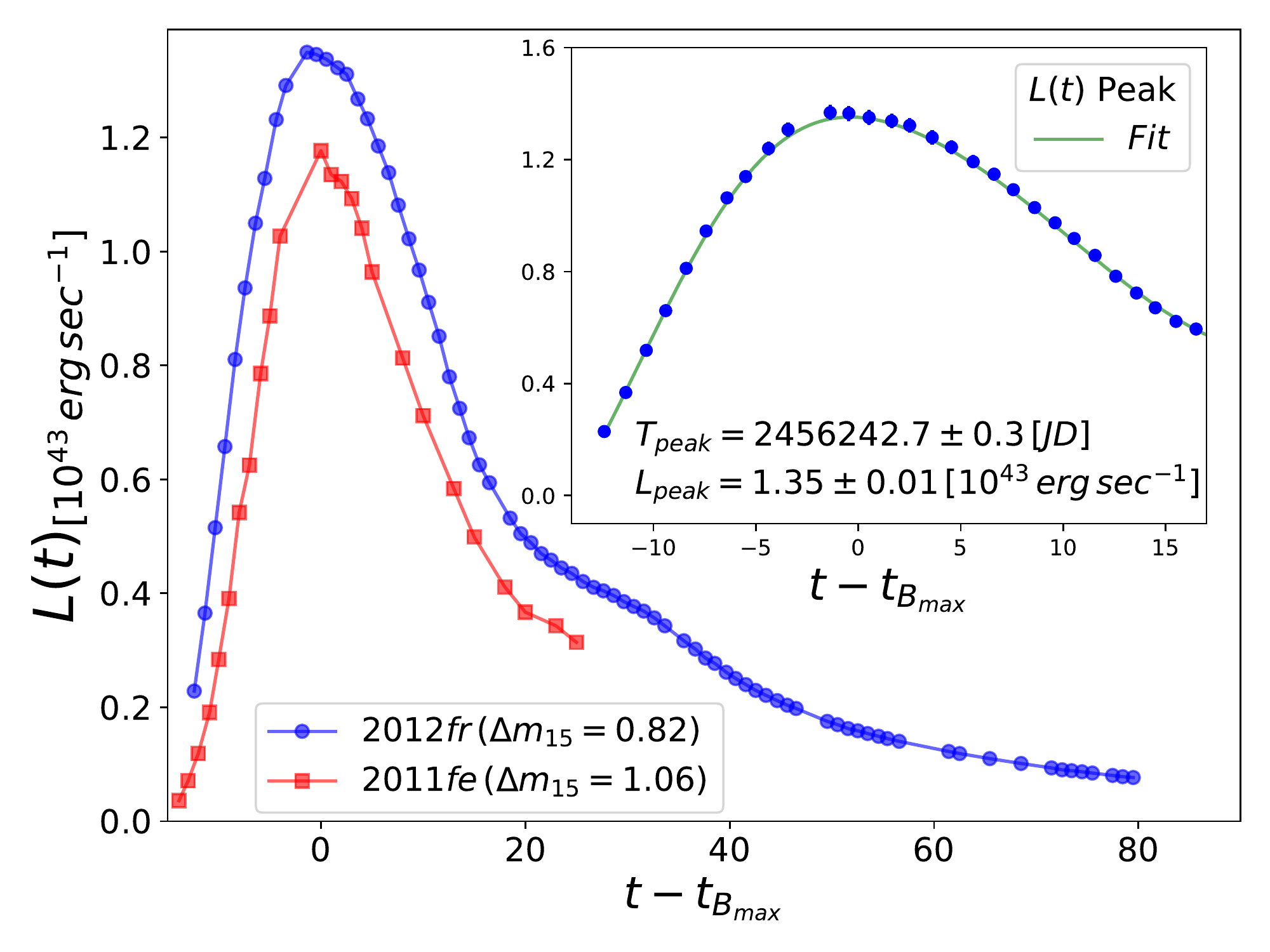}
\caption[CSP]{Bolometric light curve of SN~2012fr compared to that
of the normal SN~Ia 2011fe \citep{Pereira13}.  The bolometric light curve shown for SN~2012fr is the average
of the Trapezoidal Rule Integration and Spectral Template Fitting methods (see Appendix~\ref{sec:bolo_lc_details}).
Top right: a Gaussian process smooth curve is fitted to the bolometric data to recover the time and amplitude of peak.}
\label{fig:uvoir}
\end{figure}

Details of this calculation are given in 
Appendix~\ref{sec:bolo_lc_details}, and the final bolometric light curve is plotted in 
Figure~\ref{fig:uvoir}.  Shown for comparison is the bolometric light curve 
derived by \citet{Pereira13} for SN~2011fe, 
one of the few other SNe~Ia to have been well-observed in the UV, optical, and NIR.

\begin{figure}
\centering
\leavevmode
\includegraphics[width=16cm]{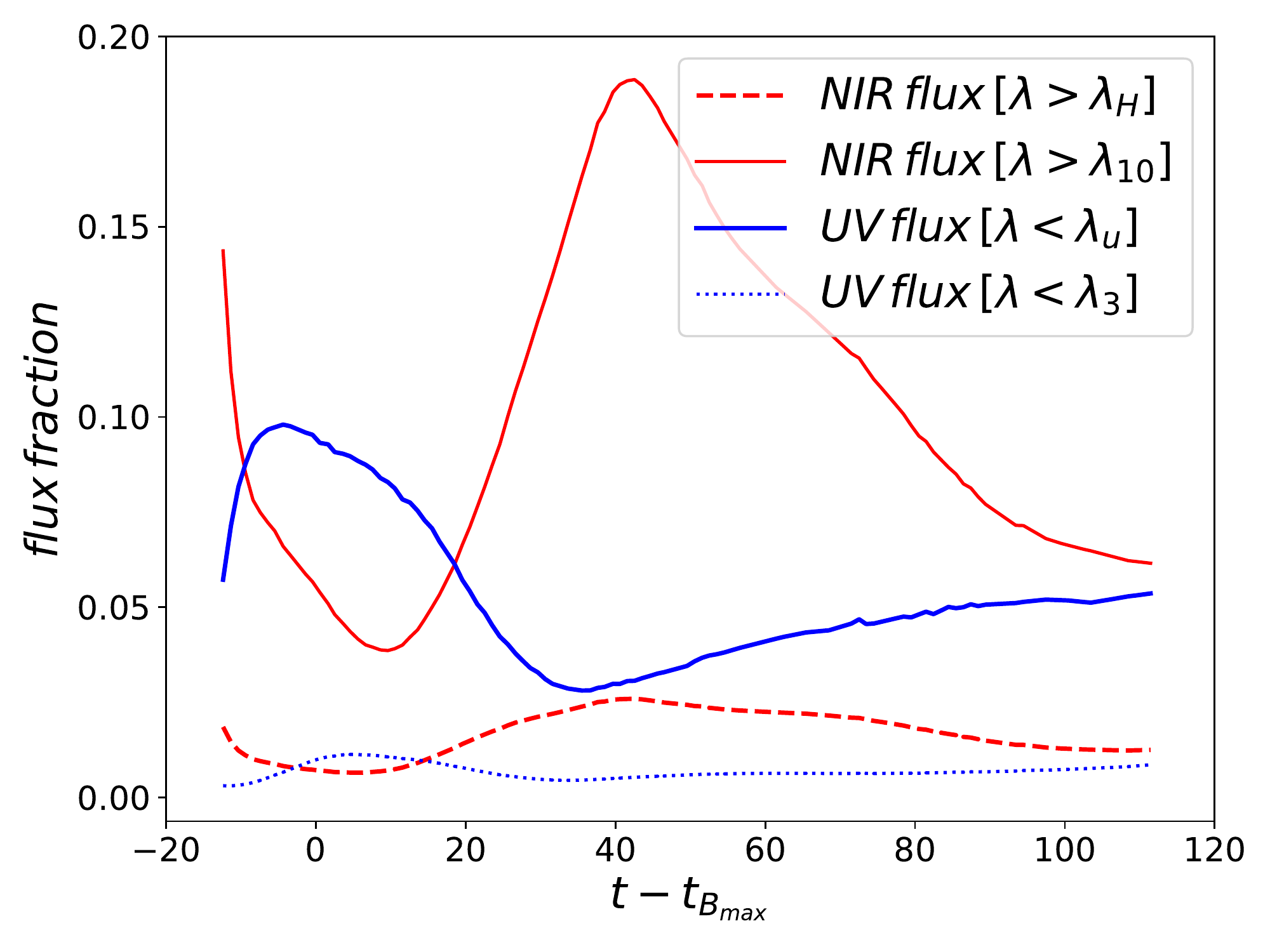} 
\caption{Estimated ratio of UV and NIR fluxes to the total flux for SN~2012fr. 
Dotted and dashed lines represent the ratios for estimated fluxes beyond the observed optical and NIR domain,
i.e. fluxes for $\lambda$ less than 3,000~\AA\, and for $\lambda$ greater than the
effective wavelength of the $H$~band.  For comparison purposes, the red solid line gives the flux beyond 10,000 \AA,
and the blue solid line shows the UV flux for wavelengths less than $u$-band effective wavelength.}
\label{fig:frac}
\end{figure}

At peak, SN~2012fr reached a maximum luminosity of 
L$_{Bol} = (1.35 \pm 0.14) \times 10^{43}$~ergs~s$^{-1}$,
which is on the bright end of the normal SNe~Ia distribution
\citep[cf. Figure 6 of][]{scalzo14}. 
The relative fractions of the NIR ($\lambda > 10,000$~\AA) integrated flux
is plotted as a function of light curve phase in Figure \ref{fig:frac},
showing that the NIR contribution to the bolometric
light curve of SN~2012fr is nearly 15\% at $-12$~days.  It 
falls to a minimum of $\sim$4\% a few days after \tmax, and then rises again steeply to 19\% at
+40~days, at which point it begins to slowly decrease again.  The latter behavior, that is similar
to what is shown in Figure~3 of \citet{scalzo14}, is responsible for the
prominent shoulder in the light curve seen between $+20$ and $+45$~days 
in Figure~\ref{fig:uvoir}.    Figure \ref{fig:frac} shows that the
UV contribution ($\lambda < \lambda_u$) to the bolometric light curve of SN~2012fr is generally of less importance
than the NIR except around the \tmax epoch, when both contributions are similar,
peaking around 10\% a few days before \tmax and contributing $> 5$\% of the
integrated flux only during the early epochs, i.e., before 20~days after \tmax.  
This behavior is consistent with previous attempts
to quantify the UV contribution to the bolometric luminosity \citep[e.g., see][]{suntzeff03}.

\subsection{Rise Time}
\label{sec:risetime}

SN~2012fr was discovered by the TAROT collaboration on 27.05~October~2012 (UT),
15.6~days before \tmax.  As reported by \citet{klotz12}, 
the SN was not visible to  a limiting magnitude 
of $R > 15.8$ in an image of NGC~1365 taken three days earlier on 24.05~October~2012 
(UT) with the same telescope.  A more stringent
non-detection of $R > 19.3$ on 24.02~October~2012 (UT) was obtained by 
J.~Normand from stacked images taken with an 0.6~m telescope at the Observatoire 
des Makes \citep{klotz12}.  Thus, the rise time to \tmax was constrained to be somewhere
between 18.6 and 15.6~days.

Fortunately, the LSQ images presented in this paper provide a much tighter constraint on
the rise time since the SN was clearly visible on 26.19~October~2012 (UT), but was 
absent in an image of similar depth obtained  25.34~October~2012 (see Figure~\ref{fig:ds9}).  Thus, 
the SN was detected less than a day after explosion, which occurred some time
between 17.3 and 16.5~days before \tmax.

\begin{figure}
\centering
\leavevmode
\includegraphics[width=17cm]{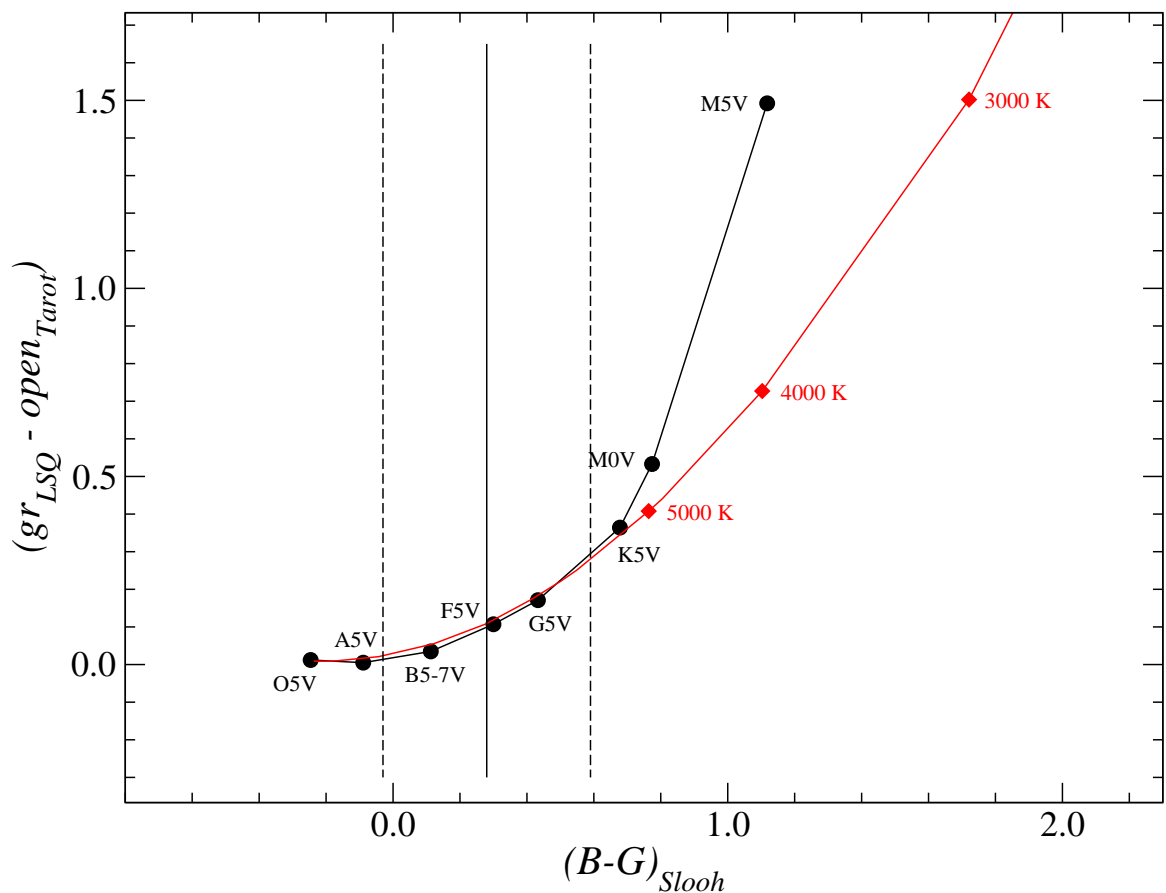}
\caption{Color-color plot for converting TAROT open filter magnitudes 
to the natural system magnitudes for the LSQ $gr$ filter using the $(B-G)_{Slooh}$
measurement obtained less than two hours after the TAROT observation.  The black curve
show synthetic photometry carried out using the \citet{pickles98} stellar atlas, and
the red line corresponds to synthetic photometry of black bodies
covering a range of temperature.  The solid vertical line and the dashed lines on
either side indicate the $(B-G)_{Slooh}$ color measurement of $0.28 \pm 0.31$ mag.}
\label{fig:tarot_to_gr}
\end{figure}

\begin{figure}
\centering
\leavevmode
\includegraphics[width=13cm]{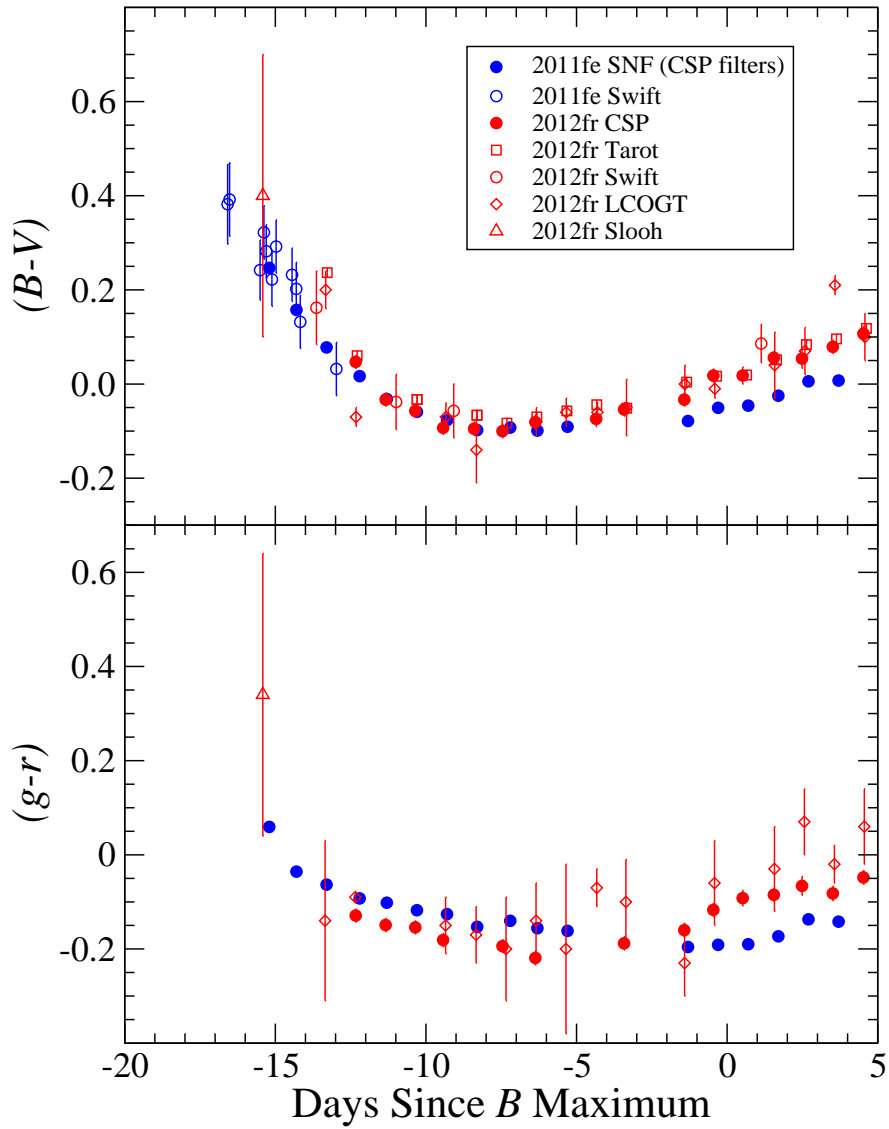}
\caption{Early-time $(B-V)$ and $(g-r)$ color evolution of SN~2012fr.  Shown for comparison is  
the color evolution of SN~2011fe.  The observations for SN~2012fr are taken from the CSP, Tarot,
Swift, and LCOGT \citep{graham17}.  The observations of 
SN~2011fe are from Swift \citep{brown12} and synthetic photometry in the CSP filter bandpasses of the 
spectrophotometry of \citet{Pereira13}.  The abscissa is corrected for time
dilation.}
\label{fig:color_evolution}
\end{figure}

\begin{figure}
\centering
\leavevmode
\includegraphics[width=17cm]{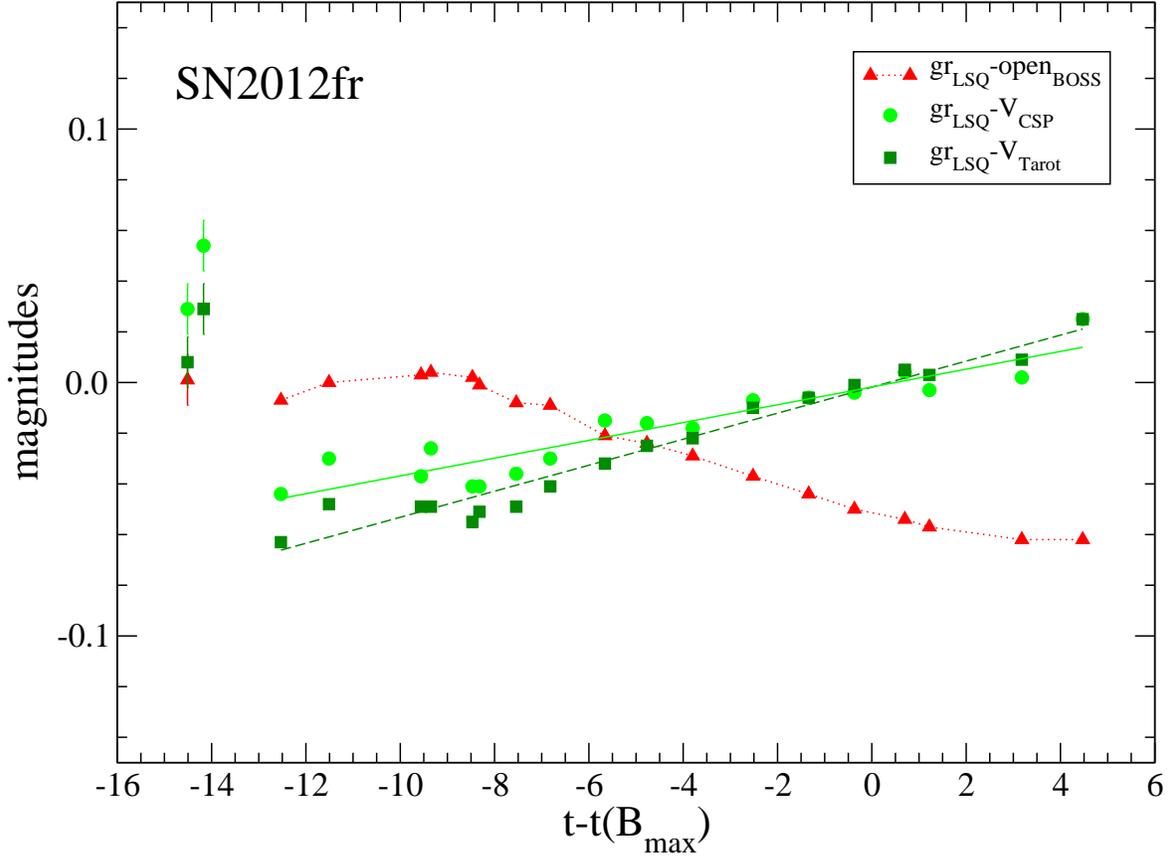}
\caption{Magnitude differences between synthetic photometry of the \citet{childress13}
spectra of SN~2012fr in the CSP $V$, TAROT $V$, and BOSS open filters and synthetic
photometry of the same spectra in the LSQ gr filter.  The spectra were first color matched to 
the CSP $BVgri$ photometry to improve their spectrophotometric precision.  The single BOSS point
at $-14.5$~days was derived from the first spectrum obtained of the SN at $-14.5$~days 
\citep{childress12}.  Likewise, the CSP $V$ and TAROT $V$ points at $-14.5$ and $-14.2$~days
were calculated from the first two spectra obtained by \citep{childress13}. 
The larger errors on these points reflect the fact that these spectra 
could not be color matched since the CSP photometry did not begin until $-12.4$ days. 
}
\label{fig:mag_conversions}
\end{figure}

Our observations of SN~2012fr present a rare opportunity to study the early rising light curve of 
SN~2012fr.  However, we first must S-correct the various measurements to the 
same filter bandpass.  Since the earliest detection and non-detection of SN~2012fr were 
made in the LSQ~gr filter, we choose to convert the Slooh, TAROT, BOSS, and CSP photometry to
LSQ~gr magnitudes.  The steps required are:

\begin{itemize}

\item{\bf Slooh $L$ filter}

\hspace{.8cm}As discussed in Appendix~\ref{sec:filter_functions}, the Slooh $L$-band photometry is essentially in the
same natural system as the $gr_{LSQ}$ observations.  Hence, no S-correction is 
required for this measurement, which was obtained less than two hours after the
TAROT discovery image.

\item {\bf TAROT open filter}  

\hspace{.8cm}The TAROT discovery image was obtained within approximately 1-2 days of explosion.  
This is more than a day before the first spectroscopic 
observation.  Fortunately, the Slooh $B$ and $G$ filter observations, obtained less than 
two hours after the TAROT discovery image, provide color information that can be
used to estimate the S-correction under the assumption that the spectrum at this
epoch can be approximated by a black body.  Figure~\ref{fig:tarot_to_gr} shows the magnitude 
difference $(gr_{LSQ} - open_{TAROT})$ as a function of $(B-G)_{Slooh}$ as 
derived from synthetic photometry of main sequence stars 
from the \citet{pickles98} stellar library spectra
and black bodies of varying 
temperature.  From the photometry given in Appendix~\ref{sec:filter_functions}, 
$(B-G)_{Slooh} = 0.28 \pm 0.31$~mag.  Using the black body curve, this implies
$(gr_{LSQ} - open_{TAROT}) = 0.09^{+0.20}_{-0.08}$~mag.  

As a sanity check on this result, we plot the $(B-V)$ color evolution for SN~2012fr 
in Figure~\ref{fig:color_evolution}.  For comparison, observations of SN~2011fe
are also shown.  Combining the $B_{Slooh}$ and $G_{Slooh}$
magnitudes measured for the SN and the color-color plots
in Appendix \ref{sec:filter_functions} gives $(B-V) = 0.40 \pm 0.31$~mag.
This measurement is plotted in Figure~\ref{fig:color_evolution}, and appears
to be generally consistent with expectations if SN~2011fe is a valid comparison.
Nevertheless, it should be kept in mind that the correction of the TAROT photometry to 
the LSQ system derived in the previous paragraph is strictly only valid for an object with 
a stellar or black body spectrum.  Strong features such as the  \ion{Ca}{2} 
and \ion{Si}{2} HVFs observed in the earliest spectra of SN~2012fr could affect 
the S-correction.

\item {\bf BOSS open filter}

\hspace{.8cm}The first BOSS open filter observation was made
14.2~days before \tmax.  This is only $\sim0.3$~days after the first spectrum of the SN obtained
by \citet{childress12}, and so we have used this spectrum \citep[published by ][]{childress13} to
calculate the S-correction.  We find $(gr_{LSQ} - open_{BOSS}) = -0.02 \pm 0.01$~mag, where the
error reflects the uncertainty in the spectrophotometric calibration of the spectrum.  S-corrections
for the BOSS open filter observations at $-12.2$ and $-7.1$~days were obtained from synthetic photometry
of the \citet{childress13} spectra after color matching the spectra to the
CSP photometry using first- or second- order polynomials (see Figure~\ref{fig:mag_conversions}).  Finally, the S-correction for the
BOSS open filter observation at $-13.2$~days was obtained by interpolating the S-corrections for the
$-14.2$ and $-12.2$~day spectra.

\item {\bf CSP and TAROT V filters} 

\hspace{.8cm}Although the $V$ filter is narrower than the LSQ $gr$ filter,
they are well matched in central wavelength.  The CSP $V$ observations began 12.4~days before
\tmax, and the TAROT $V$ filter imaging started one night earlier at $-13.5$~days.  S-corrections
for both filters derived from the color-matched \citet{childress13} spectra are plotted in 
Figure~\ref{fig:mag_conversions}.

\end{itemize}

\begin{figure}
\centering
\leavevmode
\includegraphics[width=13cm]{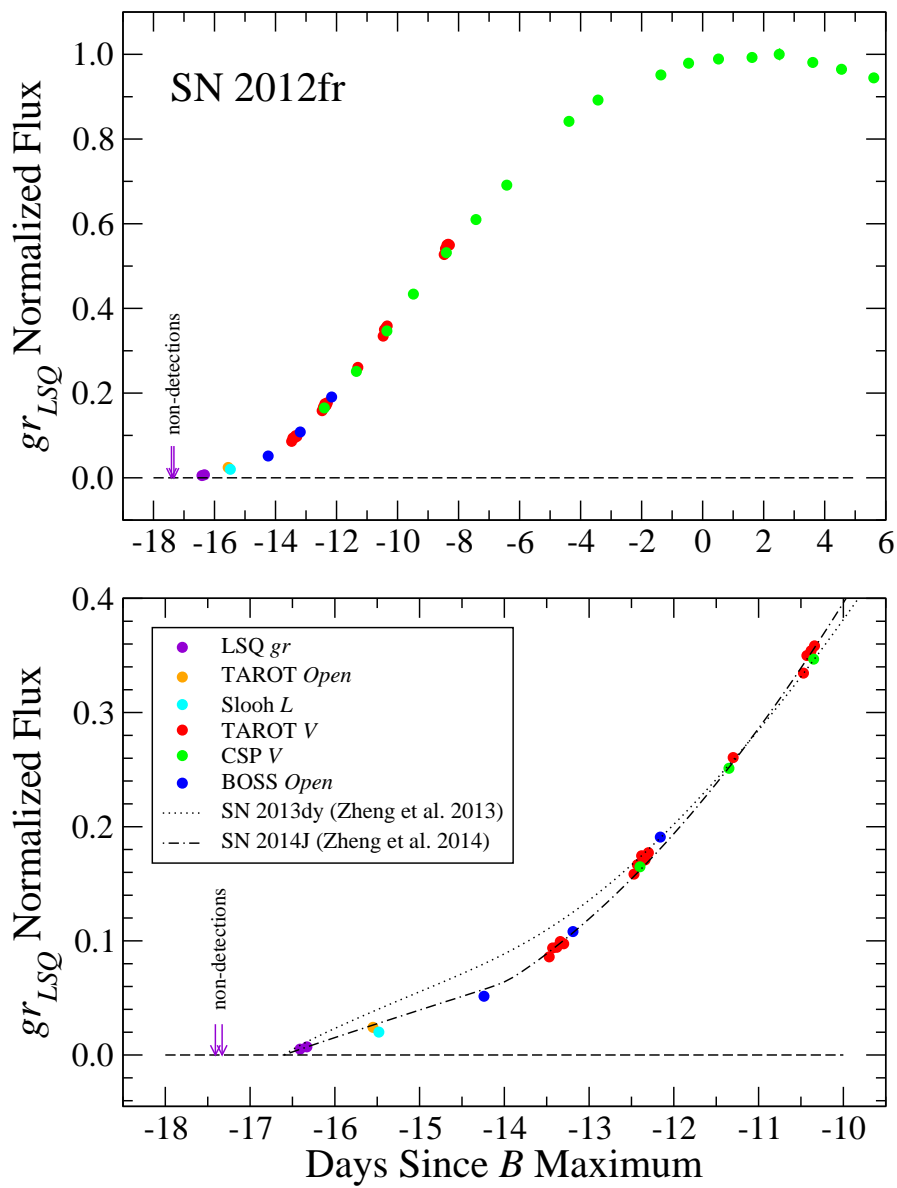}
\caption{(Top) Relative fluxes (normalized to maximum) plotted as a function of the time with respect to \tmax\
during the rising phase of the $gr_{LSQ}$-band light curve of SN~2012fr and extending to a few days past maximum.  
(Bottom) Enlargement of the same observations during the seven days following explosion.  Error bars are not visible 
since they are the same size or smaller than the symbols used to plot the data.  The broken power-law fits of the
early light curves of SNe~2013dy and 2014J \citep{zheng13,zheng14} are 
plotted for comparison.}
\label{fig:rise2}
\end{figure}

The resulting LSQ $gr$ light curve is given in Table~\ref{tab:rise} and plotted in normalized flux 
units in Figure~\ref{fig:rise2}.  Also indicated in Figure~\ref{fig:rise2} are the LSQ non-detections 
and the epoch of the first spectrum.  From the first detection of the SN at
$-16.4$~days with respect to  \tmax\ to the first BOSS observation at $-14.2$~days, the light curve
rises very close to linearly.
After the first BOSS observation,
the light curve rises more steeply to a second nearly linearly phase that lasts from approximately
$-11.5$ to $-6.5$~days.
The non-detection at $-17.33$~days and the first detection at $-16.41$~days  
constrain the time of explosion to have occurred at JD~$2456226.23 \pm 0.46$,
which is $16.87 \pm 0.46$~days 
before \tmax.  The bolometric maximum was reached on JD~$2456242.7 \pm 0.3$
(see Table~\ref{tab:pars}).  Hence, the rise to bolometric maximum took a total of $\sim16.47 \pm 0.55$~days.

\subsection{$^{56}$Ni Mass}

With a well-sampled bolometric light curve and a precise measurement of the bolometric rise time in hand, 
the amount  of $^{56}$Ni  synthesized during the explosion can be estimated using
Arnett's Rule \citep{arnett82}, which relates  
the bolometric rise time, $t_r$, and peak bolometric luminosity, $L_{peak}$, to the energy deposition, $E_{Ni}$,
within  the expanding ejecta supplied by the radioactive decay chain  
$^{56}$Ni  $\rightarrow$  $^{56}$Co  $\rightarrow$  $^{56}$Fe \citep[see, e.g.,][]{stritzinger05b}.
 This is formulated as follows:

\begin{eqnarray}
\label{eqn:arnett}
L_{peak} = \alpha E_{Ni}(t_r).
\end{eqnarray}
\noindent

\noindent Arnett's Rule is derived from semi-analytical solutions to the radiative transfer problem of the expanding SN~Ia ejecta  
and explicitly assumes equality between energy generation and luminosity, i.e. the factor $\alpha$ $=$ 1. 
A number of years ago, \citet{branch92} surveyed the explosion models then available in the literature, concluding
that $\alpha = 1.2\pm0.2$ was a more appropriate value to use.  While this value is still commonly employed for
determining $^{56}$Ni masses \citep[see][and references therein]{scalzo14},  \citet{hoeflich96} found 
that $\alpha$ ranged from 0.7 to 1.4 for a large variety of explosion models, with an average value of $1.0 \pm 0.2$.
\citet{stritzinger05b} cited radiative transport calculations for two modern 3D deflagration models from the 
MPA group that were both consistent with $\alpha = 1.0$.

An expression for $E_{Ni}(t_r)$ is provided by \citet[][see his Equation 18]{nadyozhin94}
which, after plugging in various constants, yields for 1 $M_{\odot}$ of 
$^{56}$Ni the following relation:

\begin{eqnarray}
\label{eqn:dep}
E_{Ni} (1~M_{\odot}) = \left[ 6.45 \times e^{-t_r/8.8} + 1.45 \times e^{-t_r/111.3}\right] 10^{43}\ \mathrm{ergs\ sec^{-1}}.
\end{eqnarray}

\noindent  Combining Equations~\ref{eqn:arnett} and \ref{eqn:dep},
we obtain the following simple relation to estimate the $^{56}$Ni mass: 
\begin{eqnarray}
\label{eqn:nimass}
M_{Ni} = \frac{L_{peak}}{\alpha\,E_{Ni} (1~M_{\odot})} M_{\odot}.
\end{eqnarray}
\noindent Given the non-linearity of equation~\ref{eqn:nimass}, we computed the error in $M_{Ni}$ by
simulating $10^5$ computations using  randomly drawn uncorrelated values for $\alpha$,
$E(B-V)_{host}$, $t_r$, the distance modulus, and the peak
bolometric flux, $F_{peak}$,  assuming that these parameters are described by
Gaussian distributions with mean and standard deviation vectors of
$\mu = [1.0,0.03,16.47,31.27,F_{peak}]$ and $\sigma = [0.2,0.03,0.60,0.05,0.05F_{peak}]$, respectively.
Arnett's parameter, $\alpha$, the host-galaxy reddening, and the distance modulus are the dominant
error sources, while the rest of the parameters only exert a mild effect. 
The error in the host-galaxy reddening alone translates to an error of 10\% in $M_{Ni}$.  The uncertainty in
$\alpha$ has an even larger effect and introduces a slight asymmetry to the marginalized distribution of the $^{56}$Ni mass.
The final value derived from this analysis is
$M_{Ni}=0.60_{-0.14}^{+0.16} M_{\odot}$.

\cite{childress15} give an independent estimate of $M_{Ni}$ for SN~2012fr
based on measurements of the [Co III] $\lambda$5893 emission in nebular phase spectra.  Adjusting
their value for the distance modulus for NGC~1365 adopted in the present paper and assuming
the same reddening of $E(B-V)_{host} = 0.03 \pm 0.03$ mag gives a value of
$M_{Ni} = 0.61 \pm 0.07~M_{\odot}$, fully consistent with our estimate from the 
bolometric light curve.  

We note that \citet{zhang14} derived a much higher $^{56}$Ni mass 
from a bolometric light curve constructed from their own optical photometry, 
the same SWIFT ultraviolet observations used in the present paper, and NIR corrections taken from
SN~2005cf \citep{wang09b}.  Adjusting their quoted value of $0.88 \pm 0.08~M_{\odot}$ 
to the same distance modulus and host-galaxy reddening that we assume in this paper gives
a value of $0.84 \pm 0.08~M_{\odot}$.
From private communication with J.~Zhang, the peak bolometric luminosity of
\citet{zhang14} was corrected to $L_{peak}=1.65\times10^{43}$ erg/sec. The mismatch is mainly due to 
an overestimation of the NIR contribution. After this correction and putting both measurements at
the same distance modulus, the difference amounts to 7\%,
half of which can be explained for our differences in the $u$-band domain flux.

\section{DISCUSSION}
\label{sec:discussion}

\subsection{Early Light Curve}

The earliest emission of a SN~Ia probes the location in the ejecta of the $^{56}$Ni that
powers the light curve, and thus is an important diagnostic of the explosion physics
\citep{piro13}.  Early studies of the rise times of SNe~Ia  
\citep{Riess99,aldering00,goldhaber01,conley06,hayden10,ganes11,gonzalez12}
assumed or concluded that at the earliest epochs the flux increase is proportional to $t^2$.  
This so-called ``fireball'' model is physically motivated by the idea
that the luminosity of a young SN~Ia in homologous expansion is most sensitive to its
changing radius, and less so to changes in the temperature and photospheric
velocity.  More recently, \citet{firth15} also invoked a model with a single power law, but
allowing the exponent to be a free parameter.  In a variation of this, \citet{shappee16} also
employed a power law model, but allowing the exponent to be different for each filter.

Only recently have observations of individual SNe~Ia allowed stringent tests of the 
validity of the power law model.  In two events, SNe~2011fe \citep{nugent11}
and 2012ht \citep{yamanaka14}, the fireball model (i.e, $n = 2$) seems to provide a good
fit to the earliest measurements.  The most convincing of these two cases is
SN~2011fe, which was discovered in M101 by the Palomar Transit 
Factory (PTF) with a $g$-band magnitude of 17.35 \citep{nugent11}.  PTF images 
obtained the previous night showed no source at a limiting magnitude of $g \geq 21.5$.
\citet{nugent11} found that the fireball model provided a consistent fit to the first three nights
of $g$-band measurements, and used it to infer a date of explosion only 11~hours before
discovery.  
However, as shown by \citet{piro14}, the bolometric light curve of SN~2011fe seems not to follow
the fireball model.

\citet{olling15} presented observations of the light curves of three SN~Ia
followed nearly continuously by Kepler.  The rising portion of the light curve of the
brightest of these SNe, KSN~2011b, was found to be well fitted by a single power
law, but with an exponent of $n = 2.44 \pm 0.14$.

Nevertheless, observations of SNe~2013dy \citep{zheng13} and 
2014J \citep{zheng14,goobar15} have demonstrated quite clearly that the single power law model does not
apply to all SNe~Ia.  For both of these objects, high-cadence (essentially daily) 
observations both before and after explosion revealed that the early-time light curves 
were well described by a varying power law exponent, with the flux rising nearly linearly during 
the first day, and transitioning over the next 2--4~days to a relation closer to the $t^2$ law. 
Based on the results for these two events, \citet{zheng14} speculated that the
varying power law behavior may be common to SNe~Ia, and that previous results
favoring the $t^2$ law may have been due to a lack of high-cadence observations
constraining the shape of the light curve at the earliest epochs.
This conclusion is supported by recently-published observations of the early light curve 
of iPTF16abc \citep{miller18} that also display a nearly linear rise during the first three days 
following explosion.

Figure~\ref{fig:rise2} shows a comparison between our observations of the rising light
curve of SN~2012fr with the ``broken power law'' model used by 
\citet{zheng13,zheng14} to fit the light curves of SNe~2013dy and 2014J.  The time of first 
light for these fits has been adjusted to coincide with the value we have determined for
SN~2012fr.  Note that for
both of the latter SNe, the fits were made to unfiltered photometry, and so some caution should be
taken in comparing these results with our $gr_{LSQ}$ light curve of SN~2012fr.  Nevertheless,
the resemblance is remarkable, and supports the idea that this behavior of a nearly linear rise
initially, followed by a steeper increase in luminosity may be common among SNe~Ia.

\begin{figure}[p]
\centering
\includegraphics[width=6.0in]{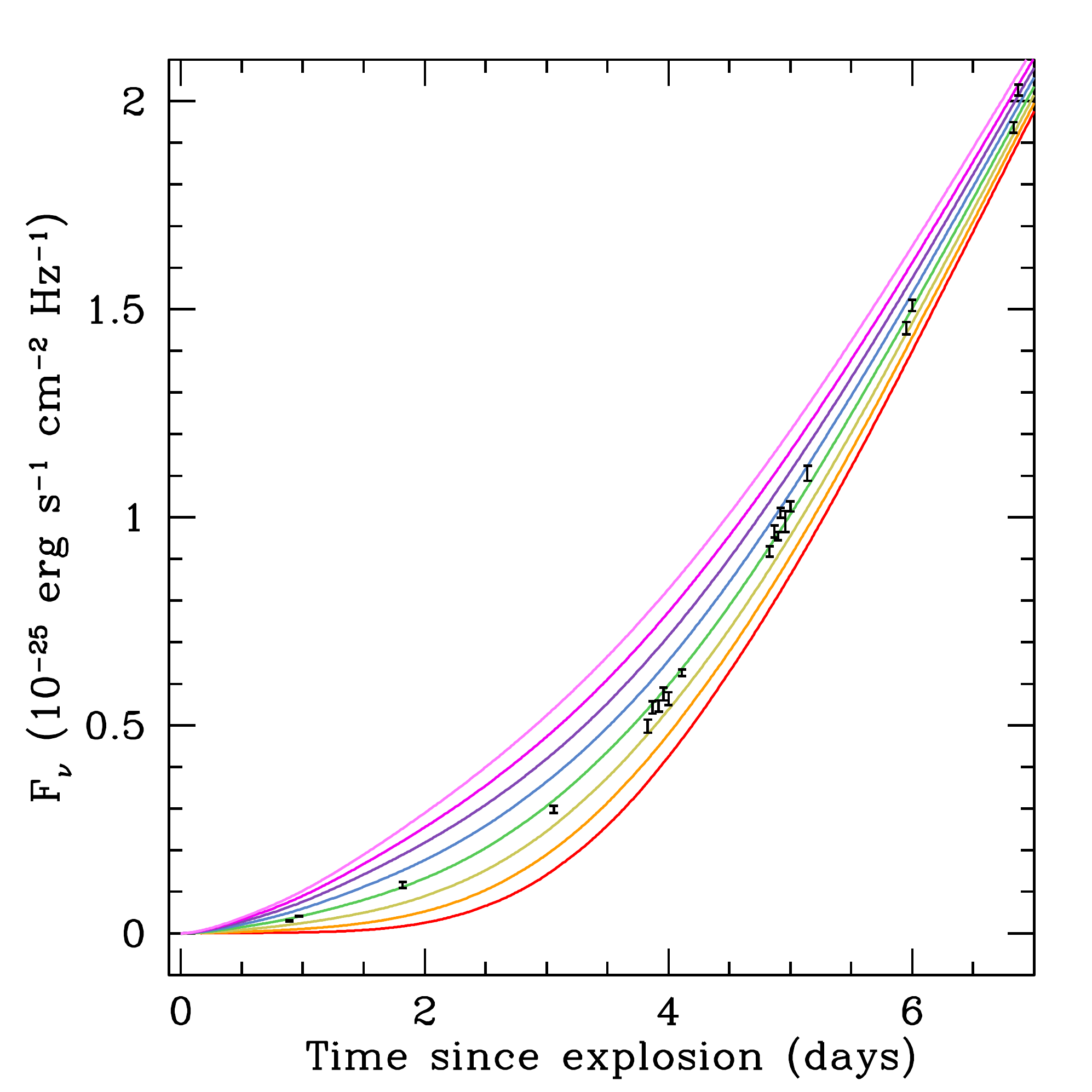}
\caption[]{Observed flux in the LSQ $gr$ band as a function of time during the first seven days
following explosion for a bare white dwarf model. The line colors indicate the level
of $^{56}$Ni mixing, which correspond to the profiles shown in Figure~\ref{fig:SNEC_mixing}.}
\label{fig:SNEC_lc}
\end{figure}

\citet{piro16} have investigated how the distribution of $^{56}$Ni in the ejecta of a SN~Ia
affects the earliest phases of the light curve using SNEC 
\citep[SuperNova Explosion Code;][]{morozova15}, an open source Lagrangian radiation 
hydrodynamics code that allowed them to initiate a shock wave within a white dwarf model, 
explode the white dwarf, and follow the early light curve evolution.
They found that models with more highly mixed $^{56}$Ni rise more quickly than do 
models with centrally concentrated $^{56}$Ni.  Using a grid of 800 models generated with 
SNEC, we have attempted to match the early light curve of SN~2012fr. A comparison of 
one of our better fitting light curves along with a range of other light curves with varying $^{56}$Ni
mixing are shown in Figure~\ref{fig:SNEC_lc} (with the best-fit light curve in green).  
This demonstrates that the steepening of SN~2012fr's light curve over the first couple days 
can naturally be accounted for by a moderately mixed $^{56}$Ni distribution. 
This corresponds to a $^{56}$Ni mass fraction of ~0.05 at roughly ~0.05$M_\odot$ below the 
surface of the exploding white dwarf. If $^{56}$Ni were not mixed out to this shallow region, 
then the theoretical light curves tend to rise too slowly in comparison to SN~2012fr. 
Whereas if the $^{56}$Ni mixing were more strong, we would not expect to see the change in 
slope in the early light curve.

\begin{figure}[p]
\centering
\includegraphics[width=6.0in]{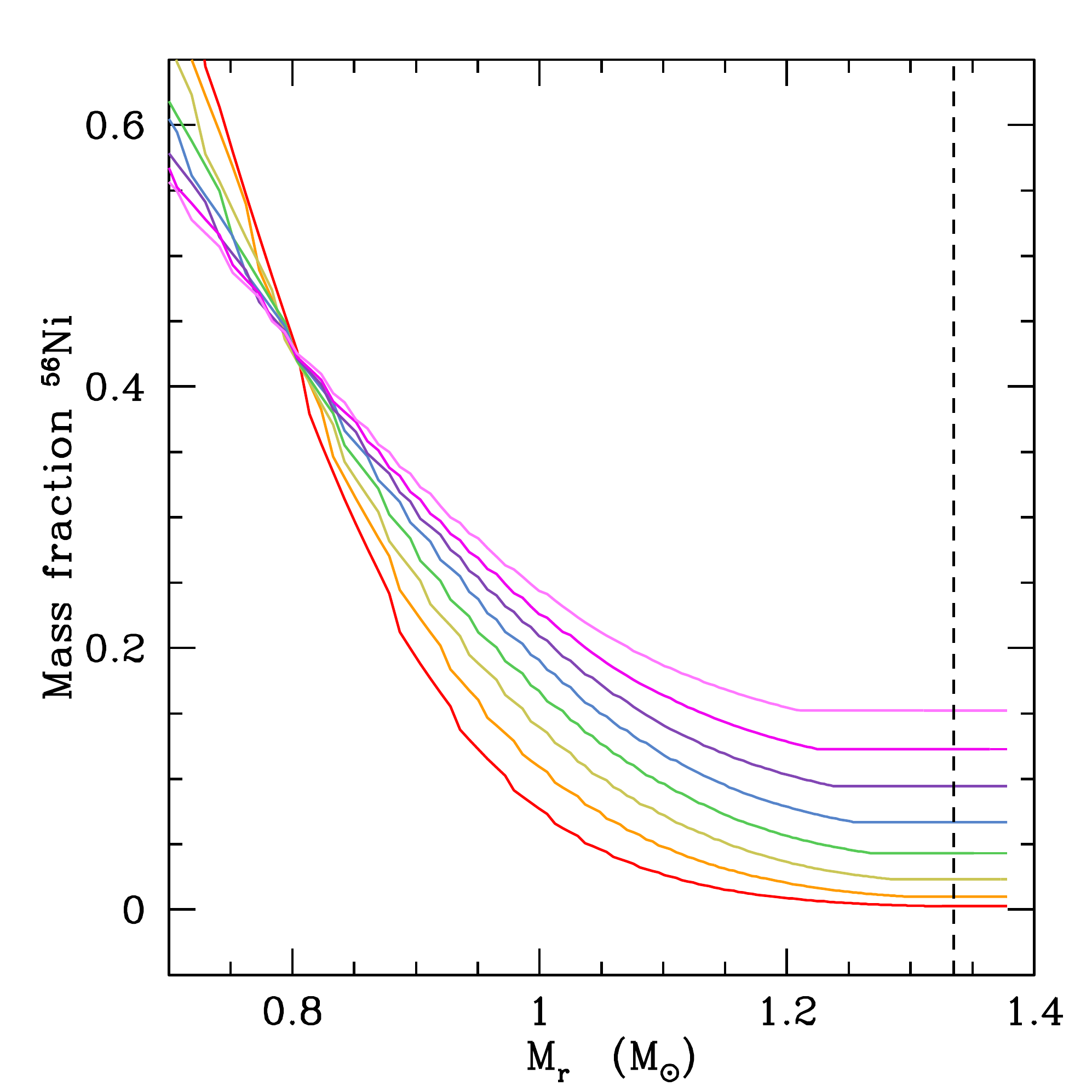}
\caption[]{Profiles of the mass fraction $^{56}$Ni as a function of the mass coordinate
in the white dwarf for the various levels of mixing corresponding to the light curves shown
in Figure~\ref{fig:SNEC_lc}.  These are constrained by the observations only out to the 
vertical dashed line.}
\label{fig:SNEC_mixing}
\end{figure}

Furthermore, this 
fit argues that the first data point for this event was  $\sim$21~hours after explosion.  
Constraining the explosion time like this is important for being able to put limits on 
interaction with a companion, as argued in \citet{shappee16}. The corresponding 
$^{56}$Ni mass fraction distributions for the models from Figure~\ref{fig:SNEC_lc} are 
shown in Figure~\ref{fig:SNEC_mixing}.

The observations only constrain the distribution of $^{56}$Ni for material interior 
to the vertical dashed line. Constraints on shallower material would require even earlier 
observations. Whether the $^{56}$Ni distribution we find for SN~2012fr is unique or not is 
not clear, since there may be degeneracies between various physical parameters when only 
photometric data is considered \citep{noebauer17}. This will be explored in more 
detail in a forthcoming work (Piro et al., in preparation). In the future, it will likely 
be useful to also have spectral information at these early epochs to uniquely determine 
the physical cause of the early light curve shape.

\subsection{Spectroscopic Peculiarities}
\label{sec:normal}

As more and more SNe~Ia have been observed, subclasses within the general phenomenon
have been identified.  \citet{benetti05} divided SNe~Ia into three groups based on the 
light curve decline rate, $\dmB$, and the post-maximum evolution of the velocity of the 
minimum of the \ion{Si}{2} $\lambda$6355 absorption.  The ``FAINT'' group, represented by the
prototypical SN~1991bg \citep{filippenko92,leibundgut93,turatto96}, consists of fast-declining events 
($\dmB > 1.5$~mag) with low  \ion{Si}{2} velocities at maximum which
decrease rapidly with time.  They then further split SNe~Ia with ``normal'' 
decline rates ($\dmB < 1.5$~mag) into two 
further categories:  the ``HVG'' group, which display a high temporal velocity gradient
($\dot{v} > 70$~\kms~day$^{-1}$) in the days following \tmax, and the
more common ``LVG'' events, showing a lower velocity gradient.
The HVG events typically also have high \ion{Si}{2} velocities at maximum.
This fact led \citet{wang09} to propose a parallel subtype classification based solely on a 
measurement of the \ion{Si}{2} $\lambda$6355 velocity within a week of maximum. 
\citeauthor{wang09} termed those having velocities 
$\gtrsim 11,800$~\kms\ at the epoch of \tmax ``high-velocity'' (``HV'') SNe~Ia.
Lower-velocity events not including the 1991bg-like and 
1991T-like spectral types \citep{branch93} were termed ``normal'' by \citeauthor{wang09}.  
As shown in Figure~2 of \citet{foley11} and Figure~6 of \citet{silverman12}, 
$\sim$80-90\% of HV events also pertain to the HVG group.   Conversely, most LVG 
objects have ``normal'' \ion{Si}{2} velocities in the \citeauthor{wang09} classification
system \citep{silverman12}.\footnote[30]{Note that the LVG subtype as defined by \citeauthor{benetti05}
includes 1991T-like objects.}

\citet{branch06,branch09} developed an independent classification system for SNe~Ia
based on a plot of the pseudo-equivalent widths of the \ion{Si}{2} $\lambda$5972 and $\lambda$6355
absorption features.  Four groups were established: ``Core Normal'' (CN), 
``Cool'' (CL), ``Broad Line'' (BL), and ``Shallow Silicon'' (SS).  Figure~\ref{fig:branch} 
displays these groups in a ``Branch diagram'' using the data and
classification criteria of \citet{blondin12}.  There is a rough correspondence 
between the BL and HV (or HVG) groups, with \citet{blondin12} and \citet{folatelli13}
independently finding that two-thirds of SNe~Ia classified as BL also belong to the HV subtype.  The CL type
correlates well with the \citeauthor{benetti05} ``FAINT'' and \citeauthor{wang09} ``1991bg'' groups,
and the LVG (or \citeauthor{wang09} ``normal'') SNe generally encompass the CN and SS classes.  
Finally, the SS group
includes 1991T-like SNe, but also the peculiar 2002cx-like events \citep{li03}.

So how does SN~2012fr fit into these classification schemes?
\citet{childress13} found that SN~2012fr lies on the border  separating the \citeauthor{branch06} 
CN and SS subclasses (see Figure~\ref{fig:branch}).  The shallow  SN~2012fr 
\ion{Si}{2} velocity gradient displayed by SN~2012fr also places it clearly
in the \citeauthor{benetti05} LVG group.  Nevertheless, the high
\ion{Si}{2} velocity at maximum of $\sim$12,000~\kms\ qualifies SN~2012fr as an
HV event in the \citeauthor{wang09} system.  This is an unusual combination of
classifications since 
5\% or less of SNe~Ia in the CN $+$ SS groups belong to the HV subtype,
and $< 10$\% of LVG SNe~Ia are also classified as HV events 
\citep{blondin12,folatelli13}.

Figure~\ref{fig:vels_11fe_12fr} shows our measurements of the  evolution 
of the expansion velocities of the \ion{Si}{2}~$\lambda$6355, 
\ion{Si}{3}~$\lambda\lambda$4564,5740, \ion{S}{2}~$\lambda\lambda$5449,5622, 
\ion{Ca}{2}~$\lambda$8662, and \ion{Fe}{2}~$\lambda\lambda$4924,5018,
absorption minima from the \citet{childress13} spectra of SN~2012fr.  Expansion
velocities of the \ion{Mg}{2}~$\lambda$10927 line measured from unpublished
CSP-II NIR spectra of SN~2012fr are also included.
For comparison, we plot the expansion velocities of the 
same features as measured from the spectra of the prototypical SN~2011fe 
published by \citet{Pereira13} and \citet{hsiao13}.  The small amount of velocity stratification of the
intermediate mass elements (IMEs) in SN~2012fr from $-5$~days onward is remarkable, 
and in stark contrast to that observed
for SN~2011fe.  \citet{childress13} concluded that either there is a shell-like density
enhancement in the ejecta at a velocity of $\sim$12,000~\kms, 
or a sharp cutoff in the radial distribution of the IMEs in the ejecta. 
As discussed
by \citet{quimby07} in the context of the slow-declining SN~2005hj --- a
\citet{wang09} ``normal'' event which showed a nearly flat \ion{Si}{2}
velocity gradient ---  a strong density enhancement is not predicted
by standard delayed-detonation and deflagration models, but 
instead is suggestive of rapidly expanding material interacting 
with overlying material.  Examples of scenarios that produce
a well-defined shell of IMEs are pulsating delayed-detonations 
\citep{khokhlov93}, or tamped detonations in a double degenerate
merger \citep{fryer10}.   This possibility will be explored in more detail in a 
future paper (Cain et al., in preparation).

In the case of SN~2012fr, \citet{childress13} argued that it is more likely that 
the small velocity evolution of the \ion{Si}{2} and \ion{Ca}{2} lines
reflects the fact that these ions are physically
confined to a narrow region in velocity space in the ejecta.   
This conclusion was based in part on the 
observation that the expansion velocity of the \ion{Fe}{2} lines in 
SN~2012fr began to slowly decrease $\sim$10 days after maximum while the \ion{Si}{2}
and \ion{Ca}{2} lines remained at a nearly constant velocity
of 12,000~\kms (see Figure~\ref{fig:vels_11fe_12fr}).  Certainly
one interpretation of Figure~\ref{fig:vels_11fe_12fr} is that the 
inner edge of the IMEs in the ejecta of SN~2012fr was located 
at a velocity of $\sim$11,000~\kms, in contrast to a typical 
SN~Ia as represented by 2011fe for which an abundance
tomography analysis indicates that the inner edge of the IMEs 
extended down to $\sim$5,000~\kms\ \citep{mazzali15}. 
This suggests a $^{56}$Ni distribution that extended out to higher velocities
than normal -- i.e., at the phase range where a typical SN~Ia has its photosphere 
still in the silicon-rich layer, the photosphere of SN~2012fr may have already receded 
into the $^{56}$Ni-rich layer due to its extended distribution.
\citet{hoeflich02} presented one-dimensional, delayed-detonation models of a Chandrasekhar mass
white dwarf that reproduce the luminosity--decline rate relation, from the sub-luminous
to the most luminous SNe~Ia,  by varying the density at which the deflagration transitions 
to a detonation, $\rho_{tr}$, from values of 8--27~$\times 10^{6}$~gm cm$^{-3}$.
The same model with a slightly higher transition density of 
$\rho_{tr} = 30 \times 10^{6}$~gm cm$^{-3}$ produces a minimum
velocity of Si/S of 12,983~\kms, $^{56}$Ni mass = 0.67~$M_{\odot}$, and $M_V = -19.35$~mag
(from private communication with Peter H\"oflich).

\citet{maund13}, \citet{childress13} and \citet{zhang14} 
have pointed out that SN~2012fr shares certain properties with luminous
events such as SN~1991T \citep{filippenko92,phillips92,ruiz92}, 
SN~1999aa \citep{li01b,garavini04}, and the above-mentioned 
SN~2005hj: slow-declining light curves,
weak \ion{Si}{2} $\lambda$6355 absorption, and shallow 
photospheric velocity gradients. From this coincidence and
the strong presence of HVFs during the pre-maximum phases,
\citet{zhang14}  proposed that SN~2012fr may 
represent a subset of the 1991T-like SNe Ia viewed at an
angle where the ejecta has a clumpy or shell-like structure.
However, 
SN~2012fr differs from 1991T-like events in three important ways: 
the high photospheric velocity of the \ion{Si}{2} $\lambda$6355 
absorption at maximum light, the strong \ion{Si}{2} absorption 
observed at early epochs, and the lack of strong features due to
\ion{Fe}{3} at maximum.

\subsection{A Distinct Subclass?}
\label{sec:subclass}

\begin{figure}[p]
\centering
\includegraphics[width=6.0in]{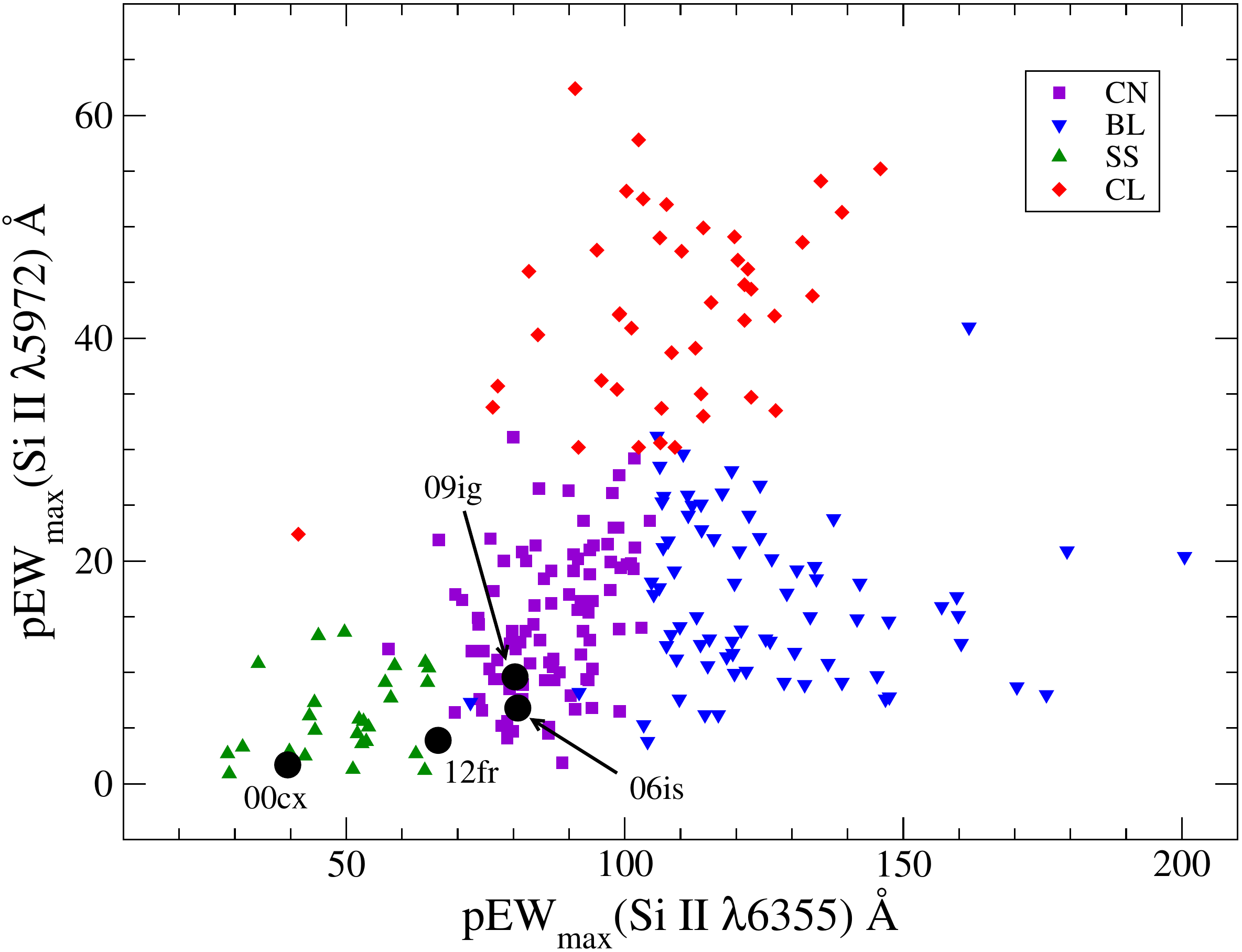}
\caption[]{Plot of the pseudo equivalent width of the \ion{Si}{2}~$\lambda$5972 absorption vs. that of the 
\ion{Si}{2}~$\lambda$6355 absorption within five days from maximum-light, illustrating the four subtypes
of SNe~Ia defined by \citet{branch06}.  Core Normal (CN; violet squares), Shallow Silicon 
(SS; green upward-looking triangles), Broad Line (BL; blue downward-looking triangles), and 
Cool (CL; red diamonds) subtypes taken from the data of  \citet{blondin12} are plotted.  The positions
of the shallow velocity gradient SNe~2000cx, 2006is, 2009ig, and 2012fr are indicated by black
circles.}
\label{fig:branch}
\end{figure}

\begin{figure}[p]
\centering
\includegraphics[width=6.5in]{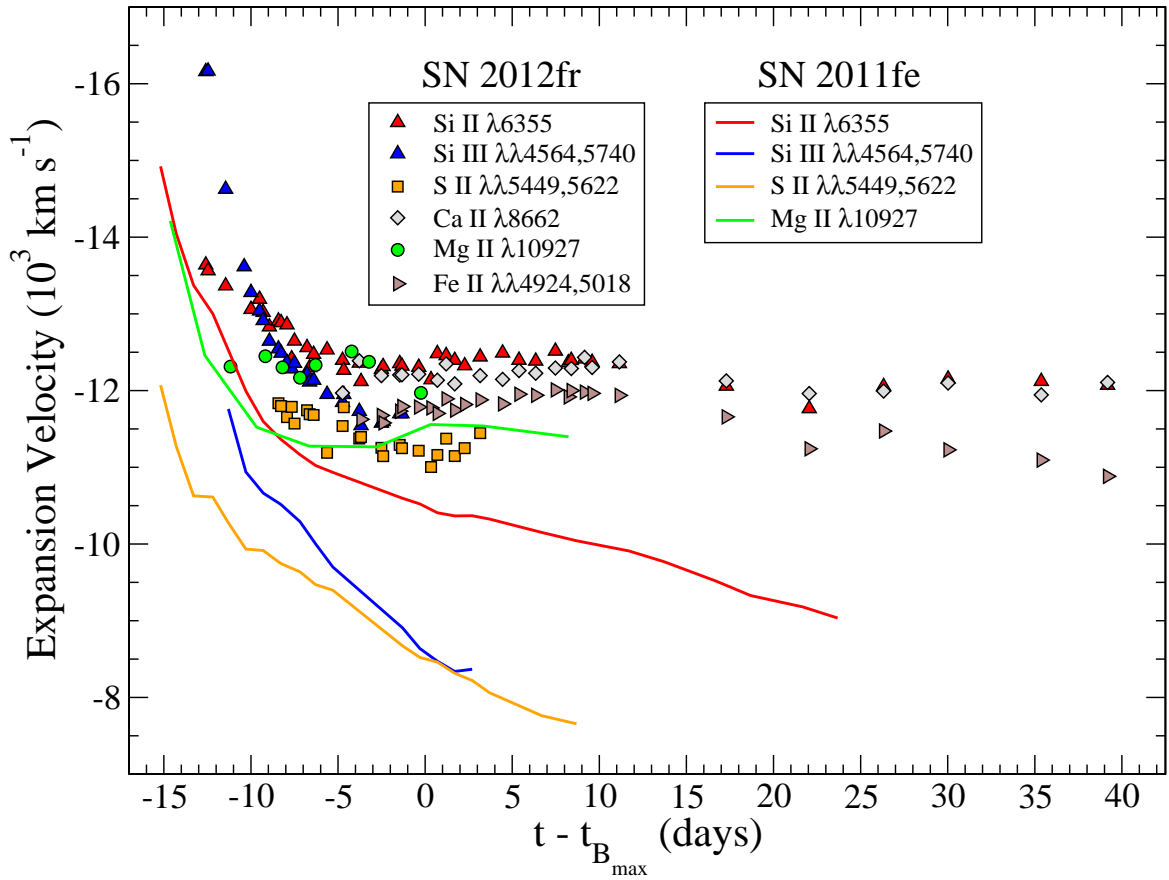}
\caption[]{Comparison of the velocity stratification of different ions for SNe 2011fe and 2012fr.
The measurements correspond to the absorption minima of each line.}
\label{fig:vels_11fe_12fr}
\end{figure}

\begin{figure}[p]
\centering
\includegraphics[width=5.7in]{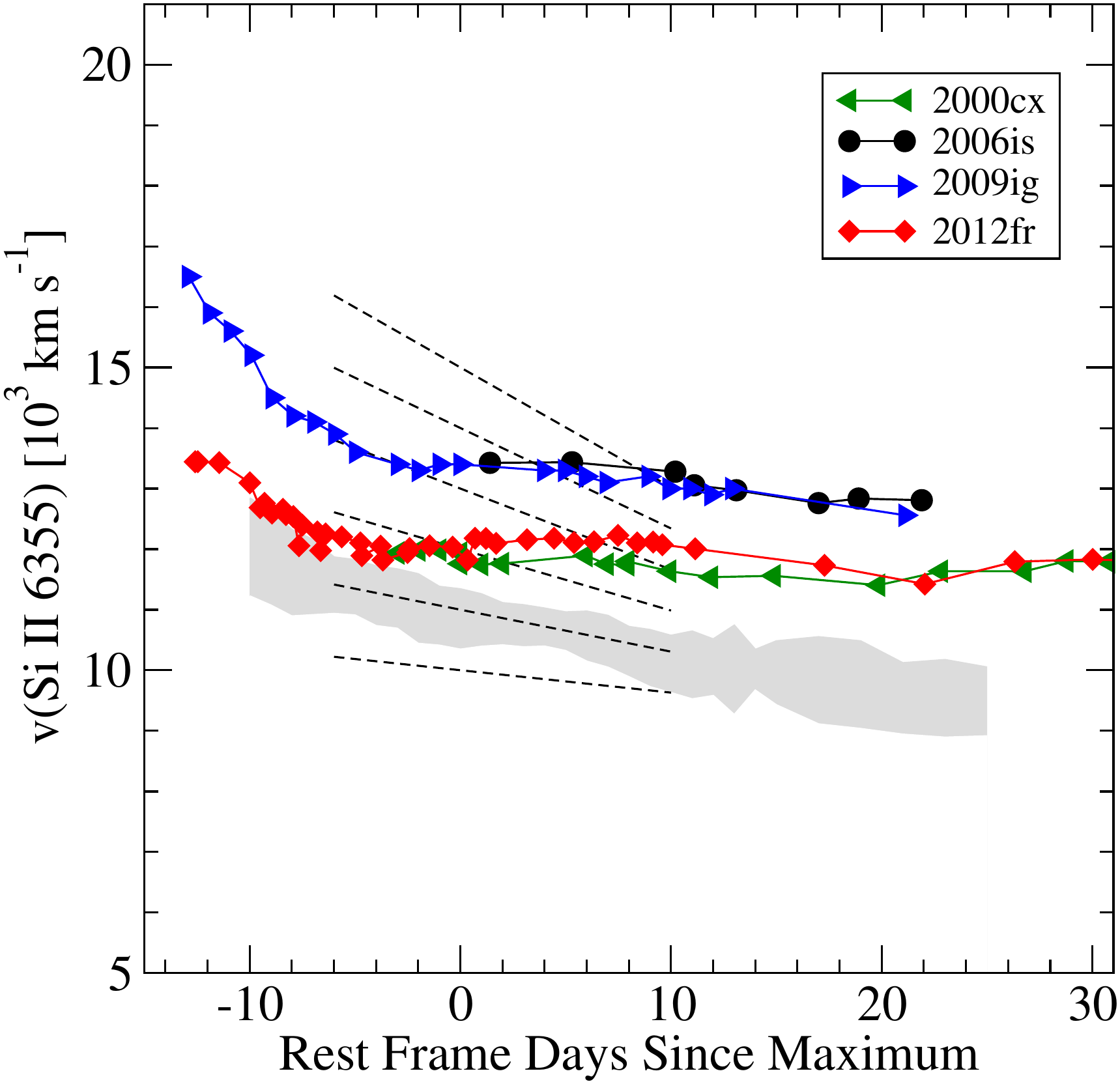}
\caption[]{Evolution of the photospheric \ion{Si}{2}~$\lambda$6355 velocity for the shallow velocity gradient
SNe~2000cx, 2006is, 2009ig, and 2012fr.  The shaded area shows the average and 1-$\sigma$
dispersion of Si II velocities of ``normal'' SNe~Ia reproduced from \citet{folatelli13}, 
while the dashed lines represent a subset of the 
family of functions that describe the velocity evolution of normal and high-velocity gradient SNe~Ia
\citep{foley11}.  The measurements for SNe~2000cx, 2009ig, and 2012fr are taken
from \citet{li01}, \citet{marion13}, and \citet{childress13}, while those for SN~2006is were
made using the spectra of \citet{folatelli13}.}
\label{fig:velscomp}
\end{figure}

Figure~\ref{fig:velscomp} illustrates the unusual nature of the \ion{Si}{2} $\lambda$6355 
photospheric velocity evolution of SN~2012fr beginning around 15~days before \tmax.
Shown for comparison is the 1-$\sigma$ dispersion about the average of the velocities 
for ``normal'' SNe~Ia in the \citeauthor{wang09} system \citep{folatelli13}, 
while the dashed lines correspond to a subset of the family of functions derived by 
\citet{foley11} to describe the velocity evolution of LVG and HVG SNe~Ia.  The 
extraordinarily shallow velocity evolution of SN~2012fr is unmatched but by only a
few other HV events.  The evolution of the \ion{Si}{2} velocity for the two best 
observed of these --- SN~2006is \citep{folatelli13} and SN~2009ig \citep{marion13}  --- is 
shown for comparison in Figure~\ref{fig:velscomp}.  The positions
of these two SNe in the ``Branch diagram'' are also indicated in Figure~\ref{fig:branch}.

\begin{figure}
\centering
\leavevmode
\includegraphics[width=6.5in]{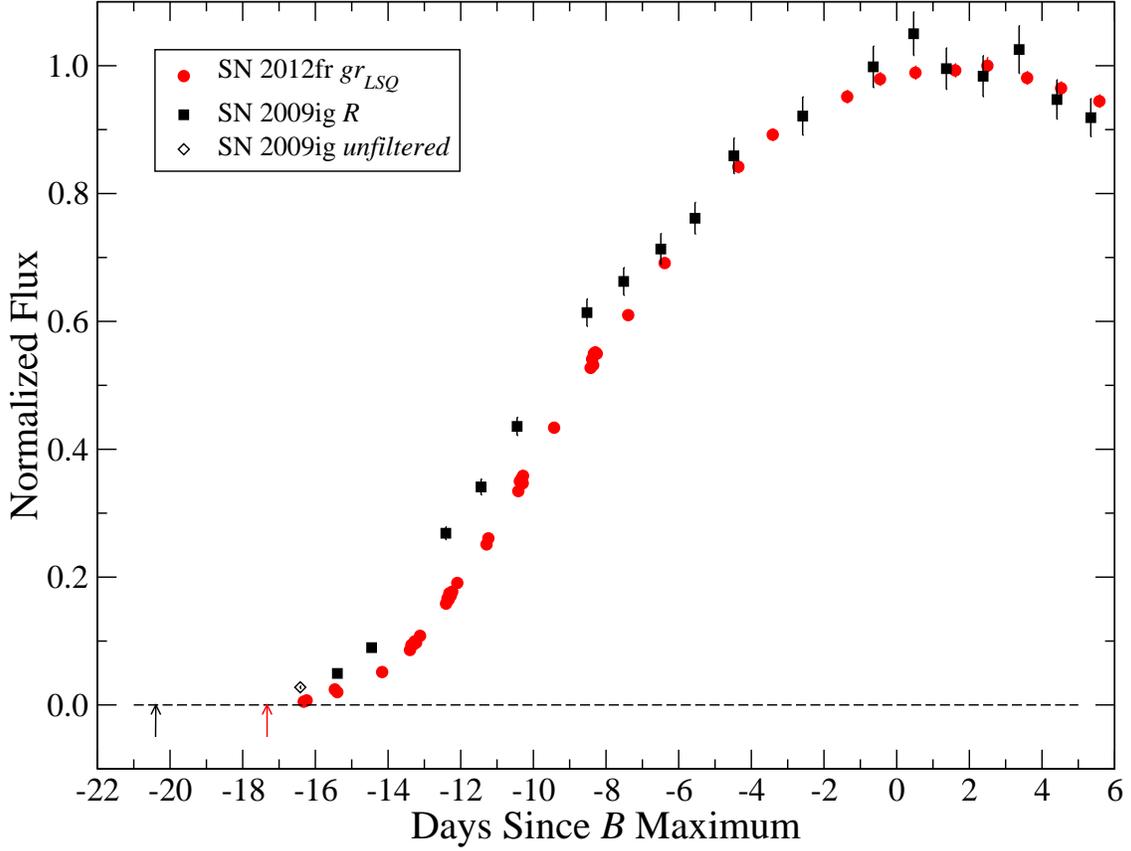}
\caption[]{Relative fluxes (normalized to maximum) plotted as a function of the time with respect to \tmax\ 
for the $R$-band light curve of SN~2009ig and the $gr_{LSQ}$ light curve of SN~2012fr.
The photometry for SN~2009ig is taken from \citep{foley12}.  The unfiltered first observation 
was approximately S-corrected assuming a color $(B-V) = 0.6 \pm 0.1$ and a sensitivity function similar to those 
of the Tarot and BOSS unfiltered photometry.  The black arrow shows the last non-detection before discovery
for SN~2009ig; the red arrow indicates the same for SN~2012fr. The abscissa is corrected for time
dilation.}
\label{fig:09ig_12fr_rise}
\end{figure}

SN~2009ig was extensively observed by \citet{foley12} and \citet{marion13}.  This 
slow-declining SN ($\dmB =  0.89$~mag)
was caught very early and displayed strong HVF absorption in \ion{Ca}{2} and
\ion{Si}{2} in the first spectra obtained $\sim$14~days before \tmax.
The earliest spectra were dominated by the HVFs, with the first
detection of photospheric features beginning 12~days before \tmax.
Both the HVF and photospheric absorption remained visible until
6~days before \tmax, when the HVF absorption was no longer
detectable. The optical spectral evolution of the
SN~2009ig was remarkably similar
to that of SN~2012fr \citep{childress13}, with the only 
significant differences being (1) the higher velocity, broader HVF absorption
in the earliest (day $-14$) spectrum of SN~2012fr, (2) the longer persistence of the 
HVF \ion{Ca}{2} and \ion{Si}{2} absorption in SN~2012fr,
and (3) the narrower \ion{Ca}{2}, \ion{Fe}{2}, \ion{S}{2}, and \ion{Si}{2}
photospheric absorption in SN~2012fr from maximum onward (seen most 
clearly in the Ca II triplet).
The photometric evolution of both SNe was also striking similar, with the
most significant difference being in the $I$ band where 
the secondary maximum of SN~2012fr occurs several days later than 
that of SN~2009ig. Like SN~2012fr, SN~2009ig was discovered $< 1$~day
after explosion \citep{foley12}.  The $R$-band light curve
of SN~2009ig is compared with the $gr_{LSQ}$ light curve of SN~2012fr in 
Figure~\ref{fig:09ig_12fr_rise}, from which it is clear that the rise time to \tmax
of SN~2009ig was somewhat longer than that of SN~2012fr.  The unfiltered
observation corresponds to the discovery image of SN~2009ig.  Unfortunately, 
the last non-detection was four days before discovery, 
making it difficult to determine with certainty that the early rise of SN~2009ig 
showed the same ``broken power law'' morphology observed for SN~2012fr.
However, we note that the shape of $R$-band light curve of SN~2009ig
closely mimics that of the $gr_{LSQ}$ light curve of SN~2012fr for 
much of the rise to maximum, which suggests a similar morphology at the
earliest epochs following explosion. 

SN~2006is, has been discussed by \citet{folatelli13}.
Photometrically, it was also quite a slow
decliner, with $\dmB =  0.80$~mag \citep{stritzinger11}.
Unfortunately, spectroscopic observations did not
begin until maximum, and so it is unknown whether this SN
also displayed strong HVF absorption at early epochs.  Nevertheless, as shown
in Figure~14 of \citet{folatelli13}, the
maximum-light spectra of SNe~2006is and 2009ig were remarkably
similar.

From the above discussion, we conclude that SNe~2006is, 2009ig, and 2012fr 
were similar events sharing the following characteristics:

\begin{itemize}
\item All three were slow decliners ($\dmB = $~0.8--0.9 mag).
\item All three occupied a similar region of the 
``Branch'' diagram near the edge of the CN and SS distributions
(see Figure~\ref{fig:branch}).
\item All three displayed unusually shallow \ion{Si}{2} velocity gradients consistent
with LVG events in the \citet{benetti05} classification scheme, but at 
velocities~$\gtrsim 12,000$~\kms\ that place them in the \citet{wang09} HV class.
\end{itemize}

\noindent Moreover, SNe~2009ig and 2012fr displayed remarkably strong HVFs that persisted
to maximum-light.  Unfortunately, SN~2006is was not discovered early enough to say if it
also shared this characteristic.

Very few other SNe~Ia have displayed this particular combination of properties,
the notable exception being the peculiar SN~2000cx which
was extensively observed by \citet{li01} and \citet{candia03}.
In addition to the high, nearly constant evolution of the
\ion{Si}{2} velocities (see Figure~\ref{fig:velscomp}), SN~2000cx displayed remarkably strong \ion{Ca}{2} 
HVFs at velocities $> 20,000$~\kms\ that persisted to maximum-light 
\citep{branch04,thomas04}.   Interestingly, the evolution
of the expansion velocities of the IMEs as deduced
from the \ion{Si}{2}~$\lambda$6355 and 
\ion{S}{2}~$\lambda\lambda$5449,5622 
lines \citep{li01} closely resembled that observed for SN~2012fr.
Optical and NIR photometry of SN~2000cx also revealed
certain anomalies in its photometric behavior:
(1) an asymmetric $B$-band light curve, with a relatively fast rise from $-10$~days to 
maximum, but then a slow post-maximum decline ($\dmB =  0.93$~mag);
(2) a weaker and earlier-occurring $I$-band secondary maximum than would be
expected for such a slow decline rate, and
(3) a $(B-V)$ color evolution displaying several peculiarities including a brief ``plateau''
phase of nearly constant color beginning $\sim$1~week after maximum, and a strikingly
bluer color than predicted by the Lira relation from $\sim$35--90 days after
maximum.

\begin{figure}
\centering
\leavevmode
\includegraphics[width=6.5in]{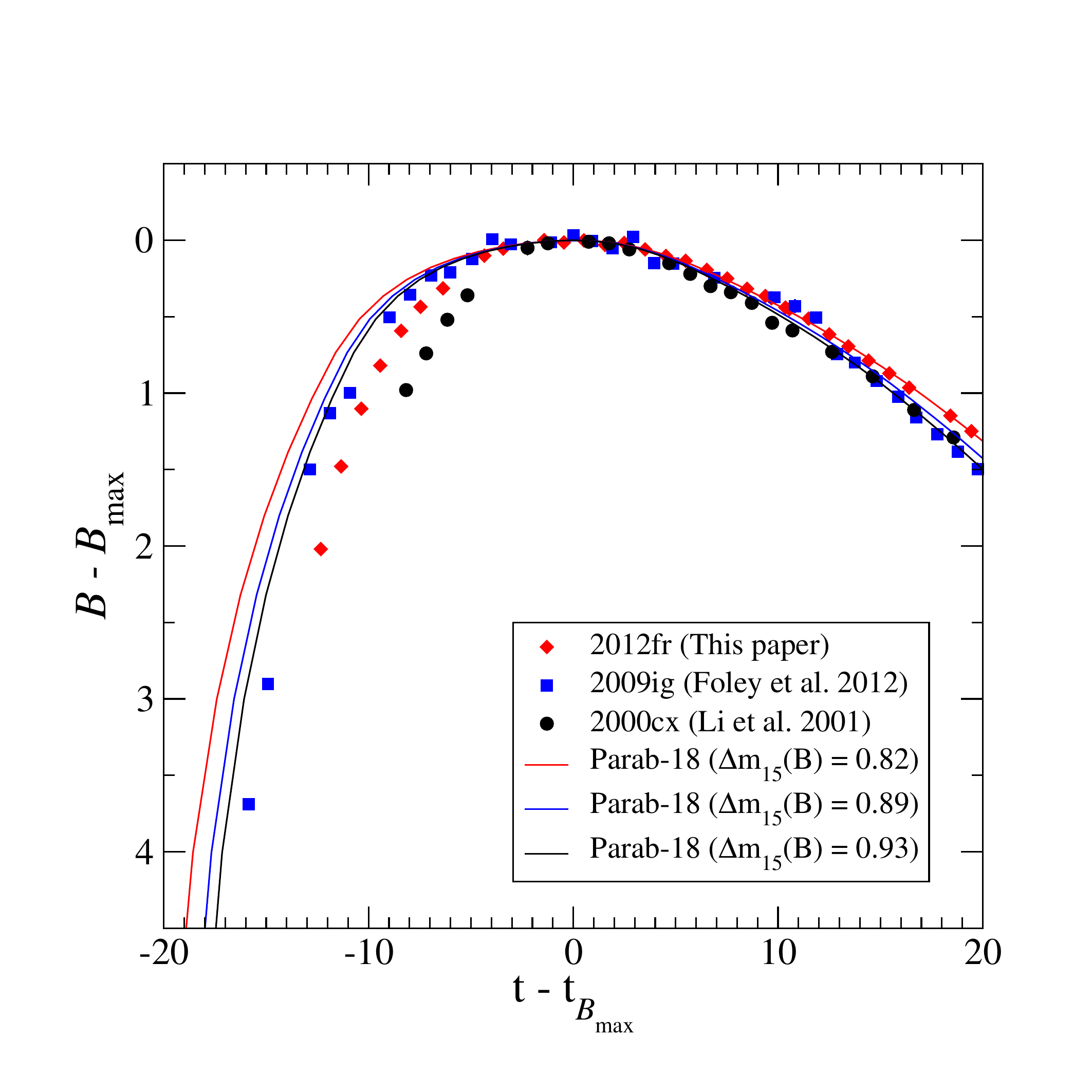}
\caption{Comparison of $B$-band light curves for SN~2000cx 
($\dmB = $~0.93~mag), SN~2009ig ($\dmB = $~0.89~mag), 
and SN~2012fr ($\dmB = $~0.82~mag).  The abscissa is corrected for time
dilation.  Shown for comparison is the $B$-band template of \citet{goldhaber01} 
stretched to the decline rates of the three SNe.}
\label{fig:lcB}
\end{figure}

\begin{figure}
\centering
\leavevmode
\includegraphics[width=6.5in]{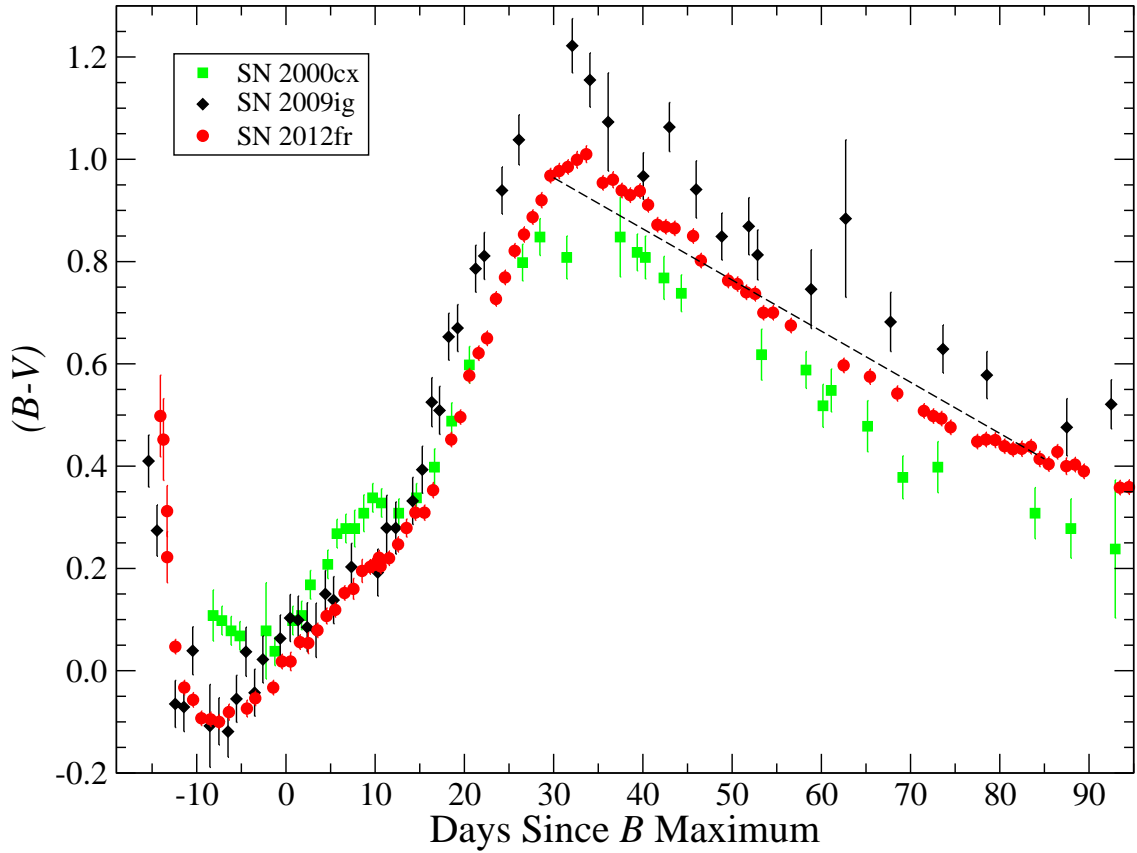}
\caption{Comparison of the $(B-V)$ color evolution of SN~2000cx (green squares), 
SN~2009ig (black diamonds), and SN~2012fr (red circles). The abscissa is corrected for time
dilation.  The data for SN~2000cx are from \citet{li01}, for SN~2009ig are from 
\citet{foley12}, and for SN~2012fr are from this paper.  No correction has been made for
host-galaxy reddening for any of the SNe.  The dashed line is the average
Lira relation from \citet{burns14}.}
\label{fig:09ig_12fr_00cx_B-V}
\end{figure}

It is tempting to postulate that SNe~2006is, 2009ig, and 2012fr were closely 
related to SN~2000cx, and to its apparent twin, SN~2013bh \citep{silverman13}. 
One difference is that SN~2000cx was a fairly extreme SS event in the
\citet{branch06,branch09} classification scheme (see Figure~\ref{fig:branch}).  
Interestingly, SN~2012fr and, to a lesser extent, SN~2009ig, 
also resembled SN~2000cx in displaying a $B$-band light curve that
rose rather quickly to maximum, but then declined more slowly than average
after maximum.  This is illustrated in Figure~\ref{fig:lcB}.  Here the normalized
$B$ light curves are plotted with respect to \tmax, with time
dilation taken into account.  Also plotted is the \citet{goldhaber01} $B$-band
template ``Parab-18''  stretched to match the observed decline rates of each
SN.  The unusually rapid rise to maximum of SN~2000cx stands out
clearly in this figure, and is mimicked to a large extent by SN~2012fr.
SN~2009ig also initially appeared to rise somewhat more quickly than the 
\citeauthor{goldhaber01} template, but this difference would not have been so 
obvious if this event had not been discovered so early.   \citet{li01} also called
attention to a peculiarity in the $(B-V)$ evolution of SN~2000cx, which showed
a phase of nearly constant color at $(B-V) \sim 0.3$~mag between 6--15~days
past \tmax.  As seen in Figure~\ref{fig:09ig_12fr_00cx_B-V},
the $(B-V)$ evolution of SN~2012fr showed a similar, nearly 
constant color of $(B-V) \sim 0.2$~mag for a few days centered 
around day $+10$.  Unfortunately, the photometry of SN~2009ig is not 
sufficiently precise to discern whether it also displayed such a peculiarity,
but Figure~\ref{fig:09ig_12fr_00cx_B-V} shows that a change in the slope of 
the $(B-V)$ evolution of this SN occured around day $+10$,
as it did for SN~2012fr.

\section{CONCLUSIONS}
\label{sec:summary}

We have presented densely-sampled, high-quality, six-band optical photometry 
of the Type Ia SN 2012fr in the Fornax Cluster member NGC~1365.  
The data span epochs from 13~days before to 140~days after the epoch of \tmax
with typical errors below 2\%.  We also present similarly high-quality NIR and 
UV photometric data sets.  Based on these observations, we conclude the
following:

\begin{itemize}
\item SN~2012fr was a slow declining ($\dmB =  0.82 \pm 0.03$ mag),
luminous event.  From the observed colors at maximum, the evolution of the $(B-V)$ 
color at later epochs, and the lack of detectable host-galaxy interstellar \ion{Na}{1}~D absorption
in high-dispersion spectra, we estimate that SN~2012fr suffered little or no host-galaxy 
reddening, adopting a conservative value of $E(B-V)_{host} = 0.03 \pm 0.03$~mag.  

\item Analysis of the optical and NIR light curves shows that the luminosity 
of SN~2012fr was completely normal for its decline rate.

\item Images obtained by the LSQ survey tightly constrain the epoch of explosion to
$16.87 \pm 0.46$ days before \tmax, or $16.5 \pm 0.5$~days before the 
bolometric maximum.  The luminosity of the SN increased nearly linearly
at first, transitioning to a faster rising phase $\sim$2.5~days after explosion.
This behavior is well-fitted by an explosion model with moderate mixing
of $^{56}$Ni in the ejecta.

\item The densely-sampled  bolometric light curve derived from our
UV, optical, and NIR photometry indicates that SN~2012fr peaked at a
luminosity of L$_{Bol} = 1.35 \pm 0.14 \times 10^{43}$~ergs~s$^{-1}$.
Combining this value with the measured rise time, we estimate
a $^{56}$Ni mass of $0.60 \pm 0.15~M_{\odot}$ using Arnett's rule.
This amount is consistent with an independent measurement of the
$^{56}$Ni mass from nebular spectra.

\item In spite of its normal luminosity, SN~2012fr displayed spectroscopic 
 properties that set it apart from most other SNe~Ia.  The
classification as a \citet{branch06} SS/CN event, with HV \ion{Si}{2} absorption
in the \citet{wang09} system but a shallow velocity gradient (LVG using the
\citet{benetti05} criteria) is uncommon and reminiscent of the peculiar
SN~2000cx.  Like SN~2000cx, SN~2012fr also displayed a fast rise to maximum,
and a much slower post-maximum decline.  Finally, the velocity of the inner edge of the IMEs in
SN~2012fr appears to have been $\sim$11,000~\kms, or approximately double that of a
typical SN~Ia such as SN~2011fe.  We call attention to two other SNe~Ia, 2006is and
2009ig, that showed photometric and spectroscopic properties similar to those of
SN~2012fr, and suggest that all three, along with SN~2000cx, may form a 
distinct sub-class of SNe~Ia.

\item With a Cepheid-based distance to its host-galaxy already available, SN~2012fr adds to 
the still relatively-small number of nearby SNe~Ia that are suitable for
measuring the value of the Hubble constant, $H_{\circ}$.

\end{itemize}

\acknowledgments 

We acknowledge WeiKang Zheng for his help on providing discovery images of SN~2009ig and
useful discussion.
The work of the CSP has been supported by the National Science Foundation under 
grants AST0306969, AST0607438, AST1008343, AST1613426, AST1613455, and AST1613472.
M. Stritzinger and C. Contreras acknowledge
generous support from the Danish Agency for Science and Technology and Innovation
through a Sapere Aude Level 2 grant. E.~Y.~H is supported by Florida Space Grant 
Consortium. M. Stritzinger is also supported by a research grant (13261) from VILLUM FONDEN. 

\appendix

\section{THROUGHPUT FUNCTIONS AND ZERO POINTS FOR THE LSQ, TAROT, SLOOH, AND
BOSS FILTERS}
\label{sec:filter_functions}

\subsection{La Silla-QUEST}
\label{sec:quest_phot}

The relative throughput of the LSQ $gr$ filter is plotted in panel ``a'' of Figure~\ref{fig:opt_filters}.
This was constructed by multiplying the filter and CCD quantum efficiency curves given by
\citet{baltay07, baltay13} with a reflectivity curve for aluminum and an atmospheric transmission
spectrum appropriate for the La Silla Observatory.  The transmission of the corrector
of the ESO Schmidt telescope was assumed to be flat over the spectral region covered by the filter.
LSQ images were processed by the LSQ pipeline before being given to us.

We measured instrumental PSF photometry
of SN~2012fr and field stars in the LSQ images using DAOPhot  \citep{stetson87} routines.
Final magnitudes for the local sequence stars in the natural system of the LSQ $gr$ filter were calculated 
via the following procedure:

\begin{itemize}

\item Figure~\ref{fig:gr} shows color-color diagrams of $(V_{CSP}-gr_{LSQ})$ and 
$(r_{CSP}-gr_{LSQ})$ colors versus $(g-r)_{CSP}$ derived from synthetic photometry of
main sequence stars,\footnote[31]{According to \citet{finlator00} 99\% of the field stars observed in the
Sloan Digital Sky Survey are on the main sequence.} selected from the \citet{pickles98} 
stellar library spectra.  The zero point of the synthetic photometry using the LSQ~$gr$ filter 
throughput function was set by requiring 
$gr_{LSQ} = 0.0$ for Vega ($\alpha$~Lyr).\footnote[32]{Vega spectrum from
CALSPEC: ftp://ftp.stsci.edu/cdbs/current\_calspec/ascii\_files/alpha\_lyr\_stis\_005.ascii} 
Likewise, synthetic magnitudes in the 
CSP $V$, $g$, and $r$ filters were calculated using the throughput functions and zero points given
by \citet{krisciunas17}.

\item Photometry of the local sequence stars in the field of SN~2012fr are plotted in Figure~\ref{fig:gr} 
as red circles.  The zero point for the instrumental magnitudes of the local sequence stars was found by 
shifting the photometric measurements along the y-axis to match the sequence defined by the Pickles stars.

\end{itemize}

\noindent The resulting magnitudes for the local sequence stars in the natural system of the LSQ $gr$ filter
are given in Table~\ref{tab:locseq2}.  In Table~\ref{tab:sn_nat_sys}, our measurement of the magnitude 
of SN~2012fr on 26~Oct~2012~UT in the natural system of the LSQ $gr$ filter is given.

SN~2012fr was not detected in a pair of $gr_{LSQ}$ images acquired on 25.34~Oct~2012~UT (see 
Figure~\ref{fig:ds9}).  We performed aperture photometry at the location of SN 2012fr on these images
using the IRAF {\tt apphot} package and calibrated the results using the local sequence stars.  This
procedure implies 3-$\sigma$ limits of $gr_{LSQ} > 20.38$ mag and $gr_{LSQ} > 20.34$, 
respectively.  This limit was verified by placing artificial sources with increasing magnitudes at the 
location of SN 2012fr in these images and carrying out photometry on the resulting images.   
The artificial source was recovered when it was $m$=20.34 mag, in excellent agreement with 
the measured upper limits above.

\subsection{TAROT}
\label{sec:tarot_phot}

The throughput of the TAROT ``open'' system was calculated from the detailed information
given in \citet{klotz08}.  Panel ``b'' of Figure~\ref{fig:opt_filters} shows the resulting sensitivity function
which includes atmospheric transmission typical of La Silla.
Fully reduced TAROT images were provided to us by Alain Klotz.

We computed PSF photometry on the images with DAOPhot.
Magnitudes for the local sequence stars in the natural system of the TAROT ``open'' filter were derived
in the same manner described above for the LSQ $gr$ filter.  Figure~\ref{fig:tarot} shows
 $(V_{CSP}-open_{TAROT})$ and $(r_{CSP}-open_{TAROT})$ colors for the Pickles main sequence stars 
plotted versus $(g-r)_{CSP}$, with the zero point for the $open_{TAROT}$ filter chosen to give a magnitude of 
0.0 for Vega. Photometry of the local sequence stars in the field of SN~2012fr are plotted as red points
after adjusting the zero points of the $open_{TAROT}$ instrumental magnitudes to provide the best fit to
the Pickles stars.  Final magnitudes for the local sequence stars in the natural system of the 
$open_{TAROT}$ filter are found in Table~\ref{tab:locseq2}, and the photometry of the SN is
given in Table~\ref{tab:sn_nat_sys}.

\subsection{Slooh}
\label{sec:slooh_phot}

Panel ``c'' of Figure~\ref{fig:opt_filters} shows the throughput functions calculated for the Slooh
$LRGB$ filters using information provided by E.~Conseil.  Included is the atmospheric
transmission appropriate for the Teide Observatory.
Processed Slooh images were provided by E.~Conseil.  These displayed a 
strong gradient in the sky background that was subtracted prior to computing
instrumental PSF magnitudes. 

SN~2012fr is clearly detected in the Slooh images, but with asymmetric, low signal-to-noise ratio 
PSFs unsuitable for measuring magnitudes with our standard 
photometry tools.  For the analysis of these data, it was therefore necessary to develop 
a specialized tool that produces a 3-dimensional model of the PSF using an isolated 
bright star in the images. The model is background subtracted and subsampled, and is then fitted to the 
SN and local sequence stars in the image using an MCMC procedure where the amplitude
and the center coordinates are fitted simultaneously.
An example of the PSF signal subtraction for the $B$-band image of the supernova and one of the
local standards is showed in Figure \ref{fig:psfsub}.

The $B_{Slooh}$ and $G_{Slooh}$ filters are moderately well matched to the CSP 
$B$ and $V$ filters as shown in Figures~\ref{fig:slooh}a and \ref{fig:slooh}b.  
Here $(B_{Slooh}-B_{CSP})$ and $(G_{Slooh}-V_{CSP})$ are plotted
versus $(B-V)_{CSP}$ for the Pickles stars with the zero points of the $B_{Slooh}$ and $G_{Slooh}$ 
magnitudes set assuming $B_{Slooh}=V_{Slooh}=0.0$ for Vega.  The observed colors for 
the sequence stars, plotted as red points, have been adjusted to fit the synthetic colors by 
varying the zero points of the instrumental magnitudes.  
Final magnitudes of the local sequence stars in the natural systems 
of the $B_{Slooh}$ and $G_{Slooh}$ filters derived from this procedure are given in Table~\ref{tab:locseq2}.
Photometry of the SN in these filters is listed in Table~\ref{tab:sn_nat_sys}.

We also attempted to match the $R_{Slooh}$ filter to the 
$r_{CSP}$ bandpass.  The result is shown in the color-color diagram plotted in 
Figure~\ref{fig:slooh}c where $(R_{Slooh}-r_{CSP})$ is plotted
versus $(g-r)_{CSP}$ for the Pickles stars assuming $R_{Slooh}=0.0$ for Vega.  
The observed trend measured from the photometry of the local sequence stars, plotted by the red 
points, is seen to be consistent with the expectation from the synthetic photometry
of the Pickles stars after adjusting for the zero point difference.  Magnitudes of the local 
sequence stars in the natural system of the Slooh $R$ filter are found in Table~\ref{tab:locseq2},
and photometry of the SN in this filter is given in Table~\ref{tab:sn_nat_sys}.

The $L_{Slooh}$ (``luminance'') filter throughput function closely resembles that of the LSQ $gr$ filter.
This is confirmed in the color-color diagram shown in Figure~\ref{fig:slooh}d.
Here $(L_{Slooh}-gr_{LSQ})$ is plotted versus $(g-r)_{CSP}$ from synthetic photometry of the Pickles 
main sequence stars, where the zero point for the $L_{Slooh}$ filter has been calculated assuming  
$L_{Slooh} = 0.0$ for Vega.  The slope is nearly flat over the whole color range of the Pickles
stars, consistent with a color term that is essentially zero. 
The measured colors of the local sequence stars are plotted as red points
after adjusting the zero point of the $L_{Slooh}$ instrumental magnitudes to provide the best fit to
the Pickles stars.  We therefore assume the $L_{Slooh}$ filter is in the same natural system as the 
$gr_{LSQ}$ filter.   Photometry of the SN in the $L_{Slooh}$ filter is listed in Table~\ref{tab:sn_nat_sys}.
 
\subsection{BOSS}
\label{sec:boss_phot}

\begin{figure}[h]
\centering
\includegraphics[width=6.in]{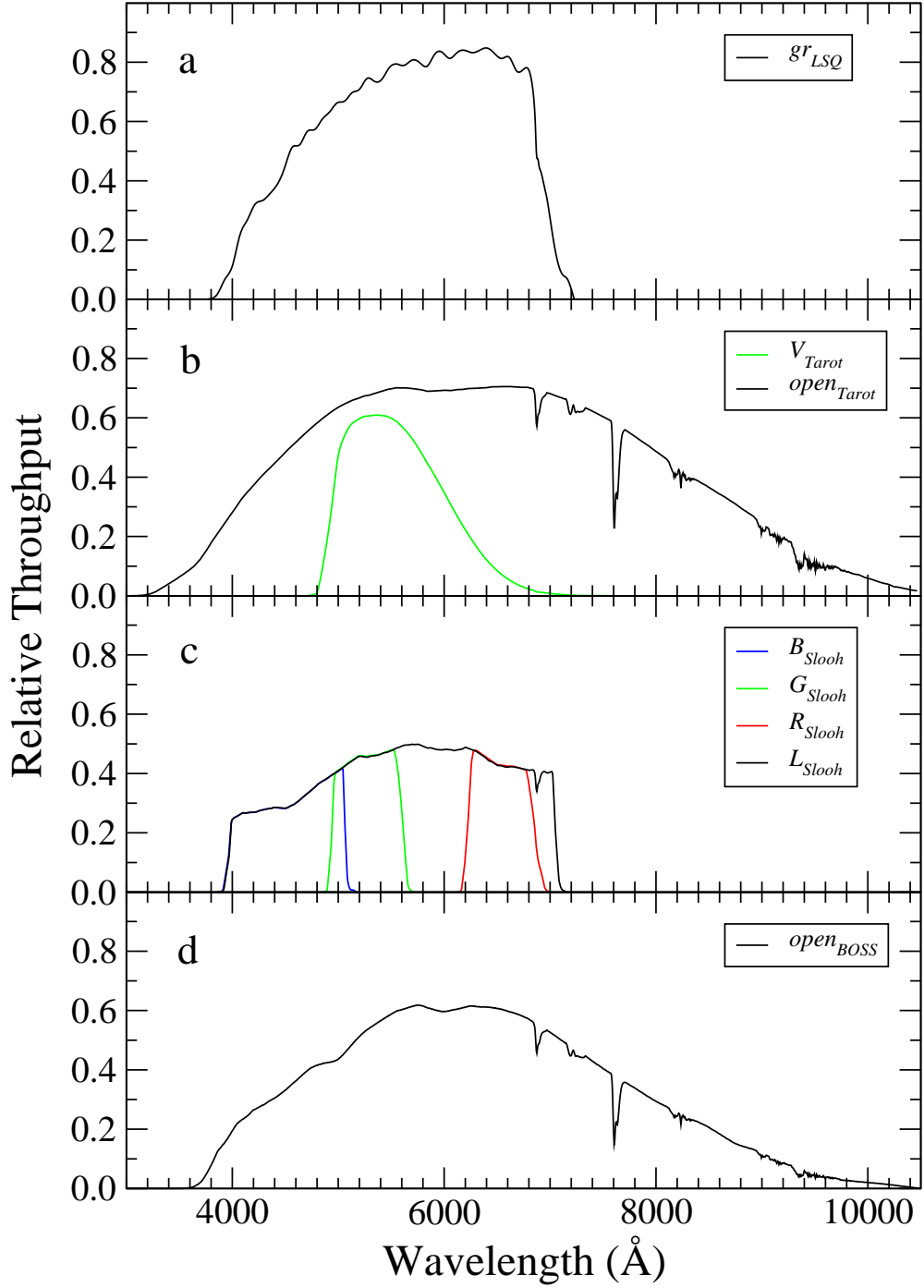}
\caption{Relative throughput functions of the LSQ, Tarot, Slooh, and BOSS filters.} 
\label{fig:opt_filters}
\end{figure}

\begin{figure}[h]
\centering
\includegraphics[width=14cm]{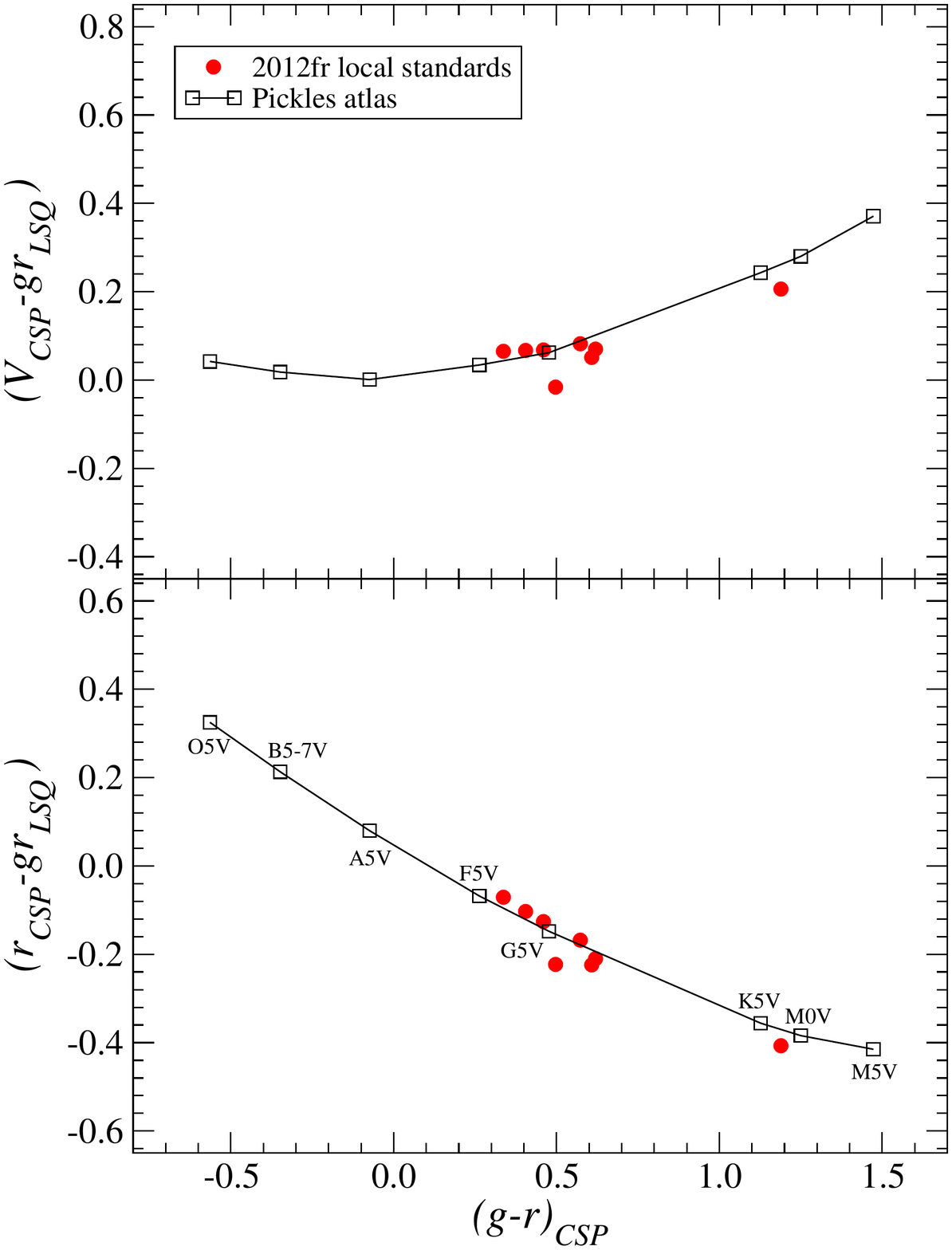}
\caption[]{Color-color plots for converting CSP $V$ and $r$ magnitudes 
to the natural system magnitudes for the LSQ $gr$ filter.  The black curves
show synthetic photometry carried out using the \citet{pickles98} stellar atlas.  The red
points correspond to observations of the local sequence stars in the field of
SN~2012fr.}     \label{fig:gr} 
\end{figure}

\begin{figure}[h]
\centering
\includegraphics[width=13cm]{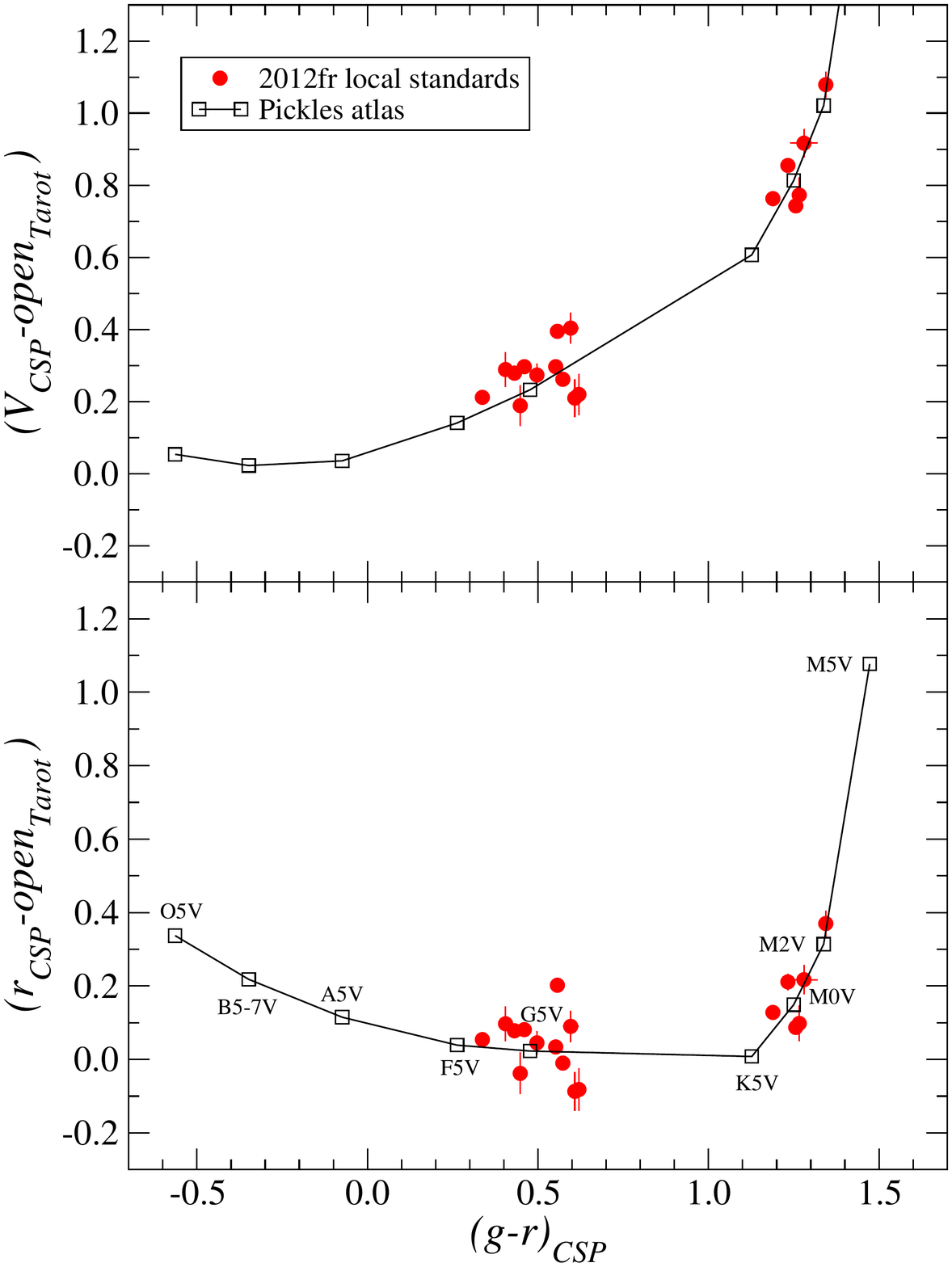}
\caption[]{Color-color plots for converting CSP $V$ and $r$ magnitudes 
to the natural system magnitudes for the TAROT $open$ filter.  The black curves
show synthetic photometry carried out using the \citet{pickles98} stellar atlas.  The red
points correspond to observations of the local sequence stars in the field of
SN~2012fr.}     \label{fig:tarot} 
\end{figure}

\begin{figure}
\centering
\leavevmode
\begin{tabular}{c}
\includegraphics[width=16cm]{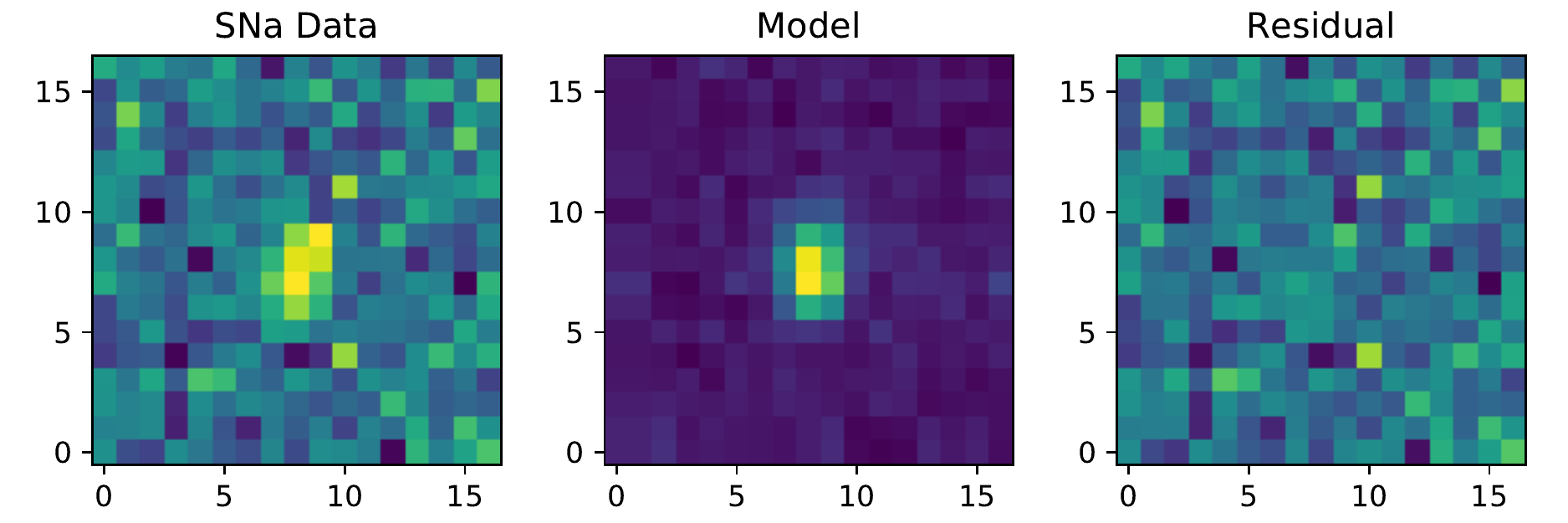} \\
\includegraphics[width=16cm]{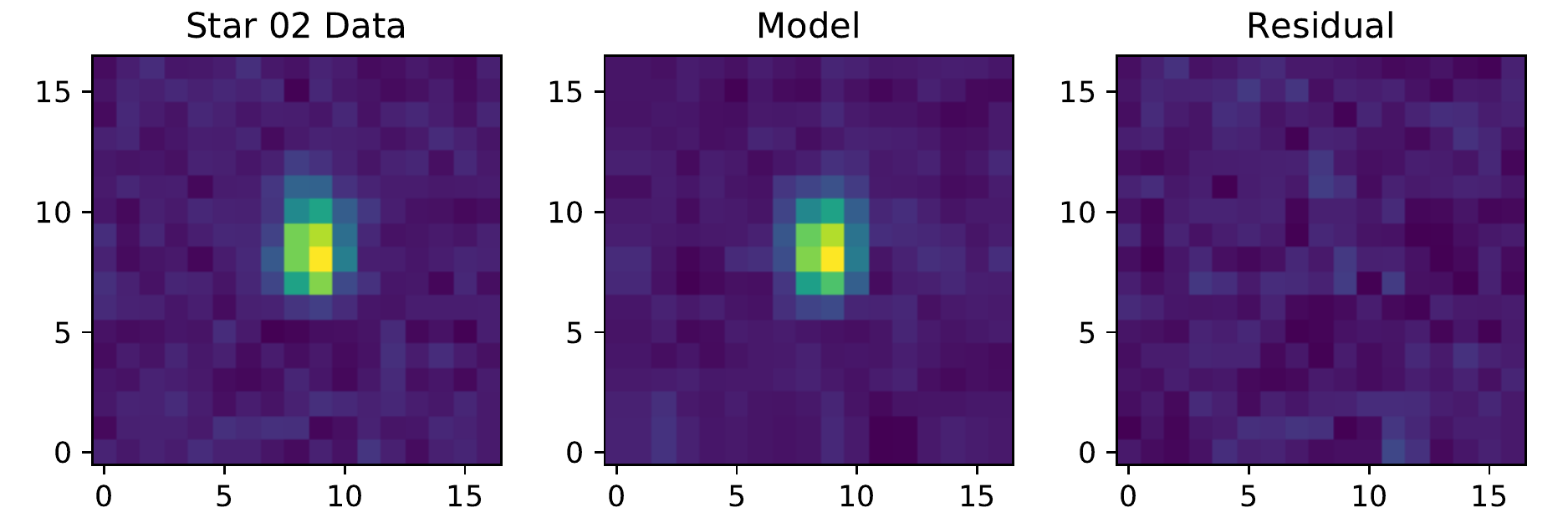}  \\
\end{tabular}
\caption{Examples of PSF subtractions for the Slooh $B$-band image, with SN~2012fr in the top row, and a star in the bottom
row.  The left image in both rows shows the original data, the center image shows the scaled PSF model, and the right images
shows the subtractions of the scaled PSF model from the original data.}
\label{fig:psfsub}
\end{figure}

\begin{figure}[h]
\centering
\includegraphics[width=13cm]{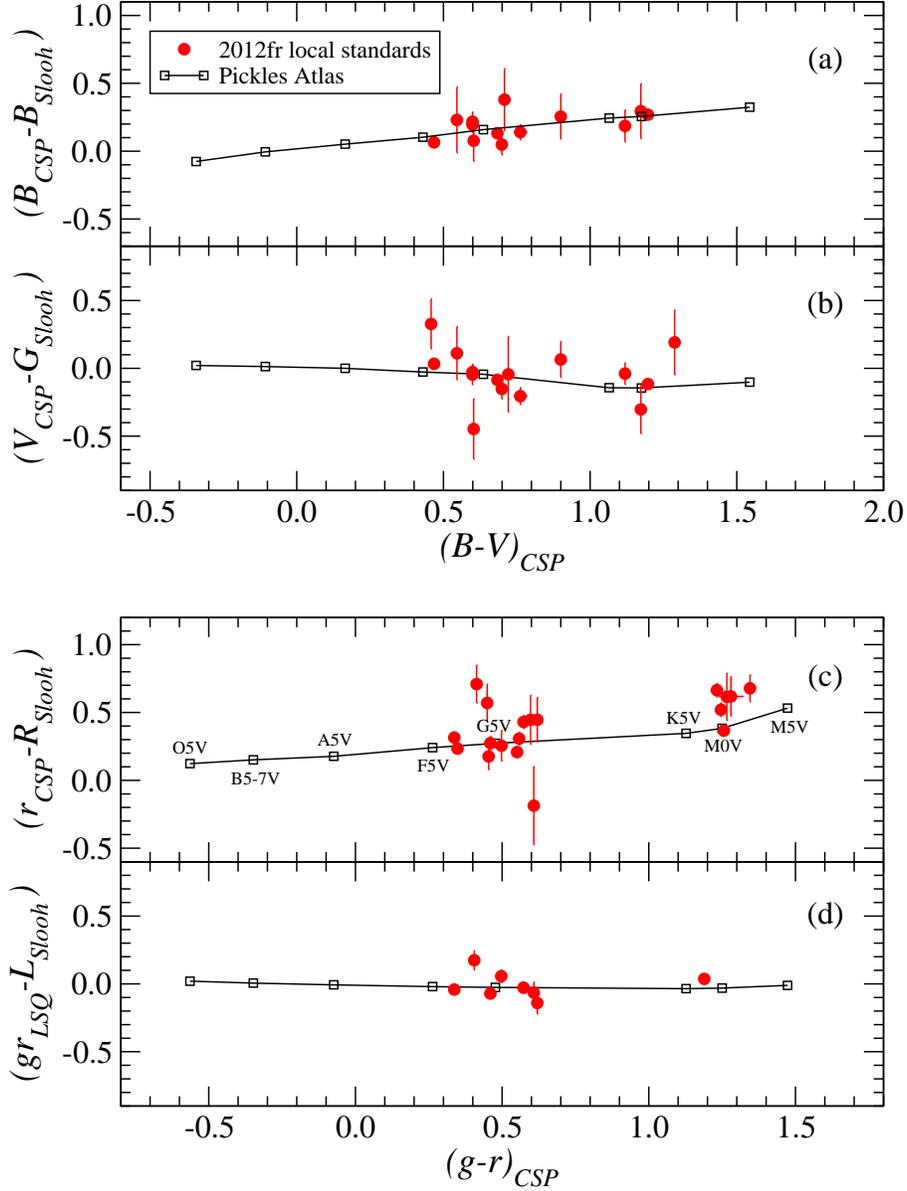}
\caption[]{Color-color plots for converting (a) $B_{CSP}$ magnitudes to natural system 
magnitudes in the $B_{Slooh}$ filter, (b) $V_{CSP}$ magnitudes to natural system 
magnitudes in the $G_{Slooh}$ filter, (c) $r_{CSP}$ magnitudes to natural system 
magnitudes in the $R_{Slooh}$ filter, and (d) $gr_{LSQ}$ magnitudes to natural system 
magnitudes in the $L_{Slooh}$ filter.  The black curves
show synthetic photometry carried out using the \citet{pickles98} stellar atlas.  The red
points correspond to observations of the local sequence stars in the field of
SN~2012fr.}     \label{fig:slooh} 
\end{figure}

The throughput of the BOSS ``open'' system was constructed from information provided by Stuart Parker, 
who observed SN~2012fr using a Celestron C14 f/10 telescope with a f/6.3 focal reducer attached before a 
Kodak KAF-3200ME CCD with coverglass.  Transmission curves for the StarBright coatings on the 
telescope optics were found on the Celestron website, but covering only the wavelength range
4000--7500~\AA.  In response to an inquiry, Celestron advised us that the ``transmission falloff in the 
IR and UV bands was pretty severe''.  The transmission of the CCD coverglass is given in a spec sheet
supplied by the manufacturer, but covering only the wavelength range from 3500--8500~\AA.  No 
information on the transmission of the focal reducer could be found.  Hence, constructing the BOSS
``open'' filter throughput required some guesswork.  Panel ``d'' of Figure~\ref{fig:opt_filters} shows our final
approximation including the transmission of the atmosphere.

\begin{figure}[h]
\centering
\includegraphics[width=14cm]{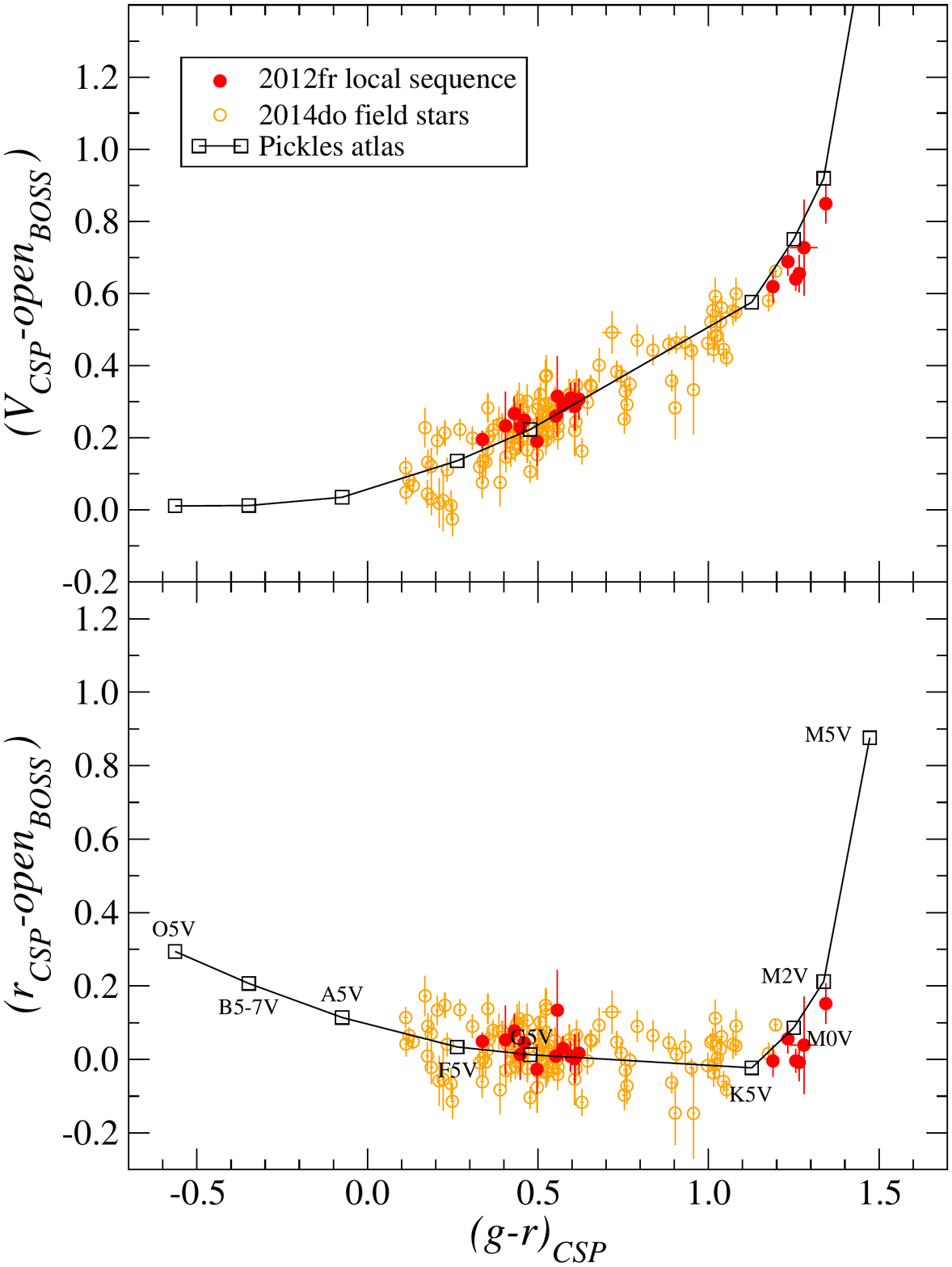}
\caption[]{Color-color plots for converting CSP $V$ and $r$ magnitudes 
to the natural system magnitudes for the BOSS $open$ filter.  The black curves
show synthetic photometry carried out using the \citet{pickles98} stellar atlas.  The red
points correspond to observations of the local sequence stars in the field of
SN~2012fr.}     \label{fig:boss} 
\end{figure}

The BOSS images were processed by Stuart Parker using CCDsoft routines.  PSF photometry
was then carried out with DAOPhot.
Final magnitudes for the local sequence stars in the natural system of the BOSS ``open'' filter were 
derived in the same manner described above for the LSQ $gr$ filter.  Figure~\ref{fig:boss} shows
the ($V-open_{BOSS}$) and ($r-open_{BOSS}$) colors for the Pickles main sequence stars plotted 
versus $(g-r)_{CSP}$, with the zero point for the synthetic magnitudes chosen to give $open_{BOSS} = 0.0$
for Vega.  The local sequence stars in the field of SN~2012fr  are plotted as solid red circles in 
Figure~\ref{fig:boss} 
after adjusting the zero points of the $open_{BOSS}$ instrumental magnitudes to provide the best fit to
the Pickles stars.  Because of the uncertainties involved in constructing the throughput curve of the BOSS 
``open'' system, these observations of the SN~2012fr local sequence are augmented by photometry 
(plotted with open red circles) of stars in the field of SN~2014do, a relatively low Galactic latitude 
transient that was observed by both Parker and the CSP-II.  In general, the observations are
well-matched by the colors derived from the synthetic photometry, providing confidence
that the throughput function shown in panel ``d'' of Figure~\ref{fig:opt_filters} is a reasonable representation of the BOSS ``open''
filter.  Final magnitudes for the local sequence stars in the natural system of the BOSS ``open'' filter
are given in Table~\ref{tab:locseq2}, and photometry of the SN in the natural system of the BOSS 
``open'' filter is given in Table~\ref{tab:sn_nat_sys}.

\section{BOLOMETRIC LIGHT CURVE CONSTRUCTION}
\label{sec:bolo_lc_details}

As discussed in detail by \citet{brown16}, producing a bolometric light
curve from photometry of an object such as a SN~Ia with a spectral energy distribution (SED) that differs
significantly from those of stars is, at best, an approximate procedure.  For this reason, we
employed two different methods which are detailed in this appendix.

\subsection{Photometric Trapezoidal Integration}

This method estimates the flux density from the observed photometry by simple trapezoidal
integration of the fluxes derived from each filter.  In addition, estimates must be made for
the UV ($\lambda < \lambda_u$) and infrared ($\lambda > \lambda_H$) contributions --- the 
former from the Swift  $uvm2$ photometry, and the 
latter from the Rayleigh-Jeans law.  The steps involved are as follows:

\begin{enumerate}

\item The NIR $YJH$ magnitudes are fitted to match optical phase coverage.

\item Optical magnitudes are interpolated for nights where observations were not available.

\item Corrections are applied for Milky Way and host-galaxy dust reddening.

\item The resulting magnitudes in the natural system are transformed to AB magnitudes.

\item The AB magnitudes are then converted to monochromatic flux densities.

\item The bolometric flux at each phase is derived by integration of the flux densities via the Trapezoidal Rule.

\item Approximations for the missing flux in the UV ($\lambda < \lambda_u$) and infrared ($\lambda > \lambda_H$) 
for each phase are applied.

\item The resulting bolometric flux vs. time is converted to luminosity assuming the Cepheid-based distance
modulus for the host-galaxy, NGC~1365.

\end{enumerate}

\noindent Note that the K-corrections can be neglected since the redshift of the host-galaxy, NGC~1365, is very 
small ($z_{helio} = 0.005457$).

We used smooth curve Gaussian process fitting to match photometry in the
NIR $YJH$ bands to the phases of the optical photometry.   For the few nights when optical observations
were not obtained, magnitudes were interpolated.  Next, the dust extinction from the Milky Way in the 
direction of SN~2012fr was corrected by subtracting the absorption values given in the third column
of Table \ref{tab:ext_ab}. 
For most of the filters, these were taken directly from NED.  For the $Y$ band, these
were calculated using the \citet{fitzpatrick99} Galactic reddening law.  Finally, the correction by host-galaxy
reddening (see \S~\ref{sec:hostred}) was performed by subtracting the values given in the final column of Table \ref{tab:ext_ab}.

The $uvm2$, $u$, $B$, $g$, $V$, $r$, $i$, $Y$, $J$, and $H$ photometry was then converted 
to AB magnitudes using the offsets given in the final column of Table \ref{tab:ext_ab}.
Once these are applied, the flux in each band is approximated by:

\begin{equation}
f_\nu(X) = 10^{-0.4 (m_{\mathrm{AB}}+48.6)}\,~\mathrm{erg\,sec^{-1}\,cm^{-2}\,Hz^{-1}} \; .
\label{eqn:refname2}
\end{equation}

The $u$-to-$H$ flux was then integrated using the Trapezoidal Rule.  
The flux beyond the $H$ band was estimated assuming that it follows the
Rayleigh-Jeans law.  This leads to the following expression for the integrated flux 
at wavelengths longward of $\lambda_{H}$:

\begin{equation}
f(\lambda > \lambda_H)=f_H\, \lambda_{H}\,\frac{1}{5}\; .\nonumber
\label{eqn:rj}
\end{equation}

We estimate the UV contribution between 1,800--3,000~\AA\ assuming the SED is flat there, 
and the contribution between 3,000~\AA\ and $\lambda_{u}$ (3,500~\AA)
by imposing the condition that the flux falls linearly to the flux estimated for $uvm2$-band from $\lambda_{u}$.
The flux below 1,800~\AA\ is assumed to be zero.  Hence,

\begin{equation}
	f(\lambda < \lambda_u)= f_{uvm2}\,(\mathrm{3,000 - 1,800~\AA}) + \frac{1}{2}(f_u+f_{uvm2})\,(\mathrm{3,500-3,000~\AA}). \nonumber
\label{eqn:fl_u}
\end{equation}

\noindent This assumption fits well with the fall off of the flux at wavelengths below the $u$ band typically 
observed in UV spectra of SNe Ia \citep[e.g., see][]{foley16}.

Finally, to calculate the bolometric luminosity in absolute terms, we assumed the
\citet{freedman01} distance modulus of $\mu = 31.27\pm0.05$~mag derived from
Cepheid variable observations.

\subsection{Spectral Template Fitting}

\begin{figure}[h]
\centering
\includegraphics[width=15cm]{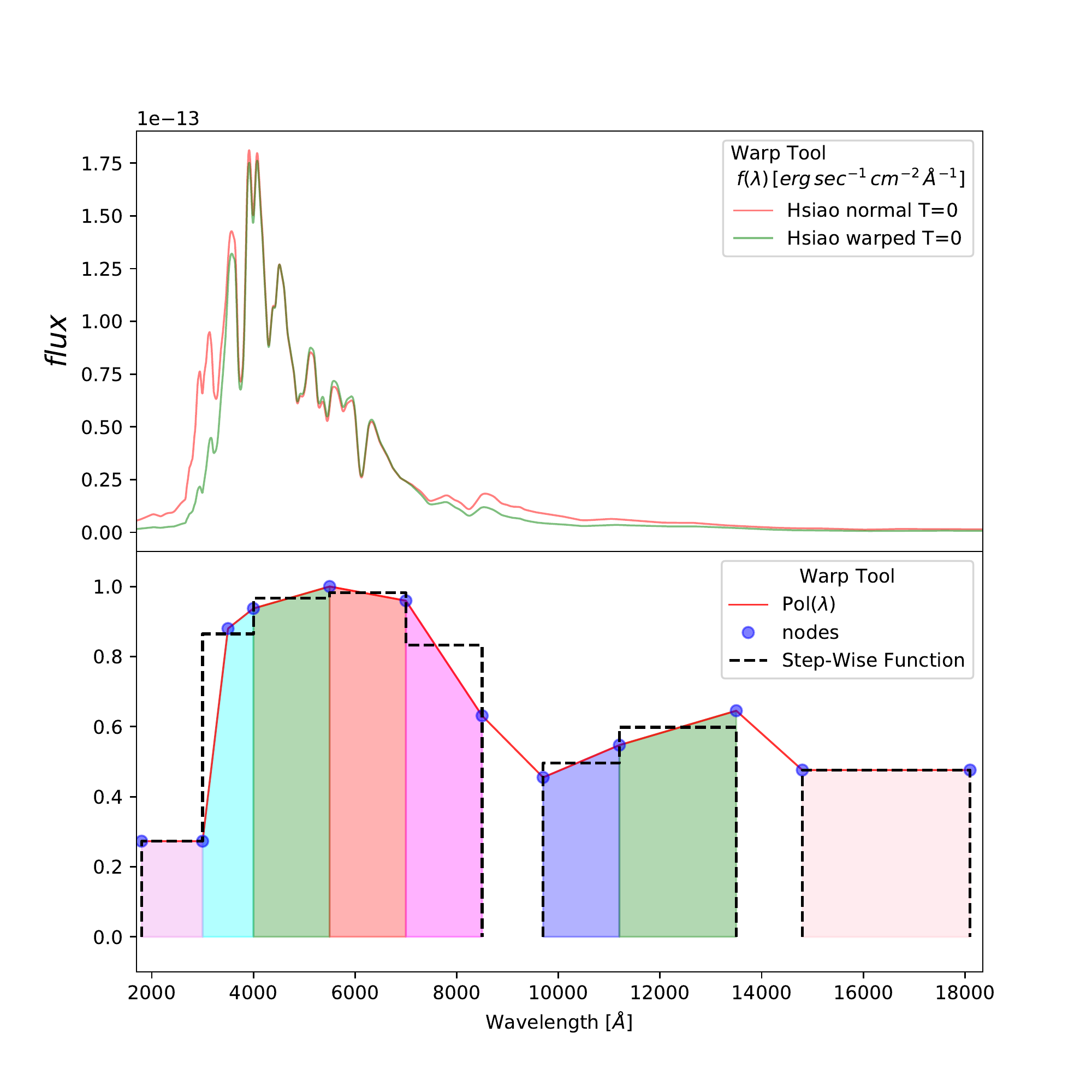}
\caption[]{Example of the two-step process devised to match the \citet{hsiao07} spectral template to
the observed CSP photometry for the same epoch.  The dashed line in the lower panel shows the first iteration
consisting of a step-wise function, $P(\lambda)$, calculated from equation~\ref{eqn:lumspec}.
In the second iteration, a continuous piece-wise function is derived from the step-wise function by forcing
the value of the nodes of the $g$-band bin to have the slope determined by the measured step-wise 
values of $u$ and $r$.  The nodes for the remaining filters are then calculated from their step-wise
values, with the shaded regions correspond to the 8 bins.
The \citeauthor{hsiao07} template spectrum is then multiplied by the final piece-wise function
as shown in the upper panel.}
\label{fig:warp}
\end{figure}

Our second method to estimate the bolometric luminosity takes the spectral template for each phase 
from \citet{hsiao07} and multiplies it by a function $P(\lambda)$ such that the synthetic photometry 
measured on the template matches exactly the real photometry. This method is limited to $t-t_{B_{max}}= +79$ 
as this is the last epoch for the spectral template. The steps can be summarized as follows:

\begin{enumerate}

\item The NIR $YJH$ magnitudes are interpolated to match optical phase coverage.

\item Optical magnitudes are interpolated for nights where observations were not available.

\item Corrections are applied for Milky Way and host-galaxy dust reddening.

\item For each photometry epoch, the \citeauthor{hsiao07} spectrum corresponding to
that phase is selected.

\item 
The template spectrum for each epoch is then matched in flux to the photometry via an
iterative procedure:

\begin{enumerate}

\item First iteration: The spectrum is divided into wavelength bins corresponding to the 
non-overlapping filters $uvm2$, $u$, $g$, $r$, $i$, $Y$, $J$, and $H$ which, in practice, 
cover almost all the UV-to-NIR wavelength 
domain except for two gaps: one between $i$ and $Y$,
and other between $J$ and $H$. Then a step-wise function $P(\lambda)$ is fitted such that:

\begin{equation}
m_X = -2.5\,log_{10}\int S_X F_{\lambda} \lambda  P_{\lambda}(\lambda) d\lambda + zp_X\;,  
\label{eqn:lumspec}
\end{equation}

where $X$ is the filter, $m_X$ is the reddening-corrected magnitude of the SN for that filter,
$S_X$ is the transmission function of the filter, $F_{\lambda}$
is the \citeauthor{hsiao07} spectrum, and $zp_X$ is the zero point previously adjusted to match our system.

\item Second iteration: The step-wise function calculated in step~5 is now converted to a piece-wise 
function (or polygonal line) with nodes at the blue and red limits of each filter, 
except for the $u$~band where we added an extra node at the effective wavelength
to account for the rapid change of flux that occurs across this filter.
The values of the nodes for the $g$-band
bin are imposed to have the slope determined by the step wise values of $u$ and $r$ measured
in the previous step. The nodes of $Y$ and $J$ adopt the slope of the two corresponding
 step-wise values measured in the previous step as well. This gives smoothness to the final function $P(\lambda)$.
All the other nodes are then  determined and may be calculated via  Equation~\ref{eqn:lumspec}.
Figure \ref{fig:warp} shows an example of this two-step procedure.

\end{enumerate}

\item The modified \citeauthor{hsiao07} spectra, $W(\lambda)=P(\lambda)F_{\lambda}$, 
are then integrated between the effective wavelengths
$\lambda_u$ and $\lambda_H$ to calculate the bolometric flux at each phase. 

\item For the infrared flux at wavelengths longward of $\lambda_{H}$, a Rayleigh-Jeans law is again assumed:
\begin{equation}
f(\lambda > \lambda_H)=f_H\, \lambda_{H}\,\frac{1}{5}\nonumber \;.
\label{eqn:rj}
\end{equation}

\item Finally the flux is converted into luminosity assuming the Cepheid-based distance
modulus to NGC~1365.

\end{enumerate}

\begin{figure}[h]
\centering
\includegraphics[width=16cm]{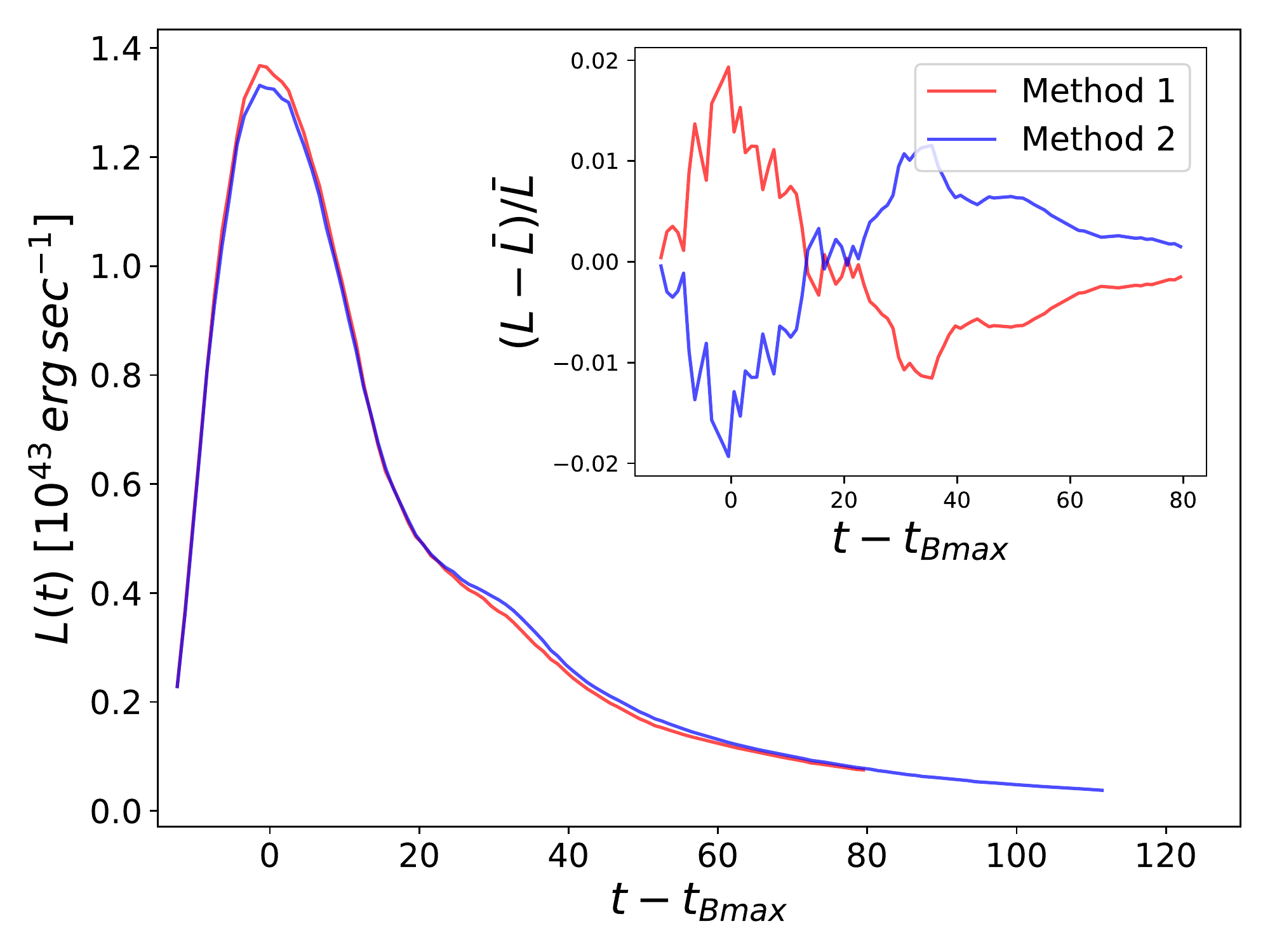}
\caption[]{Bolometric light curves calculated with the Photometric Trapezoidal Integration (method~1) and the Spectral Template Fitting  (method~2) techniques.  The inset plot shows luminosity differences of both methods with respect to the average of the two methods.}
\label{fig:bolmethods} 
\end{figure}

The upper panel of Figure~\ref{fig:bolmethods} displays the final bolometric light curves
calculated using the Photometric Trapezoidal Integration and Spectral Template Fitting methods.  
In the lower panel of the same figure, the ratio of each light curve to the average of the two is 
plotted versus phase with respect to the epoch of \tmax.
As is seen, the two methods produce light curves that are consistent at the $\pm$5\% level. 
The Spectral Template Method yields a higher luminosity at phases before $t-t_{B_{max}}\sim30$
and a lower luminosity at later epochs.
The final bolometric light curves for the two methods are given in Table~\ref{tab:lum}.


\clearpage
%

%
\clearpage
%

%
\clearpage
%

%
\clearpage
%

%
\clearpage
%

%
\clearpage
%

%
\clearpage
%
%
\clearpage
%

%
\clearpage
%

%
\clearpage
%

%
\clearpage
%

%
\clearpage
%

%
\clearpage
%

\clearpage
\clearpage

%
\clearpage
\clearpage

%
\clearpage
\clearpage
%

%
\clearpage
\clearpage
%

%
\clearpage
\clearpage
%

%
\clearpage
%

\clearpage
%

\clearpage
%

%
\clearpage
%

%
\clearpage
%

%
\clearpage
%

%
\clearpage
%

%
\clearpage
%

%
\clearpage
%

%
\clearpage
%
%

%
\clearpage
\clearpage
\begin{deluxetable}{ccccccr}
\tablewidth{0pt}
\tabletypesize{\scriptsize}
\tablecaption{UVOT Photometry of SN~2012fr in the Vega System\label{tab:uv}}
\tablehead{
\colhead{JD (days)} &
\colhead{$uvw2$ (mag)}&
\colhead{JD (days)} &
\colhead{$uvm2$ (mag)}&
\colhead{JD (days)} &
\colhead{$uvw1$ (mag)}&
\colhead{phase (days)}}
\startdata
2456229.32 & 19.139(186) & $\cdots$   & $\cdots$    & 2456229.31 & 17.795(115)&     $-$13.78  \\ 
2456231.99 & 16.798(077) & 2456232.00 & 18.815(118) & 2456231.99 & 15.090(058)&     $-$11.11 \\
2456233.93 & 15.936(074) & 2456233.94 & 17.676(101) & 2456233.92 & 14.096(046)&     $-$9.17  \\
2456235.68 & 15.625(074) & 2456235.68 & 16.977(081) & 2456235.71 & 13.559(044)&     $-$7.42  \\
2456236.17 & 15.303(064) & $\cdots$   & $\cdots$    & 2456236.16 & 13.502(042)&     $-$6.93  \\
2456237.80 & 15.099(063) & 2456237.80 & 16.603(070) & 2456237.70 & 13.332(044)&     $-$5.30   \\
$\cdots$   & $\cdots$    & 2456239.17 & 16.372(072) & $\cdots$   & $\cdots$   &     $-$3.93   \\
2456239.60 & 14.913(075) & 2456239.60 & 16.327(108) & 2456239.60 & 13.176(044)&     $-$3.50   \\
\enddata
\tablecomments{Values in parentheses are 1$\sigma$ measurement uncertainties in millimag.
The phase is relative to $t_{B_{max}}=\mathrm{JD}~2456243.1$.}
\end{deluxetable}

\clearpage
\begin{deluxetable}{ccccccccccccccccccccccc}
\rotate
\tablecolumns{23}
\tablewidth{0pt}
\setlength{\tabcolsep}{0.02in} 
\tabletypesize{\tiny}
\tablecaption{Photometry of SN~2012fr Local Sequence Stars in the Standard System.\label{tab:locseq}}
\tablehead{
\colhead{Star ID} &
\colhead{RA} &
\colhead{DEC} &
\colhead{$u$ (mag)}&
\colhead{$g$ (mag)}&
\colhead{$r$ (mag)}&
\colhead{$i$ (mag)}&
\colhead{$B$ (mag)}&
\colhead{$V$ (mag)}&
\colhead{$Y$ (mag)}&
\colhead{$J$ (mag)}&
\colhead{$H$ (mag)}}
\startdata
01 & 03:33:55.48 & -36:07:19.0 &  16.836(010) &  14.087(003) & 12.846(002) & 12.332(009) & 14.748(003)&  13.423(002) & $\cdots$    & $\cdots$   & $\cdots$ \\ 
02 & 03:33:33.67 & -36:07:14.1 &  15.314(004) &  13.879(002) & 13.343(002) & 13.160(008) & 14.293(003)&  13.560(002) &12.467(034)  &12.167(023) &11.821(035) \\ 
03 & 03:33:49.55 & -36:07:42.8 &  15.781(006) &  14.774(003) & 14.450(003) & 14.370(008) & 15.062(004)&  14.569(003) & $\cdots$    & $\cdots$   & $\cdots$ \\ 
04 & 03:33:55.66 & -36:08:38.2 &  16.881(007) &  15.385(004) & 14.833(003) & 14.641(009) & 15.803(006)&  15.050(004) & $\cdots$    & $\cdots$   & $\cdots$ \\ 
05 & 03:33:40.67 & -36:10:52.9 &  16.565(007) &  15.368(004) & 14.924(004) & 14.789(009) & 15.732(006)&  15.091(005) & $\cdots$    & $\cdots$   & $\cdots$ \\ 
06 & 03:33:57.36 & -36:09:24.0 &  18.750(074) &  17.019(011) & 16.429(007) & 16.244(014) & 17.463(014)&  16.689(014) & $\cdots$    & $\cdots$   & $\cdots$ \\ 
07 & 03:33:54.96 & -36:09:09.6 &  18.769(020) &  17.017(015) & 16.423(010) & 16.237(014) & 17.487(030)&  16.659(013) & $\cdots$    & $\cdots$   & $\cdots$ \\ 
08 & 03:33:46.83 & -36:03:04.1 &  15.247(007) &  14.249(005) & 13.916(005) & 13.790(005) &  $\cdots$  &   $\cdots$   & $\cdots$    & $\cdots$   & $\cdots$ \\
\enddata
\tablecomments{Values in parentheses are 1$\sigma$ measurement uncertainties in millimag.}
\end{deluxetable}

\clearpage
\begin{deluxetable}{crcccccccccc}
\tablewidth{0pt}
\tabletypesize{\tiny}
\tablecaption{Optical Photometry of SN~2012fr in the Natural System of the Swope\label{tab:opt}}
\tablehead{
\colhead{JD (days)} &
\colhead{phase (days)} &
\colhead{$u$ (mag)}&
\colhead{$g$ (mag)}&
\colhead{$r$ (mag)}&
\colhead{$i$ (mag)}&
\colhead{$B$ (mag)}&
\colhead{$V$ (mag)}}
\startdata
2456230.70 & $-$12.40	& 14.975(015) & 13.952(009) & 14.079(008) & 14.576(009) & 14.066(027) & 13.990(015) \\ 
2456231.70 & $-$11.40	& 14.220(017) & 13.458(011) & 13.609(010) & 14.028(011) & 13.524(025) & 13.528(011) \\ 
2456232.70 & $-$10.40	& 13.688(016) & 13.098(010) & 13.248(009) & 13.646(009) & 13.126(025) & 13.178(009) \\ 
2456233.62 & $-$9.48	& 13.351(018) & 12.841(010) & 13.022(009) & 13.369(010) & 12.846(026) & 12.931(010) \\ 
2456234.66 & $-$8.44	& 13.061(019) &  $\cdots$   & 12.778(012) & 13.136(012) & 12.621(026) & 12.707(011) \\ 
2456235.60 & $-$7.50	& 12.901(017) & 12.450(011) & 12.647(012) & 13.026(013) & 12.465(026) & 12.569(013) \\ 
2456236.69 & $-$6.41	& 12.758(017) & 12.312(012) & 12.515(012) & 12.941(013) & 12.333(028) & 12.417(013) \\ 
2456238.72 & $-$4.38	& 12.594(017) &  $\cdots$   & 12.298(010) & 12.867(013) & 12.113(030) & 12.196(015) \\
\enddata
\tablecomments{The phase is relative to $t_{B_{max}}=\mathrm{JD}~2456243.1$. Values in parentheses are 1$\sigma$ measurement uncertainties in millimag.}
\end{deluxetable}

\clearpage
\begin{deluxetable}{lllllll}
\tablewidth{0pc}
\tablecolumns{7}
\tabletypesize{\scriptsize}
\tablecaption{S-corrections to Standard Photometric Systems\label{tab:sc}}
\tablehead{%
\colhead{Filter} & \colhead{$B$} & \colhead{$V$} &
\colhead{$u^\prime$} & \colhead{$g^\prime$} & \colhead{$r^\prime$} &
\colhead{$i^\prime$} }
\startdata
2456230.70 & $-$0.040& $-$0.014&$-$0.014& $-$0.004& $-$0.000& $-$0.020 \\
2456231.70 & $-$0.036& $-$0.015& 0.002& $-$0.004& $-$0.000& $-$0.016 \\
2456232.70 & $-$0.031& $-$0.022& 0.015& $-$0.004& $-$0.000& $-$0.011 \\
2456233.62 & $-$0.031& $-$0.020& 0.015& $-$0.005& $-$0.000& $-$0.012 \\
2456234.66 & $-$0.024& $-$0.014& 0.030& $-$0.006& $-$0.000& $-$0.018 \\
2456235.60 & $-$0.026& $-$0.012& 0.023& $-$0.006& $-$0.000& $-$0.017 \\
2456236.69 & $-$0.025& $-$0.013& 0.011& $-$0.006& $-$0.001& $-$0.017 \\
2456238.72 & $-$0.020& $-$0.014($-$0.008)&$-$0.005& $-$0.005($-$0.005)& $-$0.000($-$0.002)& $-$0.017($-$0.013) \\
\enddata
\tablecomments{S-corrections, given by $M_{std}=M_{nat}+S_M$, are calculated for all phases using the \citet{hsiao07} spectral template.  
For those phases where a spectrum of SN~2012fr exists \citep{childress13}, the S-correction calculated 
from that spectrum is given in parentheses. For $B$ and $V$, the filter functions and zero points are taken 
from \citet{stritzinger05a}. For $u$, $g$, $r$  and $i$, the BD$+$17$^{\circ}$4708 spectrum from \citet{bohlin14} is used, along with the \citet{fukugita96} 
filter functions and \citet{smith02} standard magnitudes to obtain photometric zeropoints.}
\end{deluxetable}

\clearpage
\begin{deluxetable}{crcccc}
\tablewidth{0pt}
\tabletypesize{\scriptsize}
\tablecaption{Near-infrared Photometry of SN~2012fr in the Natural System of du Pont\label{tab:nir}}
\tablehead{
\colhead{JD (days)} &
\colhead{phase (days)} &
\colhead{$Y$ (mag)}&
\colhead{$J$ (mag)}&
\colhead{$H$ (mag)} &
\colhead{telescope} } 
\startdata
2456231.84 & $-$11.26& 13.675(027) & 13.607(033) & 13.712(031)  & BAA\\ 
2456232.87 & $-$10.23& 13.390(022) & 13.302(020) & 13.507(029)  & BAA\\ 
2456233.86 & $-$9.24 & 13.177(022) & 13.069(020) & 13.269(029)  & BAA\\ 
2456234.71 & $-$8.39 & 13.061(025) & 12.953(020) & 13.148(061)  & DUP\\ 
2456234.85 & $-$8.25 & 13.028(022) & 12.938(020) & 13.207(029)  & BAA\\ 
2456235.73 & $-$7.37 & 12.979(022) & 12.848(020) & 13.094(061)  & DUP\\ 
2456235.84 & $-$7.26 & 12.974(022) & 12.834(020) & 13.072(031)  & BAA\\ 
2456236.67 & $-$6.43 & 12.929(025) & 12.799(020) & 12.974(062)  & DUP\\
\enddata
\tablecomments{Values in parentheses are 1$\sigma$ measurement uncertainties in millimag. 
The phase is relative to $t_{B_{max}}=\mathrm{JD}~2456243.1$.}
\end{deluxetable}

\clearpage
\begin{deluxetable}{cccc}
\tablewidth{0pt}
\tabletypesize{\scriptsize}
\tablecaption{Baade Fourstar S-Corrections for SN~2012fr\label{tab:scorr}}
\tablehead{
\colhead{phase (days)} &
\colhead{$S_{J1_{FS}}$ (mag)}&
\colhead{$S_{J_{FS}}$ (mag)}&
\colhead{$S_{H_{FS}}$ (mag)} }
\startdata

$-$11  &    0.044  &    0.002  &   0.001 \\
$-$10  &    0.032  &    0.001  &   0.002 \\
$-$09  &    0.024  & $-$0.001  &   0.002 \\
$-$08  &    0.011  & $-$0.001  &   0.002 \\
$-$07  & $-$0.001  & $-$0.003  &   0.002 \\
$-$01  & $-$0.036  & $-$0.003  &   0.000 \\
$+$02  & $-$0.012  & $-$0.007  &   0.000 \\
$+$04  & $-$0.021  & $-$0.010  &   0.001 \\
$+$06  & $-$0.016  & $-$0.009  &   0.003 \\
$+$07  & $-$0.017  & $-$0.007  &   0.005 \\
$+$09  & $-$0.024  & $-$0.005  &   0.006 \\
$+$10  & $-$0.017  & $-$0.007  &   0.006 \\
$+$14  & $-$0.027  & $-$0.006  &$-$0.013 \\
$+$38  &    0.010  & $-$0.004  &   0.001 \\
$+$53  &$\cdots$   & $-$0.006  &$-$0.002 \\
$+$55  &    0.063  & $-$0.008  &$-$0.001 \\
$+$61  &    0.048  & $-$0.013  &$-$0.004 \\
\enddata
\tablecomments{The S-correction values were applied as: $X_{Retrocam}=X_{Fourstar}$+S-corr.
The phase was computed relative to $t_{B_{max}}=\mathrm{JD}~2456243.1$ and rounded to match the phases of \citet{hsiao07}
spectral templates.}
\end{deluxetable}

\clearpage
\begin{deluxetable}{ccccccc}
\tablewidth{0pt}
\tabletypesize{\scriptsize}
\tablecaption{SN~2012fr: Light Curve Parameters and Absolute Magnitudes.\label{tab:pars}}
\tablehead{
\colhead{} &
\colhead{$t_{max}\mathrm{(JD)}$} &
\colhead{$\Delta t_{peak}$} &
\colhead{$m_{Xpeak}$} &
\colhead{$m_{Xpeak}-A_X$} &
\colhead{$\Delta_{m15}$} &
\colhead{$M_{X\,Ceph}$} \\
\colhead{Band} &
\colhead{(days)} &
\colhead{(days)} &
\colhead{(mag)} &
\colhead{(mag)} &
\colhead{(mag)} &
\colhead{(mag)} }
\startdata
$u$ & $2456240.5 \pm 0.3$ & $-$2.8 & 12.59(02) & 12.36(15) & 0.87(03) & $-$18.91(16) \\
$B$ & $2456243.1 \pm 0.3$ & $+$0.0 & 12.04(02) & 11.84(13) & 0.82(03) & $-$19.43(14) \\
$g$ & $2456243.5 \pm 0.3$ & $+$0.6 & 11.96(02) & 11.78(12) & 0.70(03) & $-$19.49(13) \\
$V$ & $2456245.1 \pm 0.3$ & $+$2.0 & 11.99(02) & 11.84(10) & 0.67(03) & $-$19.43(11) \\
$r$ & $2456245.7 \pm 0.3$ & $+$2.6 & 12.06(02) & 11.93(08) & 0.79(03) & $-$19.34(09) \\
$i$ & $2456240.1 \pm 0.3$ & $-$3.0 & 12.84(02) & 12.74(07) & 0.59(03) & $-$18.53(09) \\
$Y$ & $2456238.0 \pm 1.0$ & $-$5.0 & 12.87(02) & 12.81(04) & 1.01(03) & $-$18.46(06) \\
$J$ & $2456239.4 \pm 1.0$ & $-$3.7 & 12.69(02) & 12.65(03) & 1.22(03) & $-$18.62(06) \\
$H$ & $2456238.6 \pm 1.0$ & $-$4.5 & 12.99(02) & 12.96(02) & 0.39(03) & $-$18.31(05) \\
$L_{UVOIR}$ & $2456242.7 \pm 0.3$ & $-0.4$ & ... & ... & ...& ...\\\\ \hline
  \multicolumn{7}{l}{$\Delta t_{peak}$ = $t_{max} - t_{Bmax}$}         \\
  \multicolumn{7}{l}{$A_X$ : Total reddening, i.e. Milky Way plus host galaxy reddening.}         \\
\enddata
\tablecomments{Values in parentheses are 1$\sigma$ measurement uncertainties in hundredths of a magnitude.}
\end{deluxetable}

\clearpage
\begin{deluxetable}{cccl}
\tablecolumns{4}
\tablewidth{0pt}
\tabletypesize{\footnotesize}
\tablecaption{$gr_{LSQ}$ Light Curve of SN~2012fr\label{tab:rise}}
\tablehead{
\colhead{} &
\colhead{} &
\colhead{} &
\colhead{Filter of} \\
\colhead{JD (days)} &
\colhead{Phase\tablenotemark{a} (days)} &
\colhead{$gr_{LSQ}$ (mag)\tablenotemark} &
\colhead{Observation}}
\startdata
2456225.72 & $-17.41$ & $>20.38\tablenotemark{b} $ & $gr_{LSQ}$ \\
2456225.84 & $-17.33$ & $>20.34\tablenotemark{b} $ & $gr_{LSQ}$ \\
2456226.69 & $-16.41$ & $17.71 \pm 0.03$ & $gr_{LSQ}$ \\
2456226.77 & $-16.33$ & $17.36 \pm 0.02$ & $gr_{LSQ}$ \\
2456227.55 & $-15.55$ & $16.03~^{+0.21}_{-0.10}$ & $open_{TAROT}$ \\
2456227.62 & $-15.48$ & $16.24 \pm 0.06$ & $L_{Slooh}$ \\
2456228.86 & $-14.24$ & $15.21 \pm 0.03$ & $open_{BOSS}$ \\
2456229.63 & $-13.47$ & $14.66 \pm 0.03$ & $V_{TAROT}$ \\
\enddata
\tablenotetext{a}{The phase was computed relative to $t_{B_{max}}=\mathrm{JD}~2456243.1$.}
\tablenotetext{b}{$3\sigma$ upper limit for non-detection.}
\end{deluxetable}

\clearpage
\begin{deluxetable}{ccccccccccc}
\tablecolumns{11}
\tablewidth{0pt}
\setlength{\tabcolsep}{0.02in} 
\tabletypesize{\tiny}
\tablecaption{LSQ, TAROT, BOSS, and Slooh Natural System Photometry of SN~2012fr Local Sequence Stars.\label{tab:locseq2}}
\tablehead{
\colhead{} &
\colhead{} &
\colhead{} &
\colhead{$gr_{LSQ}$}&
\colhead{$open_{TAROT}$}&
\colhead{$V_{TAROT}$}&
\colhead{$open_{BOSS}$}&
\colhead{$R_{Slooh}$}&
\colhead{$G_{Slooh}$}&
\colhead{$B_{Slooh}$}&
\colhead{$L_{Slooh}$} \\
\colhead{Star ID} &
\colhead{RA} &
\colhead{DEC} &
\colhead{(mag)} &
\colhead{(mag)} &
\colhead{(mag)} &
\colhead{(mag)} &
\colhead{(mag)} &
\colhead{(mag)} &
\colhead{(mag)} &
\colhead{(mag)}}
\startdata
1 & 03:33:55.48 & $-$36:07:19.0 & $\cdots$ & 12.748(007) & 13.301(002) & 12.844(034) & 12.484(014) & 14.403(028) & 13.591(016) & 13.242(006) \\
2 & 03:33:33.67 & $-$36:07:14.1 & $\cdots$ & 13.293(007) & $\cdots$ &13.323(014) & 13.134(023) & 14.127(023) & 13.659(020) & 13.511(008) \\
3 & 03:33:49.55 & $-$36:07:42.8 & 14.513(005) & 14.376(011) & 14.541(003) & 14.386(026) & 14.130(040) & 14.975(042) & 14.539(033) & 14.534(010) \\
4 & 03:33:55.66 & $-$36:08:38.2 & 14.994(005) & 14.824(015) & 15.006(005) & 14.789(035) & 14.399(048) & 15.721(077) & 15.222(074) & 15.001(015) \\
5 & 03:33:40.67 & $-$36:10:52.9 & 15.041(006) & 14.822(015) & 15.004(015) & 14.863(025) & 14.644(052) & 15.507(060) & 15.131(052) & 15.074(017) \\
6 & 03:33:57.36 & $-$36:09:24.0 & 16.654(012) & 16.505(053) & 16.608(028) & 16.421(067) & 16.620(288) & 17.028(228) & 16.909(365) & 16.600(052) \\
7 & 03:33:54.96 & $-$36:09:09.6 & 16.641(011) & 16.501(058) & 16.652(031) & 16.407(059) & 15.988(163) & 17.751(516) & 16.883(342) & 16.685(054) \\
8 & 03:33:46.83 & $-$36:03:04.1 & $\cdots$ & $\cdots$ & $\cdots$ & 13.848(018) & 13.676(026) & 14.380(137) & 14.051(027) & $\cdots$ \\
\enddata
\tablecomments{Values in parentheses are 1$\sigma$ measurement uncertainties in millimag.}
\end{deluxetable}

\clearpage
\begin{deluxetable}{cccccccccc}
\tablecolumns{10}
\tablewidth{0pt}
\setlength{\tabcolsep}{0.02in} 
\tabletypesize{\tiny}
\tablecaption{LSQ, TAROT, BOSS, and Slooh Natural System Photometry of SN~2012fr.\label{tab:sn_nat_sys}}
\tablehead{
\colhead{JD} &
\colhead{phase} &
\colhead{$gr_{LSQ}$}&
\colhead{$open_{TAROT}$}&
\colhead{$V_{TAROT}$}&
\colhead{$open_{BOSS}$}&
\colhead{$R_{Slooh}$}&
\colhead{$G_{Slooh}$}&
\colhead{$B_{Slooh}$}&
\colhead{$L_{Slooh}$} \\
\colhead{(days)}&
\colhead{(days)}&
\colhead{(mag)}&
\colhead{(mag)}&
\colhead{(mag)}&
\colhead{(mag)}&
\colhead{(mag)}&
\colhead{(mag)}&
\colhead{(mag)}&
\colhead{(mag)}}
\startdata
2456226.69 & $-16.41$ & 17.705(029) & $\cdots$ & $\cdots$ & $\cdots$ & $\cdots$ & $\cdots$ & $\cdots$ & $\cdots$  \\					
2456226.77 & $-16.33$ & 17.364(021) & $\cdots$ & $\cdots$ & $\cdots$ & $\cdots$ & $\cdots$ & $\cdots$ & $\cdots$  \\
2456227.55 & $-15.55$ & $\cdots$ & 15.940(061) & $\cdots$ & $\cdots$ & $\cdots$ & $\cdots$ & $\cdots$ & $\cdots$  \\
2456227.62 & $-15.48$ & $\cdots$ & $\cdots$ & $\cdots$ & $\cdots$ & 15.663(139) & 16.112(149) &16.426(274) & 16.238(0.064) \\
2456228.86 & $-14.24$ & $\cdots$ & $\cdots$ & $\cdots$ & 15.214(029) & $\cdots$ & $\cdots$ & $\cdots$ & $\cdots$  \\	
2456229.63 & $-13.47$ & $\cdots$ & $\cdots$ & 14.667(027) & $\cdots$ & $\cdots$ & $\cdots$ & $\cdots$ & $\cdots$  \\
2456229.67 & $-13.43$ & $\cdots$ & $\cdots$ & 14.573(021) & $\cdots$ & $\cdots$ & $\cdots$ & $\cdots$ & $\cdots$  \\
2456229.72 & $-13.38$ & $\cdots$ & $\cdots$ & 14.576(018) & $\cdots$ & $\cdots$ & $\cdots$ & $\cdots$ & $\cdots$  \\
\enddata
\tablecomments{The phase was computed relative to $t_{B_{max}}=\mathrm{JD}~2456243.1$.
Values in parentheses are 1$\sigma$ measurement uncertainties in millimag.}
\end{deluxetable}

\clearpage
\begin{deluxetable}{ccccc}
\tablecolumns{4}
\tablewidth{0pt}
\tabletypesize{\footnotesize}
	\tablecaption{Milky Way \& Host-Galaxy Extinction Corrections and AB Magnitude Offsets\label{tab:ext_ab}}
\tablehead{
\colhead{} &
\colhead{Effective} &
\colhead{Milky Way} &
\colhead{Host-Galaxy} &
\colhead{AB Magnitude} \\
\colhead{Filter} &
\colhead{Wavelength (\AA)} &
\colhead{$A_{\lambda}$ (mag)} &
\colhead{$A_{\lambda}$ (mag)} &
\colhead{Offset}
}
\startdata
	$uvm2$ & $  2360 $ & 0.140 &0.266& $+1.69$ \\
	$u$    & $  3550 $ & 0.087 &0.147& $-0.06$ \\
	$B$    & $  4300 $ & 0.074 &0.126& $-0.13$ \\
	$g$    & $  4700 $ & 0.068 &0.115& $-0.02$ \\
	$V$    & $  5400 $ & 0.056 &0.095& $-0.02$ \\
	$r$    & $  6200 $ & 0.047 &0.080& $-0.01$ \\
	$i$    & $  7500 $ & 0.035 &0.064& $+0.00$ \\
	$Y$    & $ 10000 $ & 0.020 &0.038& $+0.64$ \\
	$J$    & $ 12500 $ & 0.015 &0.026& $+0.90$ \\
	$H$    & $ 16600 $ & 0.009 &0.017& $+1.34$ \\
\enddata
\tablecomments{ For the $uvm2$-filter, the AB shift was taken from 
\url{https://swift.gsfc.nasa.gov/analysis/uvot_digest/zeropts.html}.
For the CSP-I $uBgVriYJH$ filters, the AB offsets are given in Table~16 of \citet{krisciunas17}.} 
\end{deluxetable}

\clearpage
\begin{deluxetable}{crcrrrrrrr}
\tablewidth{0pt}
\tabletypesize{\scriptsize}
\tablecaption{Bolometric Luminosity of SN~2012fr\label{tab:lum}}
\tablehead{
\colhead{JD} &
\colhead{Phase} &
\colhead{$L_1$(Total)}&
\colhead{$L_{UVOIR}$ }&
\colhead{$L_{uvm2}$} &
\colhead{$L_{(\lambda>\lambda_H)}$} & 
\colhead{$L_2$(Total)}&
\colhead{$L_{UVOIR}$ }&
\colhead{$L_{uvm2}$} &
\colhead{$L_{(\lambda>\lambda_H)}$} \\ 
\colhead{(days)} &
\colhead{(days)} &
\colhead{($10^{43}$erg/sec)}&
\colhead{($\%$)}&
\colhead{($\%$)} &
\colhead{($\%$)}&  
\colhead{($10^{43}$erg/sec)}&
\colhead{($\%$)}&
\colhead{($\%$)} &
\colhead{($\%$)}}  
\startdata
$2456230.70$ & $-12.40$  & $0.228(021)$ & $98.15$ & $0.30$ & $1.85$  & $0.229(022)$  &  $97.51$ &  $0.59$ &  $1.91$  \\ 	 
$2456231.75$ & $-11.35$  & $0.362(035)$ & $98.54$ & $0.30$ & $1.46$  & $0.368(037)$  &  $97.95$ &  $0.57$ &  $1.48$  \\ 	
$2456232.75$ & $-10.35$  & $0.512(050)$ & $98.77$ & $0.32$ & $1.23$  & $0.519(053)$  &  $98.15$ &  $0.59$ &  $1.26$  \\ 	
$2456233.70$ & $-09.40$  & $0.655(066)$ & $98.90$ & $0.35$ & $1.10$  & $0.661(068)$  &  $98.21$ &  $0.65$ &  $1.14$  \\ 	
$2456234.70$ & $-08.40$  & $0.809(082)$ & $99.00$ & $0.39$ & $1.00$  & $0.812(085)$  &  $98.24$ &  $0.73$ &  $1.03$  \\ 	
$2456235.67$ & $-07.43$  & $0.927(095)$ & $99.05$ & $0.44$ & $0.95$  & $0.945(100)$  &  $98.22$ &  $0.82$ &  $0.96$  \\ 	
$2456236.69$ & $-06.41$  & $1.036(107)$ & $99.09$ & $0.52$ & $0.91$  & $1.063(114)$  &  $98.17$ &  $0.92$ &  $0.91$  \\ 	
$2456237.60$ & $-05.50$  & $1.117(115)$ & $99.12$ & $0.58$ & $0.88$  & $1.139(122)$  &  $98.11$ &  $1.00$ &  $0.88$  \\
\enddata
\tablecomments{The phase is relative to $t_{B_{max}}=\mathrm{JD}~2456243.1$.
$L_{UVOIR}$ is the luminosity measured from the flux of 2012fr through $uBgVriYJH$-bands photometry.}
\end{deluxetable}

\end{document}